\documentclass[journal]{IEEEtran}
\IEEEoverridecommandlockouts

\newif\ifdraft
\drafttrue

\draftfalse

\usepackage[table]{xcolor}
\usepackage{balance}
\usepackage[hyphens]{url}
\usepackage{listings}

\usepackage{textcomp}
\usepackage{silence}

\newcommand{\ie}[0]{i.e.,\xspace}
\newcommand{\viz}[0]{viz.,\xspace}
\newcommand{\wrt}[0]{w.r.t.\xspace}
\newcommand{\cf}[0]{cf.\xspace}

\newcommand{\eg}[0]{e.g.,\xspace}

\definecolor{Gray}{gray}{0.9}
\definecolor{Orange}{HTML}{FFE6CC}

\newcommand{\gr}{\rowcolor[HTML]{EFEFEF}}
\newcommand{\gc}{\cellcolor[HTML]{EFEFEF}}
\newcommand{\rc}{\cellcolor{orange!20!white}}
\newcommand{\ec}{\cellcolor{yellow!20!white}}
\newcommand{\cc}{\cellcolor{red!20!white}}
\newcommand{\oc}{\cellcolor{blue!20!white}}
\newcommand{\yc}{\cellcolor{green!20!white}}

\newcommand{\rb}[1]{\colorbox{orange!20!white}{#1}}
\newcommand{\eb}[1]{\colorbox{yellow!20!white}{#1}}
\newcommand{\cb}[1]{\colorbox{red!20!white}{#1}}
\newcommand{\ob}[1]{\colorbox{blue!20!white}{#1}}
\newcommand{\yb}[1]{\colorbox{green!20!white}{#1}}
\newcommand{\mcr}[1]{
\begin{tabular}[c]{@{}c@{}}#1\end{tabular}}

\newcommand{\mcrl}[1]{
\begin{tabular}[c]{@{}l@{}}#1\end{tabular}}

\newcommand{\mcrc}[1]{
\begin{tabular}[c]{@{}c@{}}#1\end{tabular}}

\usepackage{comment}

\specialcomment{rimuovere}{}{}
\specialcomment{riscrivere}{\begingroup\color{orange}}{\endgroup}
\specialcomment{tnsm}{\begingroup\color{black}}{\endgroup}
\specialcomment{commentolungo}{\begingroup\color{gray}}{\endgroup}
\specialcomment{draft}{\begingroup\color{blue}}{\endgroup}
\specialcomment{leggibile}{\begingroup\color{teal}}{\endgroup}
\specialcomment{warning}{\begingroup\color{red}}{\endgroup}
\newcommand{\attenzione}[1]{{\color{purple}XXX #1 XXX}\xspace}
\newcommand{\commento}[1]{{\color{gray}#1}}
\newcommand{\surplus}[1]{\textcolor{gray}{#1}}

\newcommand{\inlineTODO}[1]{\textcolor{orange}{\textbf{TODO: #1}}}

\ifdraft
    \usepackage[]{hyperref} %

    \usepackage{todonotes}
    
\else 
    \usepackage[disable]{todonotes}
    
    \specialcomment{new}{}{}
	\specialcomment{warning}{}{}
    \specialcomment{draft}{}{}
	
	\excludecomment{riscrivere}
	\excludecomment{warning}
	\excludecomment{commentolungo}
	\excludecomment{rimuovere}
	\renewcommand{\attenzione}[1]{}
	\renewcommand{\commento}[1]{}
	\renewcommand{\surplus}[1]{}
	\renewcommand{\inlineTODO}[1]{}
\fi

\lstdefinestyle{customlog}{
    basicstyle=\fontfamily{zi4}\selectfont\scriptsize, %
    keywordstyle=\bfseries,
    morekeywords={Prompt, Summarize, the, errors, and, warnings, from, these, log, messages, and, identify, the, root, cause}, %
    frame=single,
    breaklines=true,
    captionpos=b,
}

\newcommand{\processWord}[1]{%
    \begingroup
    \def\processChar##1{%
        \ifx##1\relax
        \else
            \ifx##1-%
                \text{-}%
            \else
                \ifx##1é
                    \mathtt{\acute{e}}%
                \else
                    \ifx##1\'a
                        \acute{a}%
                    \else
                        \ifx##1\'i
                            \acute{i}%
                        \else
                            \ifx##1\'o
                                \acute{o}%
                            \else
                                \ifx##1\'u
                                    \acute{u}%
                                \else
                                    \ifx##1\`e
                                        \grave{e}%
                                    \else
                                        \ifx##1\`a
                                            \grave{a}%
                                        \else
                                            \ifx##1\`i
                                                \grave{i}%
                                            \else
                                                \ifx##1\`o
                                                    \grave{o}%
                                                \else
                                                    \ifx##1\`u
                                                        \grave{u}%
                                                    \else
                                                        \mathtt{##1}%
                                                    \fi
                                                \fi
                                            \fi
                                        \fi
                                    \fi
                                \fi
                            \fi
                        \fi
                    \fi
                \fi
            \fi
            \expandafter\processChar
        \fi
    }%
    \expandafter\processChar#1\relax
    \endgroup
}

\newcommand{\fmtTT}[1]{%
    \begingroup
    \def\split##1 ##2\relax{%
        \processWord{##1}%
        \ifx##2\empty%
        \else
            \:%
            \expandafter\split\expandafter##2\relax
        \fi
    }%
    $\split#1 \relax$%
    \endgroup
}

\usepackage{amssymb}
\usepackage{amsthm}

\usepackage{graphicx}
\usepackage{amsmath}
\usepackage{epsfig}
\usepackage{epstopdf}
\usepackage{paralist}

\usepackage{booktabs}
\newlength{\extraarrayvspace}

\usepackage[english]{babel}
\usepackage{comment}

\usepackage{tabularx}

\ifCLASSOPTIONcompsoc
 \usepackage[caption=false,font=normalsize,labelfont=sf,textfont=sf]{subfig}
\else
 \usepackage[caption=false,font=footnotesize]{subfig}
\fi

\usepackage[acronym,toc]{glossaries}
\newacronym{genai}{GenAI}{Generative Artificial Intelligence}
\newacronym{traffgen}{NTG}{Network Traffic Generation}
\newacronym{traffclass}{NTC}{Network Traffic Classification}
\newacronym{id}{ID}{Intrusion Detection}
\newacronym{nda}{NDA}{Network Digital Assistance for Documentation \& Operation}
\newacronym{nid}{NID}{Network Intrusion Detection}
\newacronym{nids}{NIDS}{Network Intrusion Detection System}
\newacronym{ide}{IDE}{Implicit Density Estimation}
\newacronym{ede}{EDE}{Explicit Density Estimation}

\newacronym{la}{LA}{Log Analysis}
\newacronym{nla}{NSLA}{Networked System Log Analysis}
\newacronym{lad}{LAD}{Log-based Anomaly Detection}

\newacronym{xai}{XAI}{eXplainable Artificial Intelligence}
\newacronym{dl}{DL}{Deep Learning}
\newacronym{ai}{AI}{Artificial Intelligence}
\newacronym{iat}{IAT}{Inter Arrival Time}
\newacronym{tcpws}{TCP WS}{TCP Window Size}
\newacronym{pl}{PL}{Payload Length}
\newacronym{ps}{PS}{Packet Size}
\newacronym{ttl}{TTL}{Time To Live}
\newacronym{to}{TO}{Traffic Object}
\newacronym{tp}{TP}{Traffic Prediction}
\newacronym{ac}{AC}{Attack Classification}
\newacronym{ad}{AD}{Anomaly Detection}
\newacronym{nad}{NAD}{Network Anomaly Detection}
\newacronym{tc}{TC}{Traffic Classification}
\newacronym{nta}{NTA}{Network Traffic Analysis}
\newacronym{cnn}{CNN}{Convolutional Neural Network}
\newacronym{ml}{ML}{Machine Learning}
\newacronym{ids}{IDS}{Intrusion Detection System}
\newacronym{shap}{SHAP}{SHapley Additive exPlanations}
\newacronym{eda}{EDA}{Exploratory Data Analysis}
\newacronym{ae}{AE}{AutoEncoder}
\newacronym{ece}{ECE}{Expected Calibration Error}
\newacronym{ls}{LS}{Label Smoothing}
\newacronym{fl}{FL}{Focal Loss}
\newacronym{lime}{LIME}{Local Interpretable Model-agnostic Explanations}
\newacronym{qos}{QoS}{Quality of Service}
\newacronym{dt}{DT}{Decision Tree}
\newacronym{md}{MD}{Misuse Detection}
\newacronym{bmd}{bMD}{Binary Misuse Detection}
\newacronym{mmd}{mMD}{Multi-class Misuse Detection}
\newacronym{can}{CAN}{Controller Area Network}
\newacronym{apt}{APT}{Advanced Persistent Threat}

\newacronym{ran}{RAN}{Radio Access Network}
\newacronym{iot}{IoT}{Internet of Things}
\newacronym{llm}{LLM}{Large Language Model}
\newacronym{lstm}{LSTM}{Long Short-Term Memory}
\newacronym{tsne}{t-SNE}{t-distributed Stochastic Neighbor Embedding}
\newacronym{dir}{DIR}{Packet Direction}
\newacronym{darpa}{DARPA}{Defense Advanced Research Projects Agency}
\newacronym{nlp}{NLP}{Natural Language Processing}
\newacronym{sni}{SNI}{Server Name Indication}
\newacronym{svm}{SVM}{Support Vector Machine}
\newacronym{lrp}{LRP}{Layer-wise Relevance Propagation}
\newacronym{gradcam}{Grad-CAM}{Gradient-weighted Class Activation Mapping}
\newacronym{ig}{IG}{Integrated Gradients}
\newacronym{tls}{TLS}{Transport Layer Security}
\newacronym{dnn}{DNN}{Deep Neural Network}
\newacronym{som}{SOM}{Self Organizing Map}
\newacronym{knn}{KNN}{K-Nearest Neighbors}
\newacronym{pfi}{PFI}{Permutation Feature Importance}
\newacronym{ice}{ICE}{Individual Conditional Expectation}
\newacronym{pdp}{PDP}{Partial Dependence Plot}
\newacronym{cem}{CEM}{Contrastive Explanation Method}
\newacronym{coin}{COIN}{Contextual Outlier INterpretation}
\newacronym{ciu}{CIU}{Contextual Importance and Utility}
\newacronym{brcg}{BRCG}{Boolean Decision Rules via Column Generation}
\newacronym{cade}{CADE}{Contrastive Autoencoder for Drifting detection and Explanation}
\newacronym{deeplift}{DeepLIFT}{Deep Learning Important FeaTures}
\newacronym{lemna}{LEMNA}{Local Explanation Method using Nonlinear Approximation}
\newacronym{sdn}{SDN}{Software Defined Networking}
\newacronym{fpga}{FPGA}{Field Programmable Gate Array}
\newacronym{ddos}{DDos}{Distributed Denial of Service}
\newacronym{nmm}{NMM}{Network Monitoring and Management}
\newacronym{api}{API}{Application Programming Interface}
\newacronym{ui}{UI}{User Interface}
\newacronym{cli}{CLI}{Command Line Interface}
\newacronym{rnn}{RNN}{Recurrent Neural Network}
\newacronym{gru}{GRU}{Gated Recurrent Unit}
\newacronym{mlm}{MLM}{Masked Language Model}
\newacronym{ddpm}{DDPM}{Denoising Diffusion Probabilistic Model}
\newacronym{ssm}{SSM}{State Space Model}
\newacronym{vit}{ViT}{Vision Transformer}

\newacronym{gan}{GAN}{Generative Adversarial Network}
\newacronym{vae}{VAE}{Variational Autoencoder}
\newacronym{gasf}{GASF}{Grammian Angular Summation Field}

\newacronym{jsd}{JSD}{Jensen-Shannon Divergence}
\newacronym{tvd}{TVD}{Total Variation Distance}
\newacronym{sr}{SR}{Success Rate}

\newacronym{rfc}{RFC}{Request for Comment}
\newacronym{cpsa}{CPSA}{Cryptographic Protocol Shapes Analyzer}
\newacronym{nf}{NF}{Network Function}
\newacronym{slm}{SLM}{Small Language Model}
\newacronym{gpp}{3GPP}{Third Generation Partnership Project}
\newacronym{rag}{RAG}{Retrieval-Augmented Generation}

\newacronym{fed}{FED}{Full Encoder-Decoder}
\newacronym{eo}{EO}{Encoder-Only}
\newacronym{do}{DO}{Decoder-Only}
\newacronym{sdp}{SDP}{Sequential Denoising Process}

\newacronym{mlp}{MLP}{MultiLayer Perceptron}

\newacronym{peft}{PEFT}{Parameter-Efficient Fine-Tuning}
\newacronym{ptq}{PTQ}{Post-Training Quantization}
\newacronym{lora}{LoRA}{Low-Rank Adaptation}
\newacronym{gguf}{GGUF}{GPT-Generated Unified Format}
\newacronym{nice}{NICE}{Non-linear Independent Components Estimation}
\newacronym{gpu}{GPU}{Graphics Processing Unit}

\newacronym{crps}{CRPS}{Continuous Ranked Probability Score}
\newacronym{fpr}{FPR}{False Positive Rate}

\newlength\maxlength
\newlength\thislength

\newglossarystyle{mystyle}
{%
  {%
    \setlength{\maxlength}{0.5cm}%
    \forglsentries[\currentglossary]{\thislabel}%
    {%
       \settowidth{\thislength}{\glsentryshort{\thislabel}}%
       \ifdim\thislength>\maxlength
         \setlength{\maxlength}{\thislength}%
       \fi
    }%
    \settowidth{\thislength}{\hspace{1em}\hspace{3.5em}}%
    \setlength{\glsdescwidth}{\linewidth-\maxlength-\thislength-2\tabcolsep}%
    \begin{table}
    \centering
    \small
    \caption{List of Acronyms and Abbreviations in alphabetic order.}
    \label{tab:acronyms}
    \begin{tabular}{l@{\hspace{1.em}\hspace{2em}}p{\glsdescwidth}}
    \toprule \rowcolor{Orange}
    \multicolumn{1}{l}{\textbf{Acronym}} &
    \multicolumn{1}{@{}l}{\textbf{Definition}}\\%
    \midrule
  }%
  {%
     \bottomrule
     \end{tabular}%
     \end{table}
  }%
  \renewcommand*{\glsgroupheading}[1]{}%
}

\usepackage[T1]{fontenc} 
\usepackage[utf8]{inputenc}
\usepackage{mathabx}
\usepackage{setspace}
\usepackage[square,sort&compress,comma,numbers]{natbib} %
\usepackage{footmisc}  
\usepackage{multirow}
\usepackage{bm}
\usepackage{xspace} %
\usepackage[10pt]{moresize}
\usepackage{balance}

\usepackage{breqn}
\usepackage{units}
\usepackage[inline]{enumitem}
\usepackage{stackengine}
\usepackage{wasysym} %
\usepackage{threeparttable}
\usepackage{fontawesome5}
\usepackage{svg}
\usepackage{longtable}
\usepackage{hyperref}

\hypersetup{
    colorlinks = true, %
    linkcolor = black,  %
    citecolor = black,  %
    filecolor = black,  %
    urlcolor = black,   %
    pdfborder = {0 0 0} %
}

\usepackage{color, colortbl}
\definecolor{ferrarired}{rgb}{1.0, 0.11, 0.0}
\definecolor{royalblue}{rgb}{0.26, 0.41, 1}
\definecolor{darkpastelgreen}{rgb}{0.01, 0.75, 0.24}
\definecolor{redme}{rgb}{0.78, 0.0, 0.224}
\definecolor{bluefl}{rgb}{0.01, 0.01, 0.44}
\definecolor{bluefl}{rgb}{0.01, 0.01, 0.44}
\definecolor{greenls}{rgb}{0.133, 0.55, 0.133}
\definecolor{lightgray}{gray}{0.9}

\usepackage{threeparttable}

\usepackage{amssymb}
\usepackage{pdfrender}

\usepackage{makecell}
\usepackage[most]{tcolorbox}

\ErrorFilter{latex}{Invalid UTF-8 byte}

\makeatletter

\newcommand*{\@rowstyle}{}
\newcommand*{\rowstyle}[1]{%
 \gdef\@rowstyle{#1}%
 \@rowstyle\ignorespaces%
}
\newcolumntype{=}{%
>{\gdef\@rowstyle{}}%
}
\newcolumntype{+}{%
>{\@rowstyle}%
}
\newcolumntype{C}[1]{>{\centering\arraybackslash}p{#1}}

\makeatother

\begin{document}

\title{Mapping the Landscape of Generative AI in \\ Network Monitoring and Management}

\author{Giampaolo~Bovenzi,
        Francesco~Cerasuolo,
        Domenico Ciuonzo,~\IEEEmembership{Senior Member,~IEEE}, \\
        Davide Di Monda,
        Idio Guarino,
        Antonio Montieri,
        Valerio Persico,
        Antonio Pescap\'e,~\IEEEmembership{Senior Member,~IEEE}
\thanks{G. Bovenzi, F. Cerasuolo, D. Ciuonzo, A. Montieri, V. Persico, and A. Pescap\'e are with the Department of Electrical Engineering and Information Technologies (DIETI) at the University of Naples Federico II, Italy (name.surname@unina.it).\\
D. Di Monda is both with DIETI and IMT School for Advanced Studies, Lucca, Italy (davide.dimonda@imtlucca.it). \\
I. Guarino is with the Department of Computer Science at the University of Verona, Italy (idio.guarino@univr.it).
}
}

\markboth{IEEE Transactions on Network and Service Management,~Vol.~XX, No.~X, XXXX~2025}%
{Bovenzi \MakeLowercase{\textit{et al.}}: Mapping the Landscape of Generative AI in Network Monitoring and Management}

\maketitle

\begin{abstract}
\gls{genai} models such as LLMs, GPTs, and Diffusion Models have recently gained widespread attention
from both the research and the industrial communities. 
This survey explores their application in network monitoring and management, focusing on 
prominent
use cases, as well as challenges and opportunities. 
We discuss how 
\emph{network traffic generation and classification, network intrusion detection, networked system log analysis, and network digital assistance}
can benefit from the use of \gls{genai} models.
Additionally, we provide an overview of the available \gls{genai} models, datasets for large-scale training phases, and platforms for the development of such models.
Finally, we discuss research directions that potentially mitigate the roadblocks to the adoption of \gls{genai} for network monitoring and management.
Our investigation aims to map the current landscape and pave the way for future research in leveraging  \gls{genai} for network monitoring and management. %

\end{abstract}

\begin{IEEEkeywords}
Generative AI, Networking, LLM, GPT, Diffusion Models, Traffic Classification, Intrusion Detection.
\end{IEEEkeywords}

\IEEEpeerreviewmaketitle

\glsresetall

\section{Introduction}
\label{sec:introduction}

\IEEEPARstart{B}{ecause} of breakthroughs achieved in the last decade, \emph{\gls{genai}} stands as one of the most important stepping stones toward the intelligence era.
At its core, \gls{genai} excels in ($i$) distilling features of complex data distributions (uncovering intricate patterns) and ($ii$) utilizing these features to generate new, similar, yet distinct data.
This contrasts with the usual \emph{discriminative \gls{ai}} models that focus
on analyzing, interpreting, and classifying data to solve specific inference tasks.
This two-fold ability (\ie complex analysis and generation) positions \gls{genai} as a crucial technology in advancing both scientific research and industrial applications.
Accordingly, \gls{genai} supports tools designed to generate new content---text, images, videos, and more---based on patterns and information learned from large datasets. 

At a higher abstraction level, such capabilities showcase \gls{genai} as a powerful tool to solve intelligence-level tasks that are common to different domains: content generation, data augmentation,  conversational agents and question-answering tools, human-machine interactions, and automation.
Noteworthy examples of novel \gls{genai} models are represented by \glspl{llm}, Diffusion Models, and \glspl{ssm}.
To specify, \glspl{llm} are language models built on the \fmtTT{Transformer} architecture, and they are referred to as ``large'' due to their vast number of parameters.
Hereinafter, we use the terms ``\gls{llm}'' and ``\fmtTT{Transformer}'' synonymously to indicate the AI model~\cite{yang2024harnessing}.
Notable examples for these novel \gls{genai} solutions are represented by \texttt{GPT} and \texttt{LLaMA} for \gls{llm}, \texttt{DALL$\cdot$E} and \texttt{Stable Diffusion} for Diffusion Models, and \texttt{Mamba} for \gls{ssm}.
These models
have demonstrated significant commercial value and technical potential. 
They 
show notable reasoning, generalization, and emergent abilities in different applications, like text-to-text, text-to-image, and text-to-code.
As a consequence of such potential, the global \gls{genai} market stood at just under $45$ billion USD at the end of $2023$ (doubling its value compared to $2022$), and forecasts indicate an impressive growth of $\approx 20$ billion USD per-year through $2030$~\cite{statista1}.

The rapid development of \gls{genai} has been fueled by \textbf{three main drivers}: 
($i$) the availability of \emph{large-scale data corpora}; 
($ii$) methodological advancements in the \gls{ai} field, \ie the shift toward \emph{deep and foundational generative models}; 
($iii$) technological innovations supporting model building, \ie \emph{high-performance massive \glspl{gpu}}.
Notably, despite these drivers, only a few global stakeholders (to date) are capable of training \gls{genai} models from scratch. 
Hence, pre-trained large models are beginning to be shared by the open-source part of the \gls{genai} community.%
\footnote{See \eg \url{https://llama.meta.com/}.}

On the other side, recent networking research has focused on using \gls{dl} to develop efficient tools for \emph{\gls{nmm}} to meet modern Internet traffic needs.
In this respect, \gls{genai} can empower intent-based and autonomous networks by automating the translation of user objectives into actionable network policies~\cite{huang2024digital}.
This allows networks to self-configure, self-optimize, and self-heal, improving responsiveness and resilience. 
By leveraging \gls{genai}'s predictive capabilities, networks can indeed anticipate traffic patterns and issues, ensuring seamless operation. 
This reduces manual management complexity, accelerates innovation, and enhances user experience in a dynamically changing digital landscape. 
However, the full utilization of \gls{genai} for \gls{nmm} requires shifting from common text, audio, and image generation to network-focused synthetic content---%
fulfilling the concept of ``\gls{ai}-generated everything''~\cite{du2024age}.
Despite the interest in integrating \gls{genai} into networks and the Internet (trying to echo similar breakthroughs obtained in verticals such as computer vision or \gls{nlp}) to date, general deployment issues~\cite{lavin2022technology} and unique networking challenges remain~\cite{rossi2022landing}.

\subsection{Contributions and Survey Organization}

\begin{figure}[t]
    \centering
    \includegraphics[trim=0 40 0 40, clip, width=0.90\columnwidth]{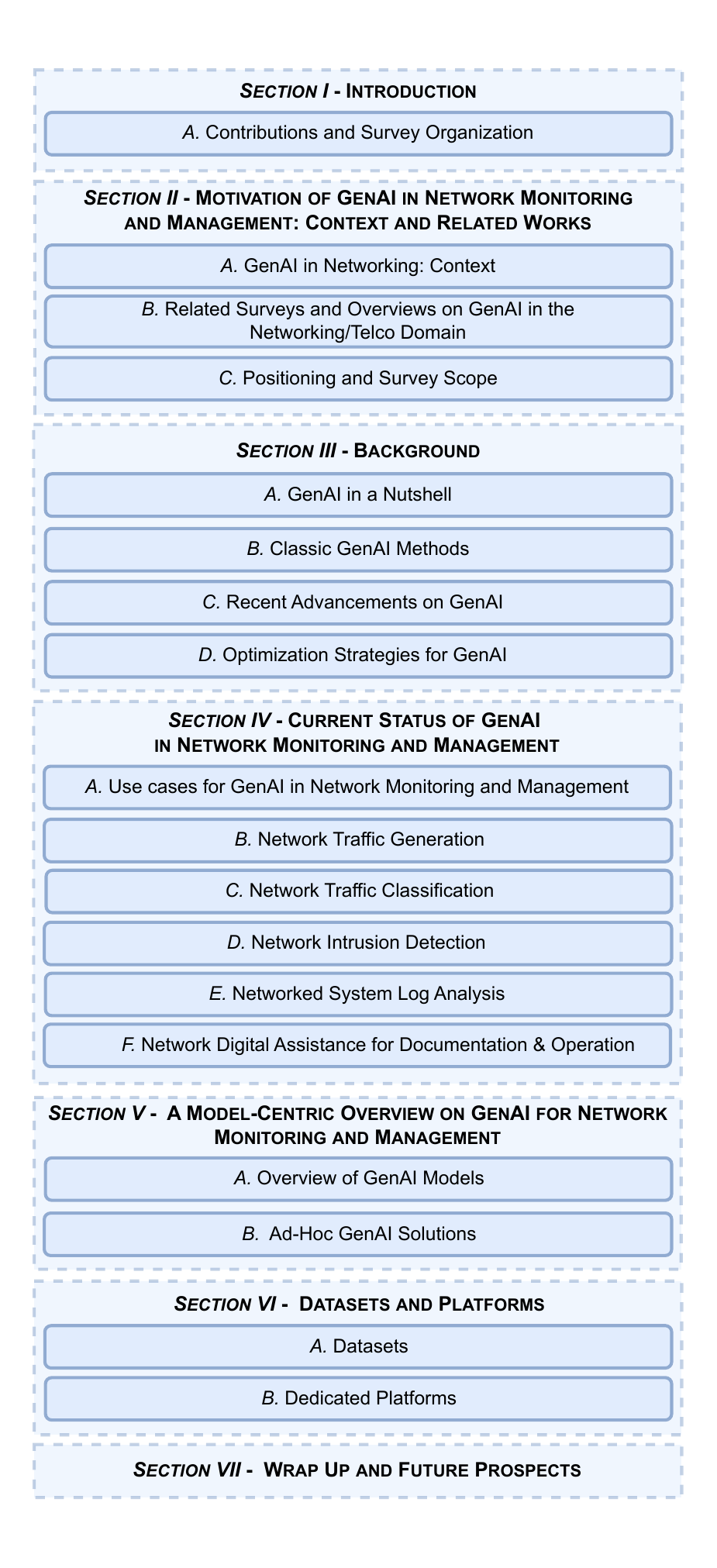}
    \caption{
    Survey organization.
    }
    \label{fig: survey_organization}
\end{figure}

This article deepens the technical understanding of \gls{genai} within the context of \gls{nmm}. 
Accordingly, the \textbf{main contributions} provided by this manuscript can be summarized as follows:

\begin{itemize}

\item we discuss the \textbf{motivation behind our ``\gls{genai} landscape mapping''} effort in the field of \gls{nmm}, highlighting the shared interest in \gls{genai} from different stakeholders, as well as the gap in the (quickly-evolving) scientific literature we aim to fill with our work (Sec.~\ref{sec:motivation_rw});

\item 
we 
present
\textbf{a categorization of novel \gls{genai} methods}, 
offering the necessary background to help readers understand the distinctive aspects of \gls{nmm}-specific research efforts and applications (Sec.~\ref{sec:background});

\item we offer a \textbf{use-case-centric viewpoint}, discussing each \textbf{practical \gls{nmm} use case and its interplay with \gls{genai}} (Sec.~\ref{sec:genai_landscape}), along with a \textbf{model-centric viewpoint} (Sec.~\ref{sec:genai_model_overview}) to obtain a nuanced perspective. 
In addition, for the newly-branded \gls{genai} solutions, we detail the proposed modifications to reference \gls{genai} architectures and their code availability.

\item we provide a comprehensive view of
the \textbf{public datasets} leveraged for \gls{genai} model lifecycle and 
the
\textbf{available computing platforms} that can support and accelerate the design of novel \gls{genai}-based \gls{nmm} solutions (Sec.~\ref{sec:dataset_code_platforms});

\item finally, 
we briefly wrap-up the current \gls{genai} \textbf{limitations} and identify potential methodological/technological \textbf{enablers} for deploying it safely and at scale in the \gls{nmm} field (Sec.~\ref{sec:future directions}).
\end{itemize}
Figure~\ref{fig: survey_organization} outlines the organization of the present survey, sketching the details of the sections constituting the manuscript.

\section{Motivation of GenAI in Network Monitoring and Management: Context and Related Works}\label{sec:motivation_rw}
\newcommand{\faLaw}{{\includesvg[height=1em]{figs/icons/law}}}
\newcommand{\faAca}{{\includesvg[height=1em]{figs/icons/academia}}}
\newcommand{\faInd}{{\includesvg[height=1em]{figs/icons/industry}}}
\newcommand{\faOrg}{{\includesvg[height=1em]{figs/icons/org}}}

\setlength{\extraarrayvspace}{-0.45ex}

\begin{table}[t]
\renewcommand{\arraystretch}{1.2}
\centering
\small
\caption{Main efforts of the stakeholders in the context of GenAI for NMM.}

\label{tab:stake}
\resizebox{\columnwidth}{!}
{
\begin{threeparttable}
\begin{tabular}{lcm{5.5cm}c}
\toprule

\textbf{Stakeholder} & & \textbf{Effort} & \textbf{Ref.} \\
\midrule

ACM SIGCOMM & \faAca & 
Scientific talks on LLMs for networking
& \cite{networkingchannel}
\\

\gr
AT\&T & \faInd & 
Release of \textit{Ask AT\&T} a GenAI digital assistant for employees and users
& \cite{att}
\\

Cisco & \faInd & 
Release of a GenAI digital assistant as support to human decision-making
& \cite{cisco}
\\

\gr
Ericsson & \faInd & 
Recognizing that GenAI will replace the traditional search process with a more intuitive and conversational experience
& \cite{ericsson}
\\

EU &\faLaw & 
EIC Accelerator for developing GenAI, with a focus on transparency and smaller models
& \cite{eic}
\\

\gr
Huawei & \faInd & 
Release of \textit{Net Master}, an LLM for network operations and maintenance tasks
& \cite{huawei}
\\

IEEE ComSoc & \faAca &
Platform for academia and industry for researching on GenAI for networking
& \cite{genainet}
\\

\gr
IETF & \faOrg &
Scientific talks on LLMs for networking GenAI
& \cite{ietf}
\\

ITU & \faOrg & 
Showcase industry- and academic-viewpoints on LLMs
& \cite{aiforgood}
\\

\gr
Nokia & \faInd & 
Recognizing the advantages of GenAI and identifying various networking use cases for its application
& \cite{nokia}
\\

Telefonica & \faInd & 
Partnership with Microsoft to integrate GenAI in its networking ecosystem 
& \cite{telefonica}
\\

\gr
TIM & \faInd & 
Integration of GenAI to support customer service and technical operations
& \cite{tim}
\\

\bottomrule

\end{tabular}
\begin{tablenotes}[flushleft]
\footnotesize
\item \faAca:~Academic Organization, \faOrg:~Organization of Multiple Entities,  \faLaw:~Government Body/Agency, \faInd:~Company. 
\end{tablenotes}
\end{threeparttable}
}
\end{table}
\setlength{\extraarrayvspace}{-0.45ex}

\begin{table*}[tp]
    \renewcommand{\arraystretch}{1.2}
    \centering
    \footnotesize
    \caption{Surveys and overviews on Generative Artificial Intelligence in related fields.
    }
    \label{tab:other_surveys}
    \resizebox{\textwidth}{!}{
    \begin{threeparttable}
    
    \begin{tabular}{rcllcc}
    \toprule
    \textbf{Work} & \textbf{Year} & \textbf{Focus} & \mcrl{\textbf{Vertical Application Fields, Use Cases,}\\ \textbf{and Networking Tasks}} & \textbf{AI Tools} & \mcrc{\textbf{\#Surveyed}\\ \textbf{Works}}\\
    \midrule
        
    \gr \citet{sai2024empowering} & 2024 & IoT (data generation) & 
    \mcrl{
    Synthetic Sensor Data\\ 
    Personalized Device Response\\
    Autonomous Control\\
    Cyber-threat Detection\\
    Predictive Maintenance\\
    Data Anonymization
    }& GANs, VAEs, LLMs & $15$\\
        
    Hassanin et al.~\cite{hassanin2024comprehensive} & 2024 & Cyber Defense & 
    \mcrl{
    Threat Intelligence\\
    Vulnerability Assessment\\
    Network Security\\
    Privacy Preservation\\
    Operations Automation}&LLMs &$149$\\
        
    \gr \citet{alwahedi2024ml} &2024 &IoT Security & \mcrl{Cyber-threat Detection\\Lightweight Encryption Optimization\\Enhancing Access Control\\Identifying Vulnerability\\Automating and Enhancing Penetration Testing}
    &ML$\dag$ &$62^*$\\

    \citet{halvorsen2024applying} &2024 &Intrusion Detection &\mcrl{Penetration Testing\\Supplementing Datasets\\Intrusion Detection Model Development} & GANs, VAEs, LLMs &$129$\\
        
\gr \citet{zhou2024large} & $2024$ &Telecommunications &
        \mcrl{
        Telecom-Domain Question Answering\\
        Troubleshooting Reports Generation\\
        Project Coding\\
        Network Configuration\\
        Network Attack Classification and Detection\\
        Telecom Text, Image, and Traffic Classification\\
        Performance Optimization\\
        Channel State Information Prediction\\
        Prediction-based Beamforming\\
        Traffic Load Prediction
        }
        & LLMs &$207$\\

        Celik et al.~\cite{celik2024dawn}& 
        2024&
        \mcrl{Wireless\\ telecommunication\\networks}
        &
        \mcrl{
        Physical Layer Design\\
        Network Organization \& Management\\
        Cross-layer Network Security\\
        Network Traffic Analytics\\
        Localization \& Positioning
        }& GANs, VAEs& $303$\\
        
        \gr 
        \citet{karapantelakis2024generative}&2024&\mcrl{Mobile\\telecommunication\\networks}&
        \mcrl{
        Improving aspects in RANs\\
        Mobile-network Management\\
        Requirements Engineering
        }& GANs, VAEs, LLMs& $117$\\
        
        \citet{huang2024digital} &2024 &Networking & 
        \mcrl{
            Threat Intelligence\\
            Vulnerability Assessment\\
            Network Security\\
            Privacy Preservation\\
            Operations Automation}& LLMs & $15$\\

        \gr\citet{liu2024large} &2024 &Networking & 
        \mcrl{Network Design\\
            Network Diagnosis\\
            Network Configuration\\
            Network Security} &LLMs &$15$\\

        \citet{huang2023large} &2023 &Networking & 
        \mcrl{Network Design\\
            Network Diagnosis\\
            Network Configuration\\
            Network Security}&LLMs &$15$\\
        
        \gr \citet{chaccour2024telecom} &2024 &Telecommunications & 
        \mcrl{
            Network Operations (i.e., log analysis)\\
            Simplified Network Interfaces (i.e., APIs generation)\\
            Synthetic Data for Digital Twins\\
            DevOps and Software Lifecycle Management
            }&GANs, LLMs&$15$\\
        
        \midrule\midrule
        
        \textit{This work} &2024 &\mcrl{Network Monitoring\\and Management} & 
        \mcrl{
        Traffic Generation\\
        Traffic Classification\\
        Intrusion Detection\\
        Log Analysis\\
        Network Digital Assistance
        }
        & LLMs, Diffusion Models, SSMs &$189$\\
        
        \bottomrule
    \end{tabular}
    \begin{tablenotes}[flushleft]
    \footnotesize
    \item \textbf{*}: only a limited number of references is related to LLMs; \bm{$\dag$}: LLMs only as a future trend.
    \end{tablenotes}
    \end{threeparttable}
    }
\end{table*}

In this section, we examine the increasing interest from both public and private stakeholders in using \gls{genai} to support \gls{nmm} processes (Sec.~\ref{subsec:genai_context}). Next, we discuss related surveys that analyze the impact of \gls{genai} methods in the networking domain \ref{subsec:related}. Finally, we outline the positioning and scope of this survey
(Sec.~\ref{subsec:positioning}).

\subsection{GenAI in Networking: Context}
\label{subsec:genai_context}
The huge and general interest in \gls{genai} solutions also maps to the networking domain, where recent initiatives reflect the endeavors of several private and public stakeholders.
Table~\ref{tab:stake} provides an overview of this interest, reporting the efforts of different stakeholders in the context of \gls{genai} for \gls{nmm}.

For instance, the current interest in \gls{genai} is witnessed by the recent establishment of IEEE ComSoc Emerging Technology Initiative on Large Generative \gls{ai} Models in Telecom (GenAINet)~\cite{genainet}.
ACM 
SIGCOMM
has already featured several online talks
in which experts have discussed the huge interest in the application of \gls{genai} to \gls{nmm} (and, in general, to networking)~\cite{networkingchannel}.
Similarly, the International Telecommunication Union (ITU) via its initiative ``AI for Good'' is showcasing both industry- and academic-oriented viewpoints, as well as first \gls{llm}-based challenges~\cite{aiforgood}.
Trending interest is also observed at the IETF, with a first side meeting
entirely dedicated to the use of \glspl{llm} in networking~\cite{ietf}.
It is further witnessed by the latest academic networking conferences and workshops that stably include the application of \gls{genai} among the topics of their call, consistently seeking contributions in this direction (\eg IEEE GLOBECOM 2024 will feature both dedicated workshops and symposia centered on \gls{genai}).%
\footnote{\url{https://globecom2024.ieee-globecom.org/call-papers}}

At the governmental level, the EU has launched a strategy for developing \gls{genai} models over the past two years, highlighted by the EIC Accelerator funding program under the Horizon Europe framework aimed at supporting start-ups and small-medium enterprises~\cite{eic}.
Specifically, one of the $2024$ challenges,
``Human Centric Generative \gls{ai} made in Europe'' ($50$ million EUR budget), aims to promote a European human-centric approach to \gls{genai}, addressing issues like transparency and trust, and seeking to 
($i$)~advance foundation language and multimodal frontier models, while also focusing on 
($ii$)~smaller foundation models with high performance in \emph{specific domains}---%
like the case of this paper.

Network providers have also attempted to capitalize on the benefits of \gls{genai}.
For instance, Bell Labs acknowledges the benefits of \gls{genai}, 
classifies several use cases
(in areas such as customer care operations, network design, network performance and optimization, and testing), 
and envisions their expected role in shaping the future of organizations and functions of Telecom service providers~\cite{nokia}.
Huawei has recently launched \emph{Net Master}~\cite{huawei},
an innovative network large model
powered by \gls{genai} 
that aims to enhance the efficiency of network operations and maintenance.
This solution is trained using Huawei's Pangu models
(\ie different foundation models tailored to different domains or specific use cases) 
and it is based on a $50$-billion-level corpus and the experience of more than $10$k networking experts. 
According to Ericsson,
in the Telecom domain, the integration of \gls{genai} capability to convert natural language to SQL and to execute complex SQL queries enables seamless interaction between users and data,
replacing the traditional search process with a more intuitive and conversational experience~\cite{ericsson}.
This capability allows Telecom companies to empower users to effortlessly access and analyze data, easily supporting data-driven decisions.
Telefonica has partnered with Microsoft to integrate Azure 
\gls{genai} in its digital ecosystem, enhancing its capabilities for key workflows, 
such as customer identity management or access to network \glspl{api}~\cite{telefonica}.
Similarly, AT\&T has launched \emph{Ask AT\&T} a \gls{genai} tool based on OpenAI’s ChatGPT, integrated within a secure AT\&T-dedicated Azure environment~\cite{att}.
This tool aims to enhance employees' productivity by translating documents, optimizing network operations, updating legacy software, and improving customer support. 
On the same line,
Cisco proposed its \gls{ai} Assistant for accessing data at large scale 
to guide and inform human decision-making and enhance productivity while guaranteeing data protection and privacy~\cite{cisco}.
Lastly, TIM
is exploring the integration of \gls{genai} across various sectors (\eg marketing, customer care, network operations).
The aim is to support customer service and technical operations through conversational interfaces,
enhance document search and summarization, assist in code generation for IT tasks, 
and improve data analysis with natural language queries~\cite{tim}.

\subsection{Related Surveys and Overviews on GenAI in the Networking/Telco Domain}
\label{subsec:related}

Given the enormous hype 
surrounding \gls{genai} techniques, 
a large number of recent surveys and tutorial-style studies aim to analyze and discuss their impact within the \emph{wide domain of networking}.
These works contribute to defining a rich but equally fragmented picture.
Indeed, the available studies are characterized by different focuses, scopes, and depths in the provided pictures of the state of the art. 
Thus, they result in identifying 
($i$)~different vertical application fields, use cases, and networking tasks that can benefit from the (rapid) progress in \gls{genai}, 
as well as ($ii$)~different families of \gls{ai} tools.
Such studies and related aspects are summarized in Tab.~\ref{tab:other_surveys} and briefly discussed in the following.

The majority of the works aim to analyze the role of \gls{genai} in the fields of the \textbf{\gls{iot}} and/or \textbf{cybersecurity}~\cite{sai2024empowering, ferrag2024generative, hassanin2024comprehensive, halvorsen2024applying, alwahedi2024ml}.
For instance, \citet{sai2024empowering}
explore the potential of combining \gls{genai} with \emph{\gls{iot}}, which enables the generation of synthetic data that can be used to train \gls{dl} models to overcome data insufficiency or incompleteness in \gls{iot} systems.
\citet{ferrag2024generative} provide a comprehensive survey of \glspl{llm} for \emph{cybersecurity} identifying $9$ application fields:
threat detection and analysis, 
phishing detection and response, 
incident response, 
security automation, 
cyber forensics, 
chatbots, 
penetration testing, 
security protocol verification, and 
security training and awareness.
Although the authors offer an in-depth analysis of the potentiality of \glspl{llm} for cybersecurity,
the potential application fields they identify
are not fully centered on networking
and do not consider several promising \gls{llm} applications in this domain.
\citet{hassanin2024comprehensive}
overview the recent progress of \glspl{llm} in \emph{cyber defense},
considering verticals that include threat intelligence, vulnerability assessment, network security, privacy preservation, and operations automation.
Moreover, \citet{halvorsen2024applying} explore the application of \gls{genai} for \emph{intrusion detection} and
discuss how \gls{genai} can support 
penetration testing, supplementing datasets, or developing 
detection models.
They claim that both the training and test phases of intrusion systems benefit from \gls{genai}.
\citet{alwahedi2024ml} aim at providing a comprehensive overview of applying \gls{ml} techniques for \emph{\gls{iot} security}.
In their future vision, the authors 
introduce the contribution of
\gls{genai} and \glspl{llm} to enhance \gls{iot} security---\eg optimization of cyber threat detection, lightweight encryption, access control, vulnerability identification, and automated penetration testing.
\emph{Unfortunately, we underline that the \gls{genai} applications and use cases discussed in the above surveys usually are not corroborated by existing state-of-the-art works.}

To the best of our knowledge, only a limited number of works~\cite{huang2023large, celik2024dawn, liu2024large, zhou2024large} aim at providing a \textbf{broader perspective} of \gls{genai} in the \textbf{networking/telecommunication field}.
In detail,
\citet{huang2023large} 
propose \emph{ChatNet}, a domain-adapted network \gls{llm} framework with access to various external network tools.
The authors discuss how \glspl{llm} promise to unify network intelligence through \emph{natural language interfaces}. 
Specifically, they remark that domain adaptation of \glspl{llm} is paramount to fill the gap between natural language and network language
and identify pre-training, fine-tuning, inference, and prompt engineering as the main enabling techniques.
\citet{liu2024large}
provide a more condensed overview of the recent advances of \glspl{llm} in networking and 
present an abstract workflow to describe the fundamental process involved in applying \gls{llm} in such a domain, including
task definition, data representation, prompt engineering, model evolution, tool integration, and validation.
Interestingly, they 
remark that \emph{network-specific \glspl{llm}} are expected to be more effective
than using \glspl{llm} originally designed for general domains to perform network-related tasks.
The works in~\cite{huang2023large} and~\cite{liu2024large} both identify
network design, 
diagnosis,
configuration, and
security
as the main vertical fields in networking  
impacted by \glspl{llm}.
On the other hand,
\citet{zhou2024large}
survey fundamentals, key techniques, and applications of \gls{llm}-enabled telecommunication networks.
Specifically, they focus on four telecommunication scenarios:
\begin{enumerate*}[label=(\emph{\roman*})]
    \item generation problems, \ie answering telecommunication-domain questions and generating troubleshooting reports, project coding, and network configuration;
    \item classification problems, \ie network-attack, telecommunication-text, image, and traffic classification;
    \item network-performance optimization, \ie automated reward function design to improve reinforcement learning applications; and 
    \item prediction problems, \ie prediction of channel state information and traffic load, and prediction-based beamforming.
\end{enumerate*}
\citet{karapantelakis2024generative} 
focus on \gls{genai} for \emph{mobile telecommunication networks} 
and consider applications lying in verticals, 
such as optimizations in \glspl{ran}, network management, and requirements engineering.
From the perspective of 
telco operations, 
\citet{chaccour2024telecom} 
identify \gls{genai} as a key to improving network operations such as predictive maintenance and real-time optimization.
They discuss use cases of \glspl{llm} and \gls{genai} for telco,
including: 
($i$) customer incident and trouble report management, proactive network management and repair, digital twin for network management, and intelligent network alert correlation---associated with \glspl{llm}; 
($ii$) generating customized network configurations, creating dynamic service descriptions, and proactive fault prediction and resolution---associated with \gls{genai} solutions beyond \glspl{llm}, \ie those performing content creation.
Finally, \citet{celik2024dawn} focus on applying \gls{genai} models within the domain of \emph{wireless communications}. The authors provide a tutorial on \gls{genai} models and a survey on their application across various wireless research areas, including: 
\begin{enumerate*}[label=(\emph{\roman*})]
    \item physical layer design,
    \item network organization and management,
    \item network traffic analytics,
    \item cross-layer network security, and
    \item localization and positioning.
\end{enumerate*}
Specifically, in the domain of network traffic analytics, the authors focus on use cases such as network traffic generation, encrypted traffic classification, traffic prediction, and traffic morphing. 
In contrast, the exploration of \gls{genai} networks security models is limited, with only a small portion addressing the enhancement of \glspl{nids}, particularly in terms of improving their robustness.
\emph{As a final remark, we note that all the works surveyed in the areas of
networking and telecommunications
mainly utilize \glspl{gan}, hence they only marginally cover the latest advancements involving more sophisticated techniques such as \glspl{llm}, Diffusion Models, and \glspl{ssm}.}

For the sake of completeness, we mention that some works~\cite{xu2024unleashing, wang2024toward} deepen \emph{how the network is expected to support \gls{genai} applications}, \eg with focus on cloud-edge-mobile infrastructure and security \& privacy concerns.
We do not consider such research paths in our study but rather consider the opposite point of view, investigating \emph{how \gls{genai} can support network-related tasks}.

\subsection{Positioning and Survey Scope}
\label{subsec:positioning}

In light of the rich but scattered literature scenario,
we position the present work against the existing
surveys and overviews in terms of the scope of the applications and tools considered,
as well as the provided outcomes of the analyses.

To the best of our knowledge, none of the considered studies surveying the impact of recent advancements in \gls{genai} primarily focuses on network monitoring and management.
In fact, the studies that are primarily centered on networking~\cite{huang2023large, huang2024digital, liu2024large} share a focus that is slightly close to ours.
However, while envisioning the great potential of \gls{genai} in networking, they lack a detailed survey and taxonomization of the current landscape, being aimed at \emph{providing only a general overview based on the analysis of a very limited number of works} (indeed, these studies reference $15$ papers each in their bibliography).
On the other hand, the studies that provide a more systematic and in-depth analysis of the literature~\cite{zhou2024large, karapantelakis2024generative} emphasize different facets of the communication networks, being oriented at capturing telecommunication aspects placed at lower layers in the communication stack---\eg \gls{ran} improvement, mobile-network management, channel state information prediction, prediction-based beamforming.
Hence, we believe \emph{they provide a view that is complementary to ours}.

In this survey,
we explore $5$ use cases:
($i$) \emph{network traffic generation}, ($ii$) \emph{network traffic classification}, ($iii$) \emph{network intrusion detection}, ($iv$) \emph{networked system log analysis}, and ($v$) \emph{network digital assistance}, which are crucial for network monitoring and management and are mostly overlooked in other such surveys.

Unlike all the related surveys, we perform an in-depth analysis of each mentioned use case aimed at identifying and providing taxonomies of the solutions proposed in the networking domain.
Specifically, we report for each task the adopted \gls{genai} architecture, its public availability, the input fed to the model, and the dataset leveraged for its pre-train or fine-tuning. 
Indeed, our study is intended for researchers and practitioners interested in capitalizing on the benefits of \gls{genai} for network monitoring and management.
Hence, we place a strong emphasis on the reproducibility of the proposals.
Therefore, we also contribute to the taxonomization of the models used for each networking application we identify.
While centered on the impact of the latest \gls{llm} wave,
our study does not simply focus on \gls{llm}-based generative solutions---such as the majority of similar surveys~\cite{hassanin2024comprehensive, zhou2024large, huang2023large, huang2024digital, liu2024large, chaccour2024telecom}. %
Instead, we analyze contributions that include the latest achievements based on Diffusion Models and \glspl{ssm}, which are often overlooked in related surveys.
On the other hand, we purposely exclude in our analysis generative algorithms such as
\glspl{gan}, \glspl{vae}, and normalizing-flows.
These methods, while significant in past years, are considered less relevant compared to the latest advancements in \gls{genai}.

\section{Background}
\label{sec:background}
\begin{figure*}[t]
    \centering
    \includegraphics[width=\textwidth, trim=0 0 10 10, clip]{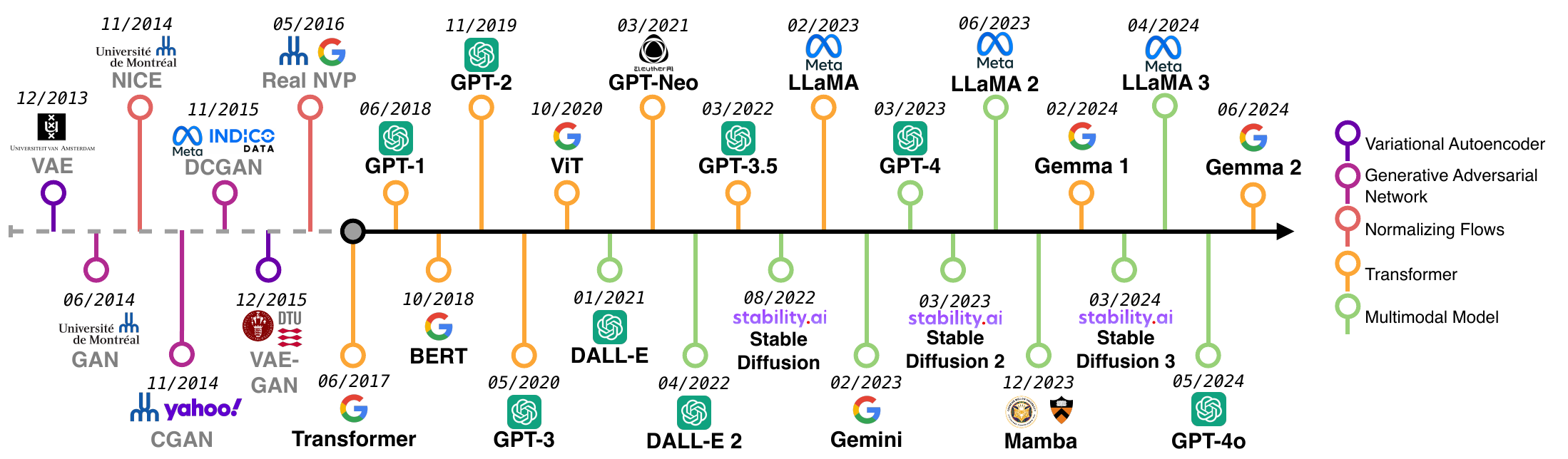}
    \caption{
    Timeline of GenAI development: while introducing VAN-based and GAN-based solutions, this work primarily focuses on developments from the Transformer onward.
    }
    \label{fig: genai_timeline}
\end{figure*}

\begin{figure}[t]
    \centering
    \subfloat[Full Encoder-Decoder (FED)\label{fig:full_enc_dec}]{\includegraphics[trim=15 10 10 5, clip, width=1.\columnwidth]{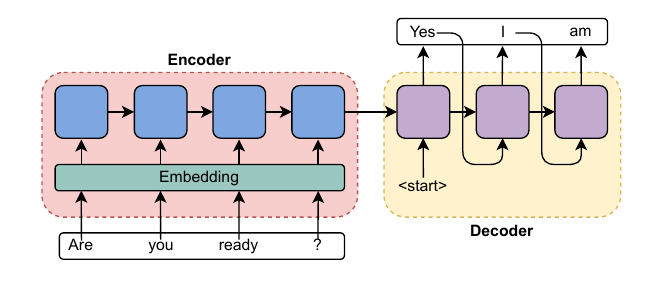}}

    \subfloat[Encoder-Only (EO)\label{fig:encoder_only}]{\includegraphics[trim=0 15 0 15, clip, width=0.7\columnwidth]{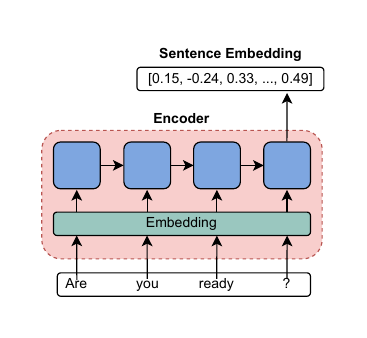}}

    \subfloat[Decoder-Only (DO)\label{fig:decoder_only}]{\includegraphics[trim=0 25 0 15, clip, width=1.0\columnwidth]{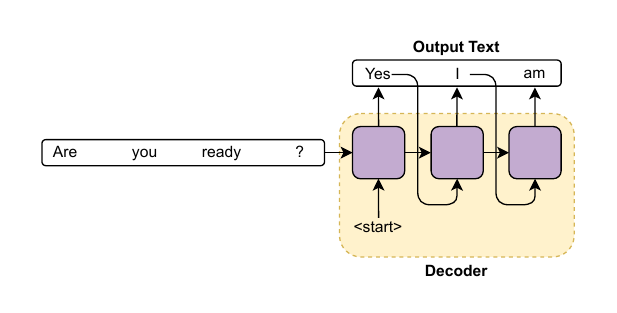}}
    
    \caption{
    Overview of the general workflow of \texttt{Transformer}-based models: (a) Full Encoder-Decoder, (b) Encoder-Only, and (c) Decoder-Only.
    }
    \label{fig:transformer}
\end{figure}
\begin{figure}[t]
    \centering
    \subfloat[Forward Diffusion\label{fig:forward_diffusion}]{\includegraphics[width=1.\columnwidth]{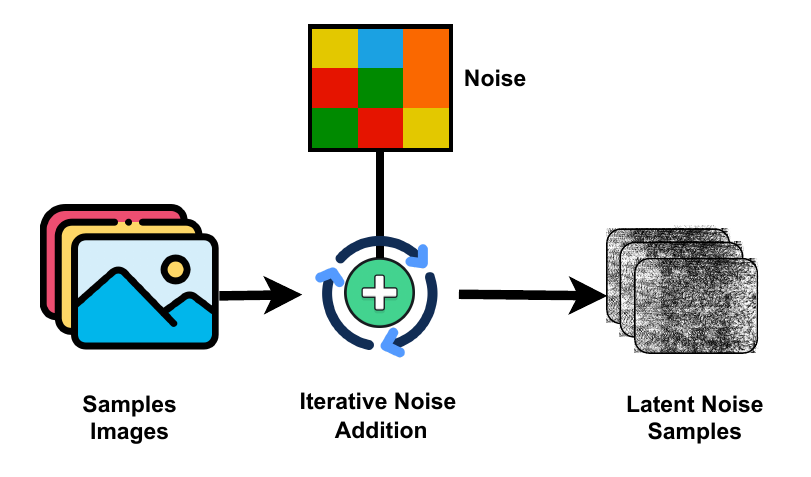}}

    \subfloat[Reverse Diffusion\label{fig:reverse_diffusion}]{\includegraphics[width=1.\columnwidth]{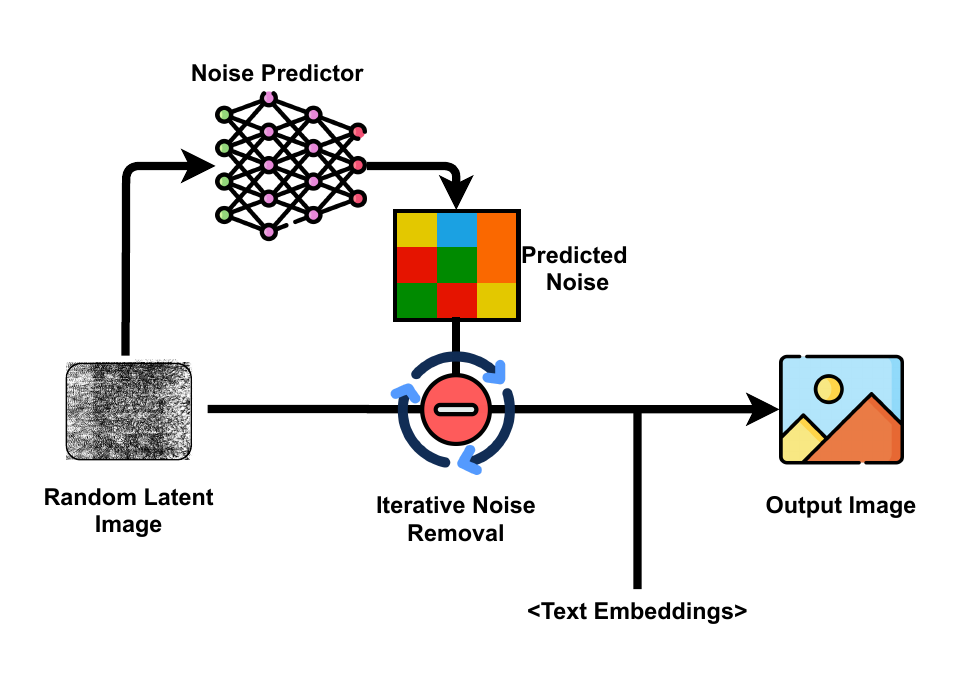}}
    
    \caption{
    Overview of the general workflow of \texttt{Diffusion~Models}, including (a) \emph{forward} and (b) \emph{reverse} diffusion processes.
    }
    \label{fig:diffusion}
\end{figure}
\begin{figure}[t]
    \centering
    \includegraphics[trim=0 25 0 25, clip, width=.8\columnwidth]{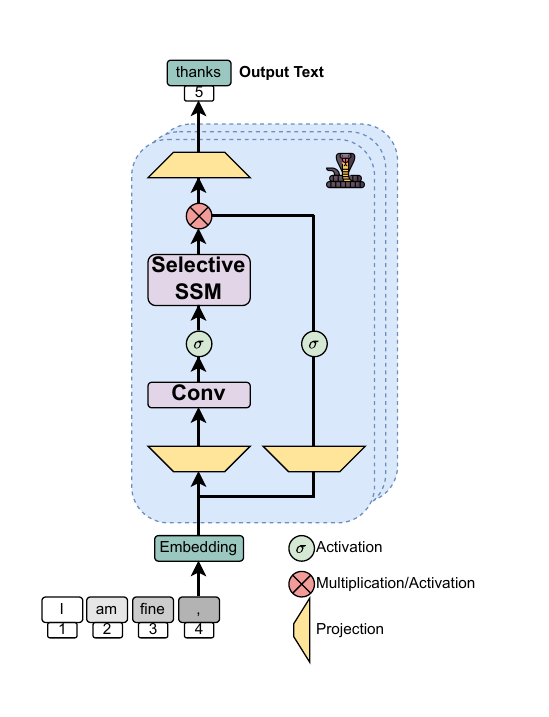}
    
    \caption{
    Overview of the general workflow of \texttt{Mamba}: 
    the \emph{Convolutional (Conv)} layer extracts relevant features from input data, focusing on spatial or temporal patterns; 
    the \emph{Selective SSM} layer filters and selects the most relevant latent states from the extracted features.
    }
    \label{fig:mamba}
\end{figure}

In this section, we first provide a formal description of \gls{genai} (Sec.~\ref{subsec:genai_nutshell}). Then, we trace the evolution of \gls{genai} models over time, from classic methods (Sec.~\ref{subsec:classic_ai}), proposed since $2013$, to the most recent advancements of the present day (Sec.~\ref{subsec:recent_ai}).
We end the section by describing the various strategies used to optimize \gls{genai} models to deal with typical \gls{nmm} use cases (Sec.~\ref{subsec:optimization}).
\subsection{GenAI in a Nutshell}
\label{subsec:genai_nutshell}
\gls{ai} models can
be classified into \emph{discriminative} and \emph{generative} models, according to the learning objective. The former makes predictions on unseen data by training on labeled data and thus can be used for various inference tasks. 
In contrast, generative models focus on synthesizing realistic content. 

From a formal viewpoint, given a set of training samples $\bm{x}_1,\ldots, \bm{x}_N$ associated to an unknown data distribution $p_d(\bm{x})$, a \gls{genai} technique learns a model to sample new (synthetic) data according to $p_{mod}(\bm{x}) \approx p_d(\bm{x})$.
This can be accomplished by either two- or one-step approaches.
In the former case, known as \gls{ede}, the model first learns an explicit distribution $p_{mod}(\bm{x}) \approx p_d(\bm{x})$ (in a tractable or approximate fashion), which is then used to sample new data.
In the latter case, known as \gls{ide}, the \gls{genai} technique directly learns a model that can sample from $p_{mod}(\bm{x}) \approx p_d(\bm{x})$ without explicitly defining it.

The design and use of \gls{genai} have a long history in \gls{nmm}: relevant methods include well-known Markov chains (tractable \gls{ede})~\cite{aceto2021characterization}, but also \glspl{vae} (approximate \gls{ede})~\cite{aceto2024synthetic}, \glspl{gan} (\gls{ide})~\cite{hui2022knowledge} and normalizing-flows (tractable \gls{ede})~\cite{gudovskiy2022cflow}.
Conversely, recent applications of generative models are \emph{\glspl{llm}}---%
based on \fmtTT{Transformer} (and variants) or 
selective \gls{ssm}---%
and \fmtTT{Diffusion Models}, which have represented a breakthrough in the realism and complexity of the content generated.
Transformer-based models enable parallelization and scalability, enhancing processing speed and contextual understanding through their self-attention mechanisms.
Moreover, Diffusion Models offer several advantages over traditional generative models, such as \glspl{vae} and \glspl{gan}, including better mode coverage and stability during training.

\subsection{Classic GenAI Methods}
\label{subsec:classic_ai}
Figure~\ref{fig: genai_timeline} reports the timeline of the development of \gls{genai}, starting from \glspl{vae} (proposed in $2013$ by~\citet{kingma2013auto} at the University of Amsterdam) until the latest models released by OpenAI in the second half of $2024$, namely \fmtTT{GPT-4o} and its successive variants and evolutions (\ie the lightweight \fmtTT{GPT-4o~mini}, and the reasoning models \fmtTT{o1-preview} and \fmtTT{o1-mini}).
We recall that \emph{this survey focuses only on works that take advantage of the most recent advances in \gls{genai}, specifically from the Transformer architecture onward}, which Google proposed in $2017$~\cite{vaswani2017attention}.
One motivation is that these solutions offer improved performance \wrt older solutions like \glspl{vae}, \glspl{gan}, and normalizing-flows, \eg \gls{nice}~\cite{dinh2014nice}.

Moreover, this choice is justified by the impressive groundbreaking impact of 
these more sophisticated \gls{genai} architectures
across various fields,
significantly improving 
generative tasks such as text generation, image synthesis, and multimodal applications.
However, we also report the \emph{classic deep generative models}---\viz~\gls{nice}, \glspl{vae}, and \glspl{gan} with their variants and hybridizations---for context and completeness.
Regarding the latter \gls{genai} architectures---%
which fall outside our scope---%
we refer the reader to these prominent surveys for a deeper background and a detailed overview of their usage for networking-related use cases: \cite{adeleke2022network, sai2024empowering, halvorsen2024applying, karapantelakis2024generative}.
\emph{The modern era of deep \gls{genai}} started with \glspl{vae} at the end of $2013$, \glspl{gan} at mid-$2014$, and \gls{nice} at the end of $2014$.
These models were the first deep neural networks capable of learning generative models for complex data, such as images.
In detail, \glspl{vae} introduced a structured and probabilistic approach to generative modeling with continuous latent spaces and improved training stability~\cite{kingma2013auto},
while \glspl{gan} presented a powerful adversarial framework that excels at generating high-quality and realistic data~\cite{goodfellow2014generative}.
Then, \gls{nice} was the first model to implement normalizing flows using neural networks, leveraging them as invertible functions to transform data from a complex distribution to a simpler one~\cite{dinh2014nice}.

Over time, diverse improvements to \glspl{vae}, \glspl{gan}, and \gls{nice} have been proposed.
Notably, the Conditional GAN (CGAN, $2014$) enables controlled data generation by incorporating additional information into the generative process. This allows for a more targeted and context-specific output~\cite{mirza2014conditional}.
Additionally, the Deep Convolutional GAN (DCGAN, $2015$) enhances the quality of generated images and improves the stability of the training process~\cite{radford2015unsupervised}.
Moreover, hybrid architectures like VAE-GAN (2015) were proposed, integrating the structured latent space of \glspl{vae} into the adversarial training of \glspl{gan}~\cite{larsen2016autoencoding}.
Lastly, the main evolution of \gls{nice} has been the Real Non-Volume Preserving (Real NVP) model---%
proposed in mid-$2016$---%
that incorporates scale transformations, allowing the model to expand or contract regions of data rather than simply rotating or translating them, leading to more accurate and expressive generated contents~\cite{dinh2016density}.

\glsreset{fed}
\glsreset{eo}
\glsreset{do}
\glsreset{sdp}
\glsreset{ssm}

\subsection{Recent Advancements on GenAI}
\label{subsec:recent_ai}

Focusing on the most recent advancements in \gls{genai},
namely from \fmtTT{Transformer} onward,
we can identify five categories of architecture divided according to the nature of the underlying layers.
We identify three variants of the \fmtTT{Transformer} architecture, namely the
\begin{enumerate*}[label=(\emph{\roman*})]
    \item \emph{full encoder-decoder}, the 
    \item \emph{encoder-only}, and the
    \item \emph{decoder-only} architectures. 
    Additionally, the other two categories are based on
    \item \emph{diffusion processes} and 
    \item \emph{state-space representations}, respectively.
\end{enumerate*}
From a general perspective, the development of \gls{genai} models has shifted in the last years toward a \emph{foundational} nature definition~\cite{bommasani2021opportunities}.
Consequently, specific training strategies are commonly employed.

\vspace{5pt}
\noindent
\textbf{Training Strategies for GenAI Models}:
Three training strategies can be adopted for \gls{genai} models.
In the case of naive ($a$) \emph{Monolithic Training}, the model is trained from scratch using a dataset tailored to the specific downstream task.
More commonly, the training follows two sequential stages:
\begin{enumerate*}[label=(\emph{\roman*})]
    \item \emph{Pre-Training}, where the \gls{genai} model is pre-trained on a large corpus of data consisting of text, images, or other input modalities, in a self-supervised or semi-supervised manner. 
    For instance, during this stage, a text-fed (resp. image-fed) model is instructed to predict masked words and the sequence of sentences (resp. to denoise or reconstruct the original picture).
    \item \emph{Fine-Tuning}, where the \gls{genai} model is then specialized for specific tasks (possibly by topping/modifying the architecture with task-specific layers). Specifically, the training parameters (or a portion of them) are jointly fine-tuned (exploiting the broader transfer learning concept), tailoring the model for the considered downstream task.
\end{enumerate*}
Consequently, the training of a model can involve either ($b$) \emph{Pre-Training \& Fine-Tuning}, \ie the model is first pre-trained on a large corpus of data (\eg a networking corpus) and then fine-tuned with a dataset related to the downstream task, or ($c$) \emph{Fine-Tuning Only}, \ie an already pre-trained model is exclusively fine-tuned for the specific downstream task.

\vspace{5pt}
\noindent
\textbf{\gls{fed}}:
This category includes the \gls{genai} architectures that reflect the typical structure of the \fmtTT{Transformer} model~\cite{vaswani2017attention}.
The \fmtTT{Transformer} represents a revolutionary approach to solving sequence processing tasks. This architecture is entirely based on the self-attention mechanism rather than using recurrences---\eg \gls{rnn}---or convolutions---\eg \gls{cnn}.
This improvement enables the \fmtTT{Transformer} to efficiently parallelize computations and significantly reduce training times while achieving state-of-the-art results.
In fact, traditional sequence processing models---%
such as \glspl{rnn} and their variants like \gls{lstm} and \gls{gru}---%
perform computations around symbol positions in input and output sequences.
This characteristic results in limited parallelization because it is inherently sequential, 
becoming a significant bottleneck for longer sequences.
To cope with this drawback, the \fmtTT{Transformer} leverages self-attention mechanisms to model dependencies between different positions in a sequence, regardless of their distance.

In general, the \fmtTT{Transformer} architecture consists of an encoder-decoder structure. The general workflow of a \gls{fed} is depicted in Figure~\ref{fig:full_enc_dec}.
The \emph{encoder} comprises a stack of identical layers, each with two sub-layers, namely a multi-head self-attention and a position-wise fully connected feed-forward network.
The \emph{decoder} is similar to the encoder but includes an additional sub-layer that performs masked multi-head attention over the encoder's output. It also modifies the self-attention sub-layer to prevent positions from attending to subsequent positions, ensuring the auto-regressive property.
Both encoder and decoder are designed with residual connections for each sub-layer followed by layer normalization.

Going into detail, the \fmtTT{Transformer} uses multi-head attention to allow the model
to learn information from different representation subspaces jointly.
In fact, instead of having a single attention function, 
the model linearly projects queries, keys, and values multiple times with different learned projections and performs the attention function in parallel. This process enhances the model's ability to focus on different parts of the input sequence.
Moreover, since the \fmtTT{Transformer} lacks the inherent sequential order provided by recurrence, it introduces positional encodings to inject information on the position of tokens in the sequence. These encodings are added to the input embeddings at the bottom of the encoder and decoder stacks, enabling the model to understand the sequence order.

The \fmtTT{Transformer} is usually leveraged for sequence-to-sequence tasks, 
\ie when the input and the output are both sequences. Examples of applications are translation (from one language to another), summarization (condensation of documents), and text generation (based on given prompts).
Notable architectures that fall into this category are \fmtTT{XLNet}, \fmtTT{T5}, \fmtTT{Gemini}, \fmtTT{Mistral}, and \fmtTT{Zephyr}.

\vspace{5pt}
\noindent
\textbf{\gls{eo}}: Compared to \gls{fed}, \gls{eo} models only leverage the encoder unit of the \gls{fed} architecture (as depicted in Figure~\ref{fig:encoder_only}). 
\gls{eo} is designed to model bidirectional relationships between tokens in an input sequence, generating either a vector representation for each token or a single vector summarizing the entire sentence. This architecture is well-suited for tasks focused on text understanding and analysis rather than generation.

Among the \gls{eo} models, \fmtTT{BERT}~\cite{devlin2018bert}--- Bidirectional Encoder Representations from Transformers---is the most representative and has served as the basis for many subsequent advances of this type of architecture. 
Developed by researchers at Google AI Language, \fmtTT{BERT} consists of multiple layers of bidirectional \fmtTT{Transformer} encoders.
\fmtTT{BERT} has been designed with a key innovation: it pre-trains deep bidirectional representations from the unlabeled text by joint conditioning on both left and right contexts in all layers. 
This procedure allows \fmtTT{BERT} to capture richer linguistic information, and it contrasts with models like OpenAI GPTs and ELMo (an \gls{lstm}-based architecture), which are unidirectional and do not fully leverage the bidirectional context. 
In detail, the bidirectional training of \fmtTT{BERT} is achieved through a \gls{mlm} objective. The \gls{mlm} randomly masks some tokens in the input sequence and predicts them using the context provided by the remaining tokens on both sides. This method allows \fmtTT{BERT} to capture the context from both directions.
To further enhance its understanding of context and sentence relationships, \fmtTT{BERT} uses a next-sentence prediction task during pre-training. This involves predicting whether a given sentence \fmtTT{B} follows sentence \fmtTT{A} in the original text, allowing the model to learn how sentences relate to each other.

The ability of \fmtTT{BERT} to understand the context from both directions and the effectiveness of its pre-training tasks enables it to achieve superior performance across a wide range of \gls{nlp} tasks. \fmtTT{BERT} has been designed for language understanding, 
\ie
encoding the input text in relation to its context for various subsequent tasks. Examples of applications are text classification (\eg spam detection), question answering, and text similarity (\eg semantic search).
Other notable architectures in this category are variants of \fmtTT{BERT}, such as \fmtTT{BERTiny}, \fmtTT{RoBERTa}, \fmtTT{DistilRoBERTa}, and \fmtTT{ViT} (Vision Transformer).

\vspace{5pt}
\noindent
\textbf{\gls{do}}:
These models exploit only the decoder component of the \gls{fed} architecture, as shown in Figure~\ref{fig:decoder_only}. 
\gls{do} models are designed for autoregressive text generation, predicting the next token based on previous tokens, thereby producing the output one token at a time.

Among them, Generative Pre-trained Transformers (\fmtTT{GPTs})~\cite{radford2018improving} are the most prominent, spearheading advancements in the field of \gls{nlp} starting with their first variant named \fmtTT{GPT-1}.
Subsequent improvements of \fmtTT{GPTs}, including \fmtTT{GPT-2}, \fmtTT{GPT-3}, \fmtTT{GPT-3.5}, \fmtTT{GPT-3.5 turbo}, \fmtTT{GPT-4}, and \fmtTT{GPT-4o}, have dramatically increased the model size and the scale of pre-training data and have included multi-modality (text and images) from \fmtTT{GPT-4} onward.
\fmtTT{GPT-4}, with its estimated $1.7$ trillion parameters%
\footnote{\url{https://the-decoder.com/gpt-4-has-a-trillion-parameters/}}%
, exemplifies the trend towards larger models and has achieved state-of-the-art results across a wide range of benchmarks without task-specific fine-tuning.

\fmtTT{GPTs} are commonly leveraged for autoregressive text generation, \ie generating text tokens conditioned on the previous token.
Examples of applications are text generation,
language modeling (\eg autocompletion), and conversational \gls{ai} or chatbots (\eg ChatGPT and Copilot).
Notable architectures in this category are \fmtTT{Falcon}, \fmtTT{LLaMA}, \fmtTT{Phi}, \fmtTT{Gemma} and improvements of \fmtTT{GPT-1}, from \fmtTT{GPT-2} to \fmtTT{GPT-4o}.

\vspace{5pt}
\noindent
\textbf{\gls{sdp}}:
\citet{ho2020denoising} proposed 
a class of generative models,
named Diffusion (probabilistic) Models,
that describes the process by which particles, information, or other entities spread through a medium over time.
In recent years, the \fmtTT{Diffusion~Model} (Figure~\ref{fig:diffusion}) has found successful applications in computer vision, as well as in audio, bioinformatics, and agent-based systems. 

The core idea involves defining a \emph{forward diffusion} process (Figure~\ref{fig:forward_diffusion}) that gradually adds noise to the data, transforming them into a simpler distribution, typically Gaussian noise. The corresponding \emph{reverse diffusion} process (Figure~\ref{fig:reverse_diffusion}) is then learned to map the noisy data back to the original data distribution.
The elegance of \fmtTT{Diffusion~Models} lies in their theoretical foundation, which leverages concepts from Markov chains (when diffusion is performed in discrete time) or stochastic differential equations (when diffusion is performed in continuous time).
This foundation allows for a rigorous treatment of the model's behavior and facilitates efficient training and sampling algorithms (through sophisticated sampling acceleration techniques). 
The resulting models, such as the \gls{ddpm} and score-based generative models, have demonstrated remarkable capabilities in generating high-quality synthetic data.
Because of the explicit definition of the forward/reverse diffusion process and the objective used to learn them (\ie a generalized evidence lower-bound~\cite{kingma2024understanding}), these models fall within the approximate \gls{ede} category.
In summary, \fmtTT{Diffusion~Models} have been used for high-quality data generation through iterative denoising. 
Examples of applications include visual and signal data processing tasks, such as image generation, audio synthesis, and video generation, contrasting with previous categories that primarily involve language and textual data-generation tasks. 
A notable model falling into this category is \fmtTT{Stable~Diffusion}.

\vspace{5pt}
\noindent
\textbf{Selective and Structured \glspl{ssm}}: %
Proposed by~\citet{gu2023mamba},
\fmtTT{Mamba} represents a significant advancement in sequence modeling, introducing a new class of selective and structured \glspl{ssm} designed to overcome the limitations of existing architectures like \fmtTT{Transformers} in handling very long input sequences. 
As depicted in Figure~\ref{fig:mamba}, \fmtTT{Mamba} integrates selective and structured \glspl{ssm} into a streamlined neural network architecture that avoids traditional attention mechanisms, achieving \emph{fast inference and linear scaling with sequence length}.

In fact, while \fmtTT{Transformers} have become the backbone of many foundation models due to their effective self-attention mechanism, they suffer from \emph{quadratic scaling} \wrt sequence length, limiting their efficiency on long sequences. Subquadratic-time architectures, including linear attention and structured-only \glspl{ssm}, have attempted to address these inefficiencies but have failed to perform in critical modalities such as language.

\fmtTT{Mamba}'s first core innovation lies in making (a part of) the parameters of \glspl{ssm} dependent on the input. This mechanism allows the model to \emph{selectively} propagate or forget information based on the current token, \ie the model can focus on relevant information or discard irrelevant or outdated information as needed.
This selective mechanism enables \fmtTT{Mamba} to handle discrete modalities effectively, providing a significant advantage over previous \glspl{ssm}.
Secondly, \fmtTT{Mamba} reduces the number of trainable parameters by assuming a \emph{structured} form for the \gls{ssm} matrices defining the information propagation.
Thirdly, to maintain efficiency, \fmtTT{Mamba} employs a ($i$) hardware-aware algorithm that computes the model recurrently without materializing the expanded state in the \gls{gpu} memory (leveraging fast memory hierarchies) and ($ii$) a parallel scan algorithm to accelerate the recursive computation of relevant quantities. This approach ensures linear scaling in sequence length and high throughput on modern hardware.

In summary, \fmtTT{Mamba} integrates selective and structured \glspl{ssm} into a simplified neural network architecture that omits attention and \gls{mlp} blocks. This streamlined design, inspired by previous \gls{ssm} architectures, offers fast training and inference with high performance in various data modalities, including language, audio, and genomics.
\fmtTT{Mamba} is designed to process and model sequences efficiently, making it ideal for applications that require handling long sequences and achieving high computational efficiency.

\subsection{Optimization Strategies for GenAI}
\label{subsec:optimization}
In this section,
we discuss the optimization strategies that have been proposed for \gls{genai} solutions,
focusing on those that have been used in \gls{nmm} use cases.
Broadly, optimization strategies can be divided into two categories~\cite{kim2024memory}, involving methods for ($i$) \emph{parameter-efficient fine-tuning} and ($ii$) \emph{post-training quantization}.

\vspace{5pt}
\noindent
\textbf{\gls{peft}}:
These methods enhance the adaptation of pre-trained \gls{genai} models to the target downstream task.
The primary objective is to minimize the computational resources required for fine-tuning while preserving inference performance.

Among recent advances proposed to optimize the fine-tuning step of \gls{genai} solutions,
the state-of-the-art approach named \gls{lora}~\cite{hu2021lora} has been recently leveraged for traffic generation purposes~\cite{jiang2024netdiffusion}.
\gls{lora} capitalizes on the intuition that changes in model weights during adaptation have a low ``intrinsic rank''.
In other words, the idea is that adapting a model to a new task does not require very complex changes, which can be efficiently represented using fewer adjustments.
In detail, \gls{lora} optimizes the fine-tuning by using rank decomposition matrices, specifically targeting the change in dense layers during training while keeping the main pre-trained weights frozen.
Thus, this method allows for efficient task adaptation by replacing some model components---%
\ie parts of the weight matrices---%
with small low-rank matrices.
This substitution reduces the need to recalculate gradients and memorize optimizer states.
The way in which \gls{lora} is designed ensures that there is no extra delay introduced during inference. Moreover, \gls{lora} is compatible with other optimization techniques, such as post-training quantization~\cite{ding2023parameter}.

\vspace{5pt}
\noindent
\textbf{\gls{ptq}}:
These methods aim to reduce the computational complexity and memory footprint of \gls{genai} models by casting the model parameters into lower precision formats \emph{after training}. Quantization facilitates faster inference and more efficient deployment on various hardware.

One of the most complete methods for enforcing \gls{ptq} that has been recently proposed is
named \gls{gguf}~\cite{gguf-website}.
\gls{gguf} is a generalized file format that has recently been adopted by~\cite{karapantelakis2024using} to enforce post-training quantization for network digital assistance purposes, namely for networking standards question answering.
\gls{gguf} has been proposed to reduce---%
in large \glspl{llm}---%
the precision of the weights and activations of the model by converting real numbers to integers, \eg $32$-bit floating-point to $8$-bit.
\gls{gguf} has been devised with two key features in mind: \emph{quantization-aware kernel optimization} and \emph{extensibility}.
On the one hand,
\gls{gguf} does not simply apply quantization to the model weights but also provides kernel optimization functionalities that consider the quantization process. This characteristic is fundamental in avoiding an inference performance decrease due to blind quantization.
On the other hand, \gls{gguf} has been designed to overcome the limits of its predecessor GGML%
\footnote{GGML is an \gls{ml} library created by Georgi Gerganov, which is why it is named ``GGML''. In addition to offering low-level \gls{ml} primitives, such as tensor types, GGML also defines a binary format for distributing \glspl{llm}.}%
, which lacks mechanisms to incorporate additional model information or add new features.
Therefore, \gls{gguf} allows the integration of new features into the file format while ensuring compatibility with models deployed in older \gls{gguf} formats, thus preserving backward compatibility for newer versions.

In general, \gls{gguf} provides several key functionalities, including single file deployment, improved model loading and saving speeds, and intuitive design and detailed information storage that facilitate extensibility. Together, these functionalities enable a more efficient and user-friendly experience in handling \glspl{llm}.

\glsresetall

\section{Current Status of GenAI in Network Monitoring and Management}
\label{sec:genai_landscape}

\begin{figure}[t]
    \centering
    \includegraphics[width=1\columnwidth, trim={0.5cm 0.5cm 0.5cm 0.5cm},clip]{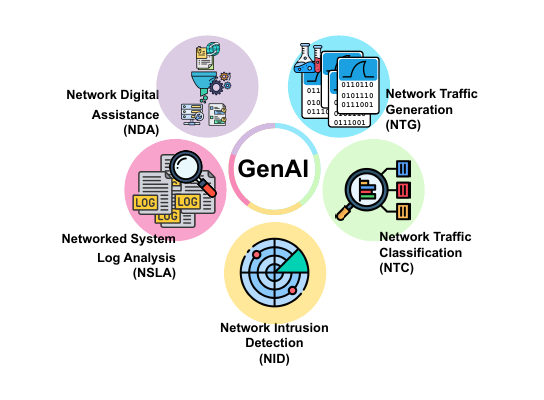}
    \caption{
    Overview of NMM use cases, leveraging GenAI, explored in this survey. These include Network Traffic Generation (NTG), Network Traffic Classification (NTC), Network Intrusion Detection (NID), Networked System Log Analysis (NSLA), and Network Digital Assistance (NDA).
    }
    \label{fig: downstream}
\end{figure}

\begin{figure*}[t]
    \centering
    \includegraphics[width=0.90\textwidth]{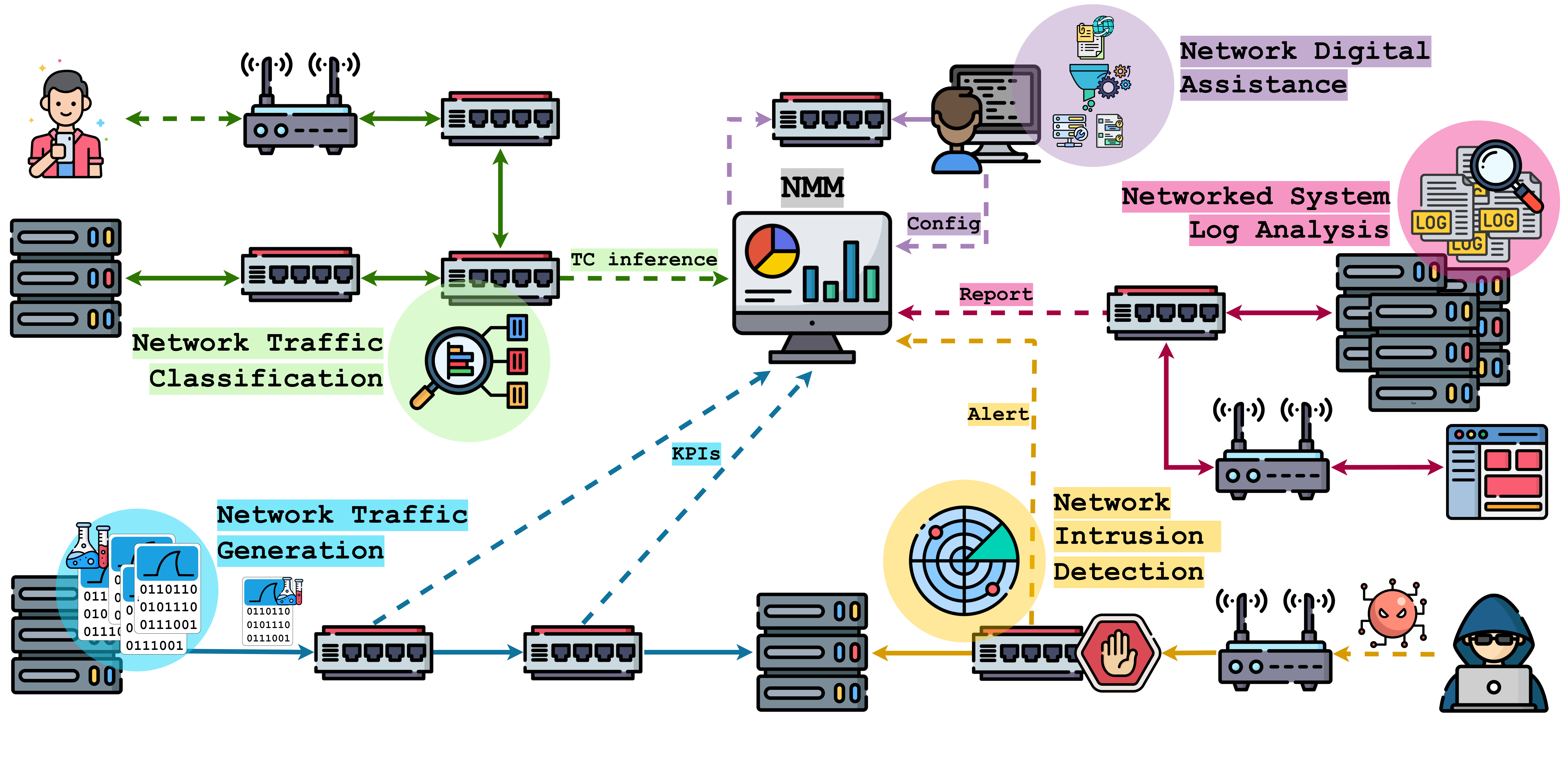}
    \caption{
    Possible interactions among 
    NTC, NTG, NDA, NSLA, and NID
    to enhance network management efficiency, highlighting the key pathways and feedback mechanisms among the different entities.
    }
    \label{fig: downstream_interaction}
\end{figure*}

This section provides an overview of the current status of \gls{genai} in the \gls{nmm} context. Accordingly, Sec.~\ref{subsec:dowstream_tasks} outlines the five \gls{nmm} use cases where \gls{genai} is currently being utilized, along with a description of the benefits \gls{genai} offers for each. Then, Secs.~\ref{sec:traffic_generation}-\ref{sec:network_management} dissect the works that employ \gls{genai} for each use case.

\subsection{Use Cases for GenAI in Network Monitoring and Management}
\label{subsec:dowstream_tasks}

\gls{genai} is actively used to address \gls{nmm} use cases in various networking domains. We have identified five key use cases where it is currently employed: ($i$) \emph{Network Traffic Generation}, ($ii$) \emph{Network Traffic Classification}, ($iii$) \emph{Network Intrusion Detection}, ($iv$) \emph{Networked System Log Analysis}, and ($v$) \emph{Network Digital Assistance for Documentation \& Configuration}.
Such use cases are summarized in Figure~\ref{fig: downstream} along with the acronyms we use in this work. 
Below, we provide detailed descriptions for each use case.

\vspace{5pt}
\noindent
\textbf{\gls{traffgen}}
refers to the process of creating synthetic network data, ranging from
the generation of
\mbox{(bi-)flow} statistical features
or sequence of features extracted from the packets within a \mbox{(bi-)flow} (e.g., packet size, inter-arrival time, and packet direction), to the generation of
the entire PCAP trace.%
\footnote{In this survey, we do not cover sensor data generation, such as temperature or pressure measurements, since this task involves modeling physical phenomena rather than actual network traffic.}

\gls{traffgen} is crucial for network traffic analysis from various applicative perspectives. It enables the simulation of different scenarios to (stress) test network infrastructure and services, validate security measures (e.g., for automated penetration tests), and 
augment training data for improving \gls{ml} models 
performance and generalization capabilities.
The key challenge of \gls{traffgen} is producing high-fidelity synthetic network samples that closely resemble real traffic.
Hence, a critical aspect 
is the validation of the synthetic traffic generated, since both \emph{effectiveness}
(in enhancing \gls{ml} or \gls{dl} models performance) 
and \emph{validity} 
(in simulating with high fidelity the real traffic) 
are desiderata of the synthetic network traffic generation task~\cite{adeleke2022network}.

\noindent
\emph{How \gls{traffgen} 
can benefit from \gls{genai}: }
\gls{genai} can significantly enhance the \gls{traffgen} task by leveraging its ability to understand and mimic natural language patterns, thus modeling network traffic as the ``language of the Internet''. Accordingly, it helps in generating realistic protocol sequences and user interactions, producing high-quality synthetic traffic that mirrors the diverse and complex traffic patterns of real environments~\cite{meng2023netgpt, jiang2024netdiffusion}.

\vspace{5pt}
\noindent
\textbf{\gls{traffclass}}
aims to categorize network traffic represented by various 
\emph{\glspl{to}}
such as packets, bursts, flows, biflows, or sessions. This process may include identifying the protocol (especially when nonstandard transport ports are used), the name of the application (\eg YouTube, Netflix, Facebook), or the type of service (\eg streaming, web browsing, VoIP) that generated the traffic.
Generally, \gls{traffclass} involves modeling target network traffic classes (\eg by using \gls{ml} or \gls{dl} algorithms) and differentiating the traffic into one of these target classes.

Since \gls{traffclass} can identify user behaviors and predict traffic categories, it is crucial in enhancing network management operations. By applying rules based on \gls{traffclass} results, network management can be adapted to address the specific needs of the network, optimizing the handling of different types of traffic. 
The main challenges affecting \gls{traffclass} include the limited availability of high-quality data to train effective models and the poor generalization capabilities shown by the state-of-the-art \gls{traffclass} techniques~\cite{pacheco2018towards, azab2022network,aceto2023aipowered}.

\noindent
\emph{How \gls{traffclass} 
can benefit from \gls{genai}: }
\gls{genai} can significantly enhance \gls{traffclass} through advanced contextual awareness and pattern recognition capabilities of pre-trained models.
These models leverage large unlabeled datasets to learn unbiased data representations, which can be easily transferred to various downstream tasks by fine-tuning on limited labeled data.
The killer idea can be the modeling of network traffic as a language, namely the language of machine-$2$-machine communication.
Thus, \gls{genai} can produce highly versatile pre-trained models that, due to their high generalizability, can be adapted to solve different \gls{traffclass} tasks with minimal effort, eliminating the need to train new models from scratch for each task~\cite{lin2022,wang2024netmamba}. 

\vspace{5pt}
\noindent
\textbf{\gls{nid}} aims to identify anomalous or malicious traffic traversing the network.
Specifically, \gls{nid} focuses on monitoring the traffic exchanged between connected entities (e.g., mobile devices, computers, servers) to secure them. 
Its main objective is to detect anomalous behaviors that may be related to security threats or intrusions by analyzing the exchanged traffic. 

\gls{nid} is crucial in identifying malicious activities by distinguishing legitimate (\viz~benign) traffic from potentially harmful (\viz~malicious) traffic. Moreover, it can be used even to identify specific attack traffic~\cite{chou2021survey}. These operations enable prompt response to threats and minimize potential damage to network infrastructure and its users. 

\gls{nid} should be seen as a specialization of \gls{traffclass} when dealing with supervised multiclass or binary classification (\viz misuse detection). In this context, \gls{nid} leverages techniques common to \gls{traffclass} to identify types of attacks based on knowledge extracted from labeled training data. However, \gls{nid} also encompasses \gls{ad}, which involves identifying outliers or abnormal behaviors that deviate from the norm. This process is typically addressed via out-of-distribution detection or one-class classification methodologies. 
For these reasons, we treated it as a separate use case in this survey also due to its importance, dedicated modeling solutions, and extensive related literature.

\noindent
\emph{How \gls{nid} 
can benefit from \gls{genai}: }
Similarly to \gls{traffclass}, \gls{genai} can enhance \gls{nid} through contextual awareness, enabling the detection of anomalies by understanding the context of network events over time. Its adaptability by means of transfer learning allows rapid adaptation to new threats, while semantic analysis identifies unusual command sequences. In addition, complex pattern recognition and unsupervised learning help detect subtle deviations and unknown threats~\cite{manocchio2024flowtransformer, ferrag2024}.

\vspace{5pt}
\noindent
\textbf{\gls{nla}} refers to solutions that automate the extraction of knowledge from network or system logs to summarize them (\ie to identify key elements) or to detect anomalies (\viz~log anomaly detection).
Network and system logs typically consist of semi-structured text/records of data that collect network or system events.
Specifically, the logs considered in this work pertain to network-related applications, such as web or email servers, or related to network entities, such as network managers.

\gls{nla} is crucial for enhancing security by identifying unauthorized access and potential security breaches. It ensures system reliability by providing hints to identify performance bottlenecks or diagnosing the root cause of the fault. Moreover, it improves software quality through log debugging or ensuring software robustness. It optimizes operations by analyzing user behaviors or auditing activities and helps maintain compliance across various domains (\eg supporting predictive maintenance)~\cite{he2021survey}.

\noindent
\emph{How \gls{nla}  
can benefit from \gls{genai}: }
Leveraging modern and advanced \glspl{llm}, \gls{genai} can efficiently parse, interpret, and summarize log data written in natural language. It extracts significant events and patterns through semantic analysis, enhancing the understanding of log data~\cite{han2023loggpt,boffa2024logprecis}.

\vspace{5pt}
\noindent
\textbf{\gls{nda}} (briefly, Network Digital Assistance) 
focuses on monitoring and controlling network operations to ensure efficient and reliable performance.
Specifically, \gls{nda} aims to maintain network reliability and availability, optimize network performance, ensure security, and enable efficient resource scheduling.
Moreover, it is fundamental for reducing downtime, preventing data leaks, and ensuring uninterrupted service delivery.
Hence, \gls{nda} is crucial for various applications. 
It facilitates interoperability issues in heterogeneous network environments characterized by multi-layer and multi-vendor infrastructures. \gls{nda} also supports advanced networking frameworks like \gls{sdn} %
and simplifies the management of ever-growing \gls{iot} environments. 
In this context, resource provisioning, device configuration, network monitoring, and software update management are essential for reducing energy consumption and strengthening the security of 
\gls{iot} devices that are resource-constrained and insecure-by-design~\cite{martinez2014network, aarikka2017network, aboubakar2022review}.

\noindent
\emph{How \gls{nda} 
can benefit from \gls{genai}: }
\gls{genai} allows extensive automation for network operations.
Specifically, \glspl{llm} provide a natural language interface that simplifies the retrieval of complex information in networking standards and documents crucial for \gls{nda}. 
This interface also facilitates the management of various network software and hardware,
enhancing operational efficiency~\cite{wang2023network}.

\vspace{5pt}
\noindent
\textbf{Relations among NMM use cases}:
In general, these five use cases are strongly related. 
Together, they improve the monitoring and control of network operations, leading to improved network performance, reliability, and security.
From a broad perspective, \gls{nmm} acts as the system's actuator,
leveraging the outputs of other tasks to make informed decisions and optimize network performance.
A graphical representation of this interaction is shown in Figure~\ref{fig: downstream_interaction}, highlighting the links between the various use cases.

Specifically, \gls{traffgen} creates synthetic traffic that can be used to evaluate potential network configurations before deployment---%
\ie \gls{nda}. 
By using reliable and accurate traffic generators, which can produce traffic that closely resembles real traffic, \gls{traffgen} can provide valuable insights into the behavior of network equipment in pseudo-real operating environments.
Additionally, by generating various types of synthetic traffic, both benign and malicious, \gls{traffgen} can be utilized to assess, in different contexts, the response of classification and intrusion detection systems---%
\ie
\gls{traffclass} and \gls{nid}, respectively.
Moreover, it can enhance their performance and generalizability by augmenting training data for \gls{ml} and \gls{dl} models.

Conversely, \gls{traffclass} and \gls{nid} offer insights into the (real) traffic traversing the network, enabling performance improvements (through specific traffic prioritization or routing rules) and enhancing security (by applying filtering rules to block anomalous or malicious traffic). Thus, \gls{traffclass} and \gls{nid} can facilitate online (re)configurations through \gls{nda}.
Additionally, data-driven \gls{traffclass} and \gls{nid} systems trained on real traffic data can provide valuable feedback on the quality of the synthetic traffic generated---%
\ie validating \gls{traffgen}.

Finally, while \gls{nla} shares similarities with \gls{traffclass} and \gls{nid} mechanisms, it
focuses on analyzing log data related to network equipment (\eg servers and routers) and the services provided (\eg web pages) rather than network traffic.
Therefore, \gls{nla} supports network management operations by providing summaries and insights from logs and by identifying anomalies in the operation of network equipment and services. This information can then be used to adjust the behavior of these network equipment and services (\ie \gls{nda}).

\begin{figure}[t]
    \centering
    
    \subfloat[NTG, NTC, and NID]{
    \includegraphics[width=\columnwidth]{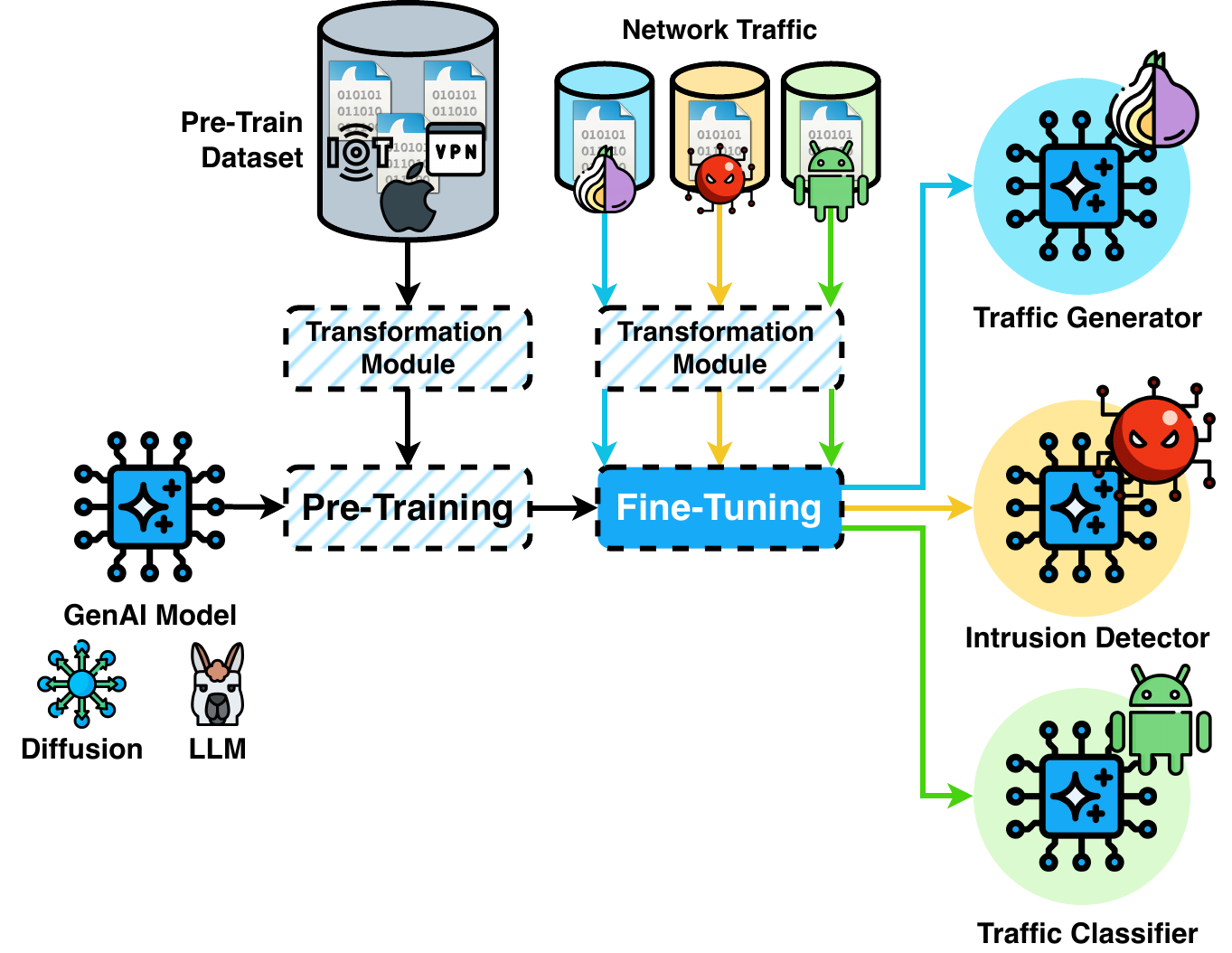}
    }

    \subfloat[NSLA and NDA]{
    \includegraphics[width=0.9\columnwidth]{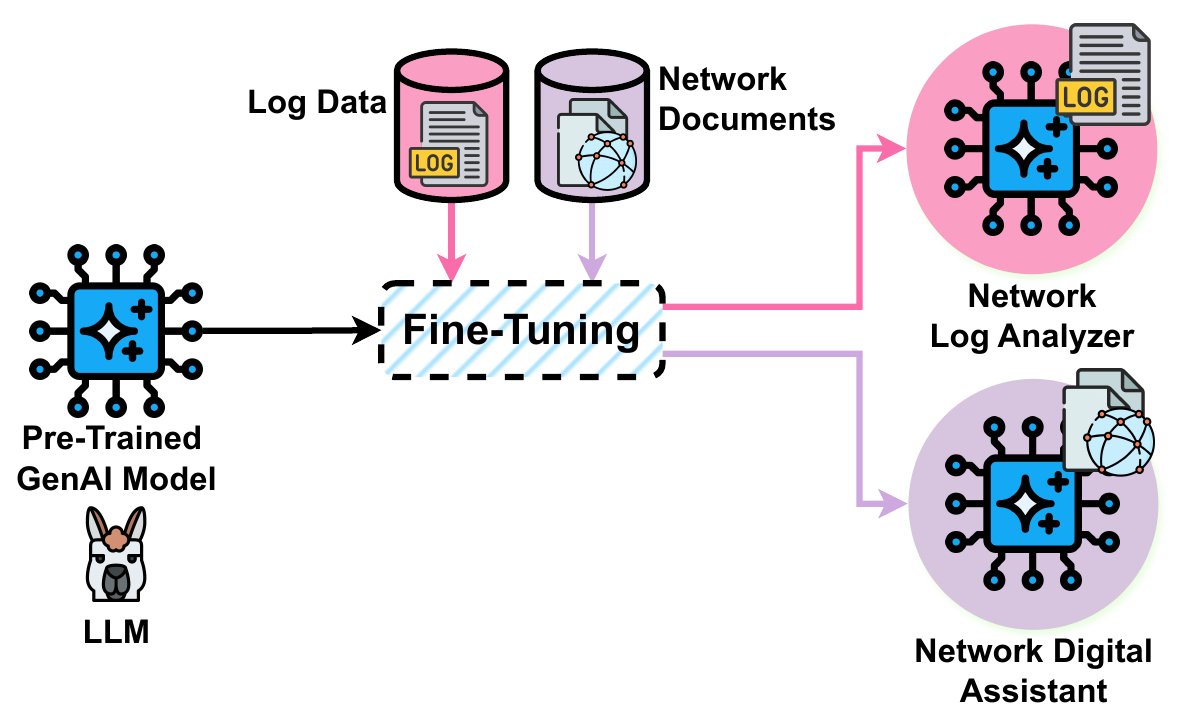}
    }

    \caption{
    Pipeline for NMM use cases with GenAI models, detailing the process for (a) NTG, NTC, NID, and (b) NSLA and NDA.
    Dashed-line blocks
    indicate optional stages.
    }
    \label{fig: usecase_detail}
\end{figure}

\vspace{5pt}
\noindent\textbf{From network traffic to GenAI:} 
Figure~\ref{fig: usecase_detail} details the generic pipeline for the use cases.
Specifically, \gls{traffgen}, \gls{traffclass}, and \gls{nid} leverage the same kind of input (\ie network traffic) to pre-train and fine-tune the \gls{genai} model.
Conversely, \gls{nla} and \gls{nda} typically employ pre-trained models, which are then fine-tuned with specific log data or different network documents to adapt to the task at hand.

\gls{genai} approaches are not designed to ingest network traffic directly. Instead, approaches based on \glspl{llm} or Diffusion Models are typically designed to process data in a text-based or image-like format. Therefore, for use cases involving direct processing of network traffic (\ie \gls{traffgen}, \gls{traffclass}, and \gls{nid}), traffic data need to be transformed into a text-based or image-like representation before using it as input to the \gls{genai} model. This transformation is performed via \emph{Datagram-to-Token} and \emph{Datagram-to-Image} operations, respectively.

\vspace{5pt}
\noindent
\emph{Datagram-to-Token}: Approaches based on \glspl{llm} typically employ a \emph{Datagram-to-Token} method to convert encrypted traffic into pattern-preserved token units for pre-training~\cite{lin2022}. 
This method involves segmenting traffic into packets and representing their characteristics as word-like tokens, similar to natural language processing. 
When packets are grouped into traffic objects, such as (bi)flows or bursts, special tokens are required to mark the packet boundaries within these traffic objects (\eg common values for these special tokens are \texttt{[SEP]}, \texttt{[MSK]}, \texttt{[PAD]}, and \texttt{[PKT]}). 
Additionally, the type of features extracted from the traffic object (or from each packet belonging to it) may require further preprocessing before being converted to tokens (\eg conversion into raw bytes, anonymization, or quantization).

\vspace{5pt}
\noindent
\emph{Datagram-to-Image}: Approaches based on Diffusion Models typically employ a \emph{Datagram-to-Image} method to convert encrypted traffic into image representations~\cite{jiang2024netdiffusion}. 
Two main Datagram-to-Image variants are commonly exploited based on the features to be used: 
($i$) raw-bytes-to-image 
and ($ii$) features-to-image. 
The former involves translating network traffic into standardized bits, where each bit corresponds to a packet header field bit. The encoded sequence of packets is then formed into a matrix, which is interpreted as an image (\eg \emph{nPrint} format~\cite{holland2021new}).
Conversely, when the model input is a time series of packet features (\eg packet sizes or inter-packet times), different transformations can be applied to encode the time series as an image, such as \emph{FlowPic}~\cite{shapira2021flowpic} or \emph{Gramian Angular Summation Field (GASF)}~\cite{sivaroopan2023netdiffus}.

\vspace{5pt}
Hereinafter, we provide a detailed overview of the existing literature for each \gls{nmm} use case.

\subsection{Network Traffic Generation}
\label{sec:traffic_generation}
\begin{table*}[t]
\renewcommand{\arraystretch}{1.2}
\centering
\normalsize
\caption{Works dealing with NTG through GenAI models (in chronological order).
}
\label{tab:traffic_generation}
\resizebox{\textwidth}{!}
{
\begin{threeparttable}
\begin{tabular}{rcccccccccccccc}
\toprule
\multirow{3}{*}{\textbf{Paper}} & \multirow{3}{*}{\textbf{Year}} & \multicolumn{2}{c}{\textbf{GenAI Model}} & \multicolumn{2}{c}{\textbf{GenAI Train}} & \multicolumn{3}{c}{\textbf{Traffic Input}} &  
\multicolumn{3}{c}{\textbf{Networking Dataset}}
& \multirow{3}{*}{\begin{tabular}{c}\textbf{Generated}\\[\extraarrayvspace]\textbf{Data}
\end{tabular}} 
& 
\multirow{3}{*}{\mcr{\textbf{Evaluation}\\[\extraarrayvspace]\textbf{Metrics}}}\\
\cmidrule(lr){3-4}
\cmidrule(lr){5-6}
\cmidrule(lr){7-9}
\cmidrule(lr){10-12}
& & \textbf{Name}&\textbf{Architecture} & \textbf{Technique} & \textbf{NetPT} & \textbf{TO} & \textbf{Data} & \textbf{Format} &
\mcr{\textbf{Monolithic Train}\\[\extraarrayvspace]\textbf{or Pre-Train}} & \textbf{Fine-Tuning} & \mcr{\textbf{Fidelity}\\[\extraarrayvspace]\textbf{Evaluation}} & & \\

\midrule
\gr
\citet{bikmukhamedov2021} & $2021$ & -- & \mcr{\fmtTT{Lightweight}\\[\extraarrayvspace]\fmtTT{GPT-2}} &
MT &
-- &
B & Packet Header Fields & Text 
& \mcr{
UNSW-IoT-Analytics\\
[\extraarrayvspace]Private\dag\\
[\extraarrayvspace]\emph{UNSWB-NB$15$}\\
[\extraarrayvspace]\emph{ISCXVPN$2016$}\\
}
& -- 
& \mcr{UNSW-IoT-Analytics\\[\extraarrayvspace]Private\dag} 
& 
\mcr{Sequence of\\[\extraarrayvspace]Packet Header Fields} & KS\\
\citet{meng2023netgpt} & $2023$ & \fmtTT{NetGPT} & \fmtTT{GPT-2} & 
PT\&FT & 
\faCheckCircle[regular] &
F/P & Raw Packet Bytes & Text 
& \mcr{ISCXVPN$2016$\\[\extraarrayvspace]USTC-TFC$2016$\\[\extraarrayvspace]CIRA-CIC-DoHBrw$2020$\\[\extraarrayvspace]PrivII 2021\dag}
& \mcr{ISCXVPN$2016$\\[\extraarrayvspace]USTC-TFC$2016$\\[\extraarrayvspace]CIRA-CIC-DoHBrw$2020$\\[\extraarrayvspace]Cybermining-$2023$\dag}
& \mcr{ISCXVPN$2016$\\[\extraarrayvspace]USTC-TFC$2016$\\[\extraarrayvspace]CIRA-CIC-DoHBrw$2020$\\[\extraarrayvspace]Cybermining-$2023$\dag}
&
Packet Header Fields & JSD\\

\gr 
\citet{sivaroopan2023netdiffus} & $2023$ & \fmtTT{NetDiffus} & \fmtTT{DDPM} &
MT &
-- &
T &  \mcr{Aggregated Traffic\\[\extraarrayvspace] or Packet Header Fields} & Image %
& \mcr{Video Streaming\dag\\[\extraarrayvspace]Deep Fingerprinting\\[\extraarrayvspace]IoT Smart-Home\dag} 
& --
& \mcr{Video Streaming\dag\\[\extraarrayvspace]Deep Fingerprinting\\[\extraarrayvspace]IoT Smart-Home\dag} 
&
\mcr{Sequence of\\[\extraarrayvspace]Aggregated Traffic or\\[\extraarrayvspace]Packet Header Fields} & FID, CA\\

Kholgh et al.~\cite{Kholgh2023PACGPT} & $2023$ & \fmtTT{PAC-GPT} & \fmtTT{GPT-3} &
PT\&FT &
\faTimesCircle[regular] & 
F & Packet Summaries & Text 
& -- 
& TON\_IoT
& TON\_IoT 
& 
\mcr{Python code for\\[\extraarrayvspace]flow traffic generation} & SR\\

\gr
\citet{jiang2024netdiffusion} & $2024$ & \fmtTT{NetDiffusion} & \fmtTT{Stable Diffusion 1.5} &
PT\&FT &
\faTimesCircle[regular] & 
B & 
Raw Packet Bytes 
& image %
& --
& Private\dag
& Private\dag
& 
\mcr{Sequence of\\[\extraarrayvspace]Raw Packet Bytes} & \mcr{JSD, TVD,\\[\extraarrayvspace]HD, CA}\\

\citet{wang2024lens} & $2024$ & \fmtTT{LENS} & \fmtTT{T5 1.1} & PT\&FT & \faCheckCircle[regular] & F & Raw Packet Bytes & Text 
& \mcr{ISCXVPN$2016$\\[\extraarrayvspace]ISCXTor$2016$\\[\extraarrayvspace]USTC-TFC$2016$\\[\extraarrayvspace]CIRA-CIC-DoHBrw$2020$ \\ [\extraarrayvspace]CIC-IoT-$2023$} 
& \mcr{ISCXVPN$2016$\\[\extraarrayvspace]ISCXTor$2016$\\[\extraarrayvspace]USTC-TFC$2016$\\[\extraarrayvspace]CIRA-CIC-DoHBrw$2020$ \\ [\extraarrayvspace]CIC-IoT-$2023$ \\ [\extraarrayvspace]Cross-Platform} 
& \mcr{ISCXVPN$2016$\\[\extraarrayvspace]ISCXTor$2016$\\[\extraarrayvspace]USTC-TFC$2016$\\[\extraarrayvspace]CIRA-CIC-DoHBrw$2020$ \\ [\extraarrayvspace]CIC-IoT-$2023$ \\ [\extraarrayvspace]Cross-Platform} 
& 
Packet Header Fields & JSD, TVD\\

\gr
\citet{qu2024trafficgpt} & $2024$ & \fmtTT{TrafficGPT} & \fmtTT{GPT-based} &
MT & --
& F/P & Raw Packet Bytes & Text 
& \mcr{ISCXVPN$2016$\\[\extraarrayvspace]USTC-TFC-$2016$\\[\extraarrayvspace]Cross-Platform\\[\extraarrayvspace]ISCXTor$2016$\\[\extraarrayvspace]CIRA-CIC-DoHBrw$2020$\\[\extraarrayvspace]CIC-IoT-$2022$} 
& -- 
& \mcr{HTTP flow\dag\\[\extraarrayvspace]DNS flow\dag\\[\extraarrayvspace]TLS flow\dag} 
&
Packet Header Fields & JSD\\

\citet{chu2024mamba} & $2024$ & -- & \fmtTT{Mamba} & MT & -- & F & Raw Packet Bytes & Text 
& \mcr{Video Streaming\dag\\[\extraarrayvspace]Video Conferencing\dag\\[\extraarrayvspace]Social Media\dag}
& --
& \mcr{Video Streaming\dag\\[\extraarrayvspace]Video Conferencing\dag\\[\extraarrayvspace]Social Media\dag}
&
\mcr{Sequence of\\[\extraarrayvspace]Raw Packet Bytes} & JSD, TVD, HD\\

\gr
\citet{zhang2024netdiff} & $2024$ & \fmtTT{NetDiff} & \fmtTT{DDPM} & MT & -- & B & Per-Flow Counters/Stats & Numeric 
& \mcr{LabeledFlows\_2017\\[\extraarrayvspace]LabeledFlows\_2019} 
& -- 
& \mcr{LabeledFlows\_2017\\[\extraarrayvspace]LabeledFlows\_2019} 
&
Per-Flow Counters/Stats  & \mcr{JSD, TVD, CRPS\\[\extraarrayvspace]CA, $R^2$}\\

\citet{li2024lightweight} & $2024$ & \fmtTT{LW-Diff} & \mcr{\fmtTT{Lightweight}\\[\extraarrayvspace]\fmtTT{Diffusion Model}} & MT & -- & F* & \mcr{Raw Packet Bytes*} & Text 
& USTC-TFC$2016$ 
& --
& USTC-TFC$2016$ 
&
\mcr{Raw Packet Bytes*}
& CA, PR, REC, F1\\

\gr
\citet{wolf2024} & $2024$ & - & \fmtTT{GPT-2} & MT & - & F & \mcr{Aggregated Traffic\\[\extraarrayvspace]or Packet Header Fields} & Text & \mcr{CSE-CIC-IDS2018\\[\extraarrayvspace]TON\_IoT\\[\extraarrayvspace]UNSWB-NB15} & - & \mcr{CSE-CIC-IDS2018\\[\extraarrayvspace]TON\_IoT\\[\extraarrayvspace]UNSWB-NB15} & \mcr{Aggregated Traffic\\[\extraarrayvspace]or Packet Header Fields} & \mcr{JSD, MAE-Corr$\star$\\[\extraarrayvspace]FPR, F1}\\

\bottomrule
\end{tabular}
\begin{tablenotes}[flushleft]
\Large
\item
\emph{GenAI Model Architecture:} \textbf{DDPM} - Denoising Diffusion Probabilistic Model;
\emph{GenAI Train}: \textbf{MT} - Monolithic Train, \textbf{PT\&FT} - Pre-Train \& Fine-Tuning, \textbf{NetPT} - Networking Pre-Train, \faCheckCircle[regular] - Present, \faTimesCircle[regular] - Absent;
\item
\emph{Traffic-Input}: \textbf{TO} - Traffic Object: \textbf{T} - Trace, \textbf{B} - Bidirectional Flow, \textbf{F} - Flow, \textbf{P} - Packet;
\item
\emph{Networking Dataset}: \dag: Private dataset, \cite{bikmukhamedov2021} considers two train scenarios: w/ all the datasets or w/o those in \emph{italic};
\item 
\emph{Evaluation Metrics}: 
\textbf{HD} - Hellinger Distance, 
\textbf{JSD} - Jensen-Shannon Divergence, 
\textbf{KS} - Kolmogorov-Smirnov, 
\textbf{TVD} - Total Variation Distance, 
\textbf{FID} - Frechet Inception Distance, 
\textbf{CRPS} - Continuous Ranked Probability Score, 
\textbf{MAE-Corr} - Mean Absolute Error of Correlation Matrices, 
\textbf{CA} - Classification Accuracy, 
\textbf{PR} - Precision, 
\textbf{REC} - Recall, 
\textbf{F1} - F1-score, 
\textbf{$\bm{R^2}$} - Coefficient of Determination, 
\textbf{SR} - Success Rate; 
$\star$: based on the attribute type, the correlation among attributes is computed via: ($i$) the Pearson Correlation Coefficient, ($ii$) the Uncertainty Coefficient, or ($iii$) the Correlation Ratio;
\item
*: Information marked with * is not explicitly reported in the reference work but has been inferred.
\end{tablenotes}
\end{threeparttable}
}
\end{table*}

\noindent
\textbf{Definition:} \gls{traffgen} entails creating synthetic data that accurately replicate real-world network traffic patterns and behaviors.

Table~\ref{tab:traffic_generation} summarizes works addressing \gls{traffgen} with the \gls{genai} model, published since $2021$.
Notably, some studies are not explicitly focused on generating synthetic network traffic.
In fact, they also tackle tasks related to \emph{traffic understanding}, such as \gls{traffclass} and \gls{nid}~\cite{meng2023netgpt, wang2024lens, qu2024trafficgpt} by opportunistically fine-tuning \gls{genai} models.

Most of the reviewed works employ \glspl{llm} (\eg \fmtTT{GPTs} and \fmtTT{T5})~\cite{bikmukhamedov2021, meng2023netgpt, Kholgh2023PACGPT, wang2024lens, qu2024trafficgpt} while other leverage Diffusion Models (\eg \fmtTT{Stable Diffusion})~\cite{sivaroopan2023netdiffus, jiang2024netdiffusion, zhang2024netdiff}.
The sole exception is the work in~\cite{chu2024mamba} that uses \fmtTT{Mamba}.
The column ``\textbf{GenAI Model - Architecture}'' outlines the architecture used by each study.

As for the
\gls{to}, the considered works segment the traffic into packets~\cite{meng2023netgpt,qu2024trafficgpt}, unidirectional~\cite{meng2023netgpt,Kholgh2023PACGPT,wang2024lens,qu2024trafficgpt, chu2024mamba, wolf2024} or bidirectional~\cite{bikmukhamedov2021, jiang2024netdiffusion, zhang2024netdiff} flows, or even consider the whole network traffic trace~\cite{sivaroopan2023netdiffus}.
As input data for \gls{genai} models (column ``\textbf{Traffic Input - Data}''), almost all works employ raw packet bytes~\cite{meng2023netgpt,jiang2024netdiffusion,wang2024lens,qu2024trafficgpt, chu2024mamba}, optionally performing IP masking operations~\cite{jiang2024netdiffusion,wang2024lens,qu2024trafficgpt}. 
Other studies leverage the sequence of packet header fields such as \glspl{ps} and \glspl{iat}~\cite{bikmukhamedov2021}, or incorporate \glspl{dir}, \glspl{iat}, and optionally \glspl{ps}~\cite{sivaroopan2023netdiffus}.
Conversely, \citet{sivaroopan2023netdiffus} and \citet{wolf2024} also employ aggregated metrics (\eg forward/backward volume and packet count, and flow duration), while \citet{Kholgh2023PACGPT} exploit packet summary extracted with the Linux's \texttt{tcpdump} tool. 
Finally, \citet{zhang2024netdiff} leverage the sequences of traffic statistics related to, for instance, the forward/backward number of packets, \glspl{iat}, and traffic volumes.

As described in Sec.~\ref{subsec:dowstream_tasks},
\glspl{llm} and \fmtTT{Diffusion Models} are not inherently designed for handling network traffic. 
Therefore, to employ these architectures, the input data need to be formatted either in a text-based~\cite{bikmukhamedov2021, meng2023netgpt, Kholgh2023PACGPT, wang2024lens, qu2024trafficgpt, chu2024mamba, zhang2024netdiff,wolf2024} or in an image-like~\cite{sivaroopan2023netdiffus, jiang2024netdiffusion} representation (see column ``\textbf{Traffic Input - Format}'').
For datagram-to-text conversion, these approaches use complex \emph{tokenization mechanisms},
to preserve the complex hierarchical structure (\ie spatial and temporal properties of packets within a flow) of real network traffic.
Additionally, \citet{qu2024trafficgpt} encodes time information into tokens, enabling the model to generate timestamp intervals for a comprehensive representation of PCAP file data.
On the other hand, \fmtTT{Diffusion Models} need an \emph{encoding-decoding strategy} to transform network traffic data into an image format and subsequently convert them back to the original traffic format. 
To perform this operation, \citet{sivaroopan2023netdiffus} exploit the \gls{gasf}~\cite{wang2015imaging} method while \citet{jiang2024netdiffusion} use nPrint~\cite{holland2021new}.

As described in Sec.~\ref{subsec:recent_ai}, the training procedure of \gls{genai} models relies on two different strategies (see column \textbf{``GenAI Train''}): 
($i$) the naive \emph{Monolithic Training (MT)} or ($ii$) \emph{Pre-Training \& Fine-Tuning (PT\&FT)}.
Most of the studies in Tab.~\ref{tab:traffic_generation} adopt the MT approach~\cite{bikmukhamedov2021, sivaroopan2023netdiffus, qu2024trafficgpt, chu2024mamba, zhang2024netdiff, li2024lightweight, wolf2024}, training the models with heterogeneous traffic data from networking datasets including ISCXVPN$2016$, USTC-TFC$2016$, and CIRA-CIC-DoHBrw$2020$. 
For models exploiting PT\&FT, while the fine-tuning is always performed on networking datasets, the column \textbf{``NetPT''} 
indicates whether the pre-training phase includes a networking-specific corpus. 
In fact, \citet{Kholgh2023PACGPT} and \citet{jiang2024netdiffusion} adapt pre-trained models not originally trained on networking data to networking-specific tasks.
Specifically, \citet{Kholgh2023PACGPT} leverage OpenAI's \fmtTT{GPT-3}, pre-trained on a mix of publicly available and licensed Internet text, while \citet{jiang2024netdiffusion} use Stability AI's \fmtTT{Stable Diffusion 1.5}, trained on the LAION dataset.
Differently, \citet{meng2023netgpt} and \citet{wang2024lens} perform the pre-training phase on networking datasets and successive fine-tuning on specific downstream datasets.
For more details on such datasets, please refer to Tab.~\ref{tab:dataset-overview}.

\gls{traffgen} approaches can also be categorized based on their output (column ``\textbf{Generated Object}''). 
Some studies focus on generating specific packet fields, such as IP addresses, ports, and packet sizes~\cite{meng2023netgpt, wang2024lens, qu2024trafficgpt, wolf2024} or aggregated traffic features, such as traffic volume, flow duration, and number of forward/backward packets~\cite{zhang2024netdiff, wolf2024}. 
It should be noted that these methods can also be applied iteratively to generate entire flows, akin to constructing sentences from individual words in the \gls{nlp} domain.
Differently, other approaches natively generate sequences of packet header fields (\eg \gls{ps}, \gls{iat}, and \gls{dir})~\cite{bikmukhamedov2021, sivaroopan2023netdiffus}, aggregated traffic features~\cite{sivaroopan2023netdiffus}, or raw traffic bytes of entire biflows~\cite{jiang2024netdiffusion, chu2024mamba}.
Interestingly, \citet{Kholgh2023PACGPT} provide Python code for interacting with the Scapy library%
\footnote{\url{https://scapy.net/}} 
to generate traffic that matches the input desiderata. This solution
is more akin to traffic replay than generation.

Finally, to evaluate the fidelity and realism of generated data, the considered works leverage two different kinds of metrics (column \textbf{``Evaluation Metrics''}): ($i$) divergence/fidelity metrics, ($ii$) \gls{ml}-based classification accuracy of related downstream tasks, or ($iii$) success rate.
In the former case, the most common metrics are the \gls{jsd}~\cite{meng2023netgpt, jiang2024netdiffusion, wang2024lens, qu2024trafficgpt, chu2024mamba, zhang2024netdiff, wolf2024} and the \gls{tvd}~\cite{jiang2024netdiffusion, wang2024lens, chu2024mamba, zhang2024netdiff} that quantify the similarity and the maximum difference between two distributions, respectively.
Both metrics range from $0$ (identical distributions) to $1$ (completely different distributions) and are commonly evaluated together. 
Hence, lower values signify synthetic traffic more similar to the real one.
Conversely, in the case of \gls{ml}-based evaluation, the synthetic traffic is used during the training or evaluation phases of different \gls{ml} models targeted for various downstream tasks.
In the case of a traffic classification task, the variation in accuracy~\cite{sivaroopan2023netdiffus, jiang2024netdiffusion, chu2024mamba, zhang2024netdiff, li2024lightweight}, F1-score~\cite{li2024lightweight, wolf2024}, Precision, or Recall~\cite{li2024lightweight} is taken as a measure of the quality of the generated data.
\citet{zhang2024netdiff} also evaluate the generated data through traffic prediction assessed via $R^2$.
In addition, \citet{wolf2024} analyze the \gls{fpr} of a model (viz. discriminator) specifically trained to distinguish between real and synthetic traffic, with higher \gls{fpr} values indicating the difficulty in distinguishing between the two types of traffic, thereby reflecting their similarity.
Lastly, \citet{Kholgh2023PACGPT} evaluate the synthetic traffic generated by their Python code based on the \gls{sr}, quantifying the proportion of successfully sent packets out of the total generated ones.

\subsection{Network Traffic Classification}
\label{sec:traffic_classification}
\newcommand{\mrw}[2]{
\multirow{#1}{*}{#2}
}

\begin{table*}[t]
\renewcommand{\arraystretch}{1.2}
\centering
\footnotesize
\caption{
Works dealing with NTC through GenAI models (in chronological order).
}
\label{tab:traffic_classification}
\resizebox{\textwidth}{!}
{
\begin{threeparttable}
\begin{tabular}{rccccccccccccc}
\toprule
\multirow{2}[2]{*}{\textbf{Paper}} & \multirow{2}[2]{*}{\textbf{Year}} & \multicolumn{3}{c}{\textbf{GenAI Model}} &  \multirow{2}{*}{\begin{tabular}{c}
     \textbf{Pre-Training}\\[\extraarrayvspace]
     \textbf{Dataset}
\end{tabular}}& \multicolumn{2}{c}{\textbf{Traffic Input}} &  \multicolumn{6}{c}{\textbf{Downstream Tasks}}\\ 
\cmidrule(lr){3-5}
\cmidrule(lr){7-8}
\cmidrule(lr){9-14}
& & \textbf{Name} & \textbf{Architecture} 
& \textbf{NetPT}
& &\textbf{TO}& \textbf{Data} &  
\textbf{Dataset}&\textbf{Encaps.}& \textbf{Serv.} & \textbf{App.} &
\textbf{Query Met.} & \textbf{Device}
\\
\midrule
\gr
\citet{lin2022}
& $2022$  & 
\fmtTT{ET-BERT} &
\fmtTT{BERT} & \faCheckCircle[regular]
&\mcr{ISCXVPN$2016$\\[\extraarrayvspace]CIC-IDS$2017$\\[\extraarrayvspace]CSTNET$\dag$}& F& L3-PAY &\mcr{ISCXVPN$2016$\\[\extraarrayvspace]ISCXTor$2016$\\[\extraarrayvspace]CSTNET-TLS$1.3$\\[\extraarrayvspace]Cross-Platform}& 
\mcr{\Circle\\[\extraarrayvspace] \Circle\\[\extraarrayvspace] \Circle \\[\extraarrayvspace] \Circle} & \mcr{\CIRCLE\\[\extraarrayvspace] \Circle\\[\extraarrayvspace] \Circle \\[\extraarrayvspace] \Circle} & \mcr{\CIRCLE\\[\extraarrayvspace] \CIRCLE \\[\extraarrayvspace] \CIRCLE \\[\extraarrayvspace] \CIRCLE} & \mcr{\Circle\\[\extraarrayvspace] \Circle\\[\extraarrayvspace] \Circle \\[\extraarrayvspace] \Circle} & \mcr{\Circle\\[\extraarrayvspace] \Circle\\[\extraarrayvspace] \Circle \\[\extraarrayvspace] \Circle} \\
\citet{meng2023netgpt}
& $2023$ & 
\fmtTT{NetGPT} & 
\fmtTT{GPT-2} & \faCheckCircle[regular] &
\mcr{ISCXVPN$2016$\\[\extraarrayvspace]USTC-TFC$2016$\\[\extraarrayvspace]CIRA-CIC-DoHBrw$2020$\\[\extraarrayvspace]PrivII $2021$$\dag$}& F/P & L3-H+PAY  &  \mcr{ISCXVPN$2016$\\[\extraarrayvspace]CIRA-CIC-DoHBrw$2020$} &
\mcr{\CIRCLE\\[\extraarrayvspace]\Circle} & \mcr{\Circle\\[\extraarrayvspace]\Circle} & \mcr{\CIRCLE\\[\extraarrayvspace]\Circle} & \mcr{\Circle\\[\extraarrayvspace]\CIRCLE} & \mcr{\Circle\\[\extraarrayvspace]\Circle} \\ 
\gr
\citet{guthula2023netfound}
& $2023$  & 
\fmtTT{netFound} & 
\fmtTT{Transformer} & 
\faCheckCircle[regular]
&Private$\dag$& F & \mcr{L4/L5 Fields \\[\extraarrayvspace]+\\[\extraarrayvspace]Metadata} & 
Private$\dag$ & \Circle & \CIRCLE & \Circle &
\Circle &\Circle \\
\mcr{\citet{sarabi2023}} & $2023$  & -- &
\fmtTT{RoBERTa} & \faCheckCircle[regular]
& CensysBQ & S & HTTP Messages & CensysBQ & \Circle & \Circle & \Circle &
\Circle & \CIRCLE \\
\gr
\citet{wang2024lens}
& $2024$  & 
\fmtTT{LENS} & 
\fmtTT{T5 1.1} & 
\faCheckCircle[regular]
&\mcr{ISCXVPN$2016$\\[\extraarrayvspace]ISCXTor$2016$\\[\extraarrayvspace]USTC-TFC$2016$\\[\extraarrayvspace]CIRA-CIC-DoHBrw$2020$\\[\extraarrayvspace]CIC-IoT-$2023$} & F & L3-H+PAY  & \mcr{ISCXVPN$2016$\\[\extraarrayvspace]ISCXTor$2016$
\\[\extraarrayvspace]CIRA-CIC-DoHBrw$2020$ \\[\extraarrayvspace]Cross-Platform}
& 
\mcr{\CIRCLE\\[\extraarrayvspace] \CIRCLE \\[\extraarrayvspace] \Circle \\[\extraarrayvspace] \Circle} & \mcr{\CIRCLE\\[\extraarrayvspace] \CIRCLE \\[\extraarrayvspace] \Circle \\[\extraarrayvspace] \Circle} & \mcr{\CIRCLE\\[\extraarrayvspace] \Circle \\[\extraarrayvspace] \CIRCLE \\[\extraarrayvspace] \Circle} & \mcr{\Circle\\[\extraarrayvspace] \Circle \\[\extraarrayvspace] \Circle \\[\extraarrayvspace] \CIRCLE}& \mcr{\Circle\\[\extraarrayvspace] \Circle\\[\extraarrayvspace] \Circle \\[\extraarrayvspace] \Circle} \\ 
\citet{qu2024trafficgpt}
& $2024$  & 
\fmtTT{TrafficGPT} &
\fmtTT{GPT-based} & 
\faCheckCircle[regular]
&\mcr{ISCXVPN$2016$\\[\extraarrayvspace]USTC-TFC$2016$\\[\extraarrayvspace]Cross-Platform\\[\extraarrayvspace]ISCXTor$2016$\\[\extraarrayvspace]CIRA-CIC-DoHBrw$2020$\\[\extraarrayvspace]CIC-IoT-$2022$}& F/P & L3-H+PAY  &\mcr{ISCXVPN$2016$\\[\extraarrayvspace]USTC-TFC$2016$*\\[\extraarrayvspace]Cross-Platform}&
\mcr{\Circle\\[\extraarrayvspace] \Circle\\[\extraarrayvspace] \Circle} &
\mcr{\Circle\\[\extraarrayvspace] \Circle\\[\extraarrayvspace] \Circle} &
\mcr{\CIRCLE\\[\extraarrayvspace] \CIRCLE\\[\extraarrayvspace] \CIRCLE} & 
\mcr{\Circle\\[\extraarrayvspace] \Circle\\[\extraarrayvspace] \Circle}  & 
\mcr{\Circle\\[\extraarrayvspace] \Circle\\[\extraarrayvspace] \Circle} \\ 
\gr
\citet{wang2024netmamba}
& $2024$  & 
\fmtTT{NetMamba} &
\fmtTT{Mamba} & \faCheckCircle[regular]
&\mcr{ISCXVPN$2016$\\[\extraarrayvspace]ISCXTor$2016$\\[\extraarrayvspace]USTC-TFC$2016$\\[\extraarrayvspace]Cross-Platform\\[\extraarrayvspace]CIC-IoT-$2022$}& F & L3-H+PAY  &   \mcr{ISCXVPN$2016$\\[\extraarrayvspace]ISCXTor$2016$\\[\extraarrayvspace]Cross-Platform} &
\mcr{\Circle\\[\extraarrayvspace] \Circle \\[\extraarrayvspace] \Circle} & \mcr{\CIRCLE\\[\extraarrayvspace] \CIRCLE \\[\extraarrayvspace] \Circle} & \mcr{\Circle\\[\extraarrayvspace] \Circle \\[\extraarrayvspace] \CIRCLE} & 
\mcr{\Circle\\[\extraarrayvspace] \Circle \\[\extraarrayvspace] \Circle} & 
\mcr{\Circle\\[\extraarrayvspace] \Circle \\[\extraarrayvspace] \Circle} \\
\citet{tao2024lambert}
& $2024$  & 
\fmtTT{LAMBERT} & 
\fmtTT{BERT} & 
\faCheckCircle[regular]
&\mcr{ISCXVPN$2016$\\[\extraarrayvspace]CIC-IDS$2017$\\[\extraarrayvspace]CSE-CIC-IDS$2018$\\[\extraarrayvspace]CSTNET-TLS$1.3$} & F & L4-PAY  & \mcr{ISCXVPN$2016$\\[\extraarrayvspace]CSTNET-TLS$1.3$} & 
\mcr{\Circle\\[\extraarrayvspace] \Circle} & \mcr{\CIRCLE\\[\extraarrayvspace] \Circle} & \mcr{\CIRCLE \\[\extraarrayvspace] \CIRCLE} & \mcr{\Circle\\[\extraarrayvspace] \Circle}& \mcr{\Circle\\[\extraarrayvspace] \Circle} \\ 

\gr \citet{li2024albert}
& $2024$
& -
& \fmtTT{ALBERT}
&  \faCheckCircle[regular]
& CSTNET-TLS1.3
&  F
&  L3-H+PAY
& \mcr{ISCXVPN2016\\[\extraarrayvspace]CSTNET-TLS1.3\\[\extraarrayvspace]EduTLS$\dag$\\[\extraarrayvspace]Cross-Platform} 

& \mcr{\CIRCLE\\[\extraarrayvspace] \Circle \\[\extraarrayvspace] \Circle \\[\extraarrayvspace] \Circle}
& \mcr{\Circle\\[\extraarrayvspace] \Circle \\[\extraarrayvspace] \CIRCLE \\[\extraarrayvspace] \Circle}  
& \mcr{\Circle\\[\extraarrayvspace] \CIRCLE \\[\extraarrayvspace] \Circle \\[\extraarrayvspace] \CIRCLE}
& \mcr{\Circle\\[\extraarrayvspace] \Circle \\[\extraarrayvspace] \Circle \\[\extraarrayvspace] \Circle}
& \mcr{\Circle\\[\extraarrayvspace] \Circle \\[\extraarrayvspace] \Circle \\[\extraarrayvspace] \Circle}
\\ 
\bottomrule
\end{tabular}
\begin{tablenotes}[flushleft]
\footnotesize
\item\emph{GenAI Model Architecture}: \textbf{NetPT} - Networking Pre-Train, \faCheckCircle[regular] - Present, \faTimesCircle[regular] - Absent; \emph{Traffic-Input}: \textbf{TO} - Traffic Object: \textbf{F} - Flow, \textbf{P} - Packet, \textbf{S} - Service, \textbf{L3/L4} - ISO/OSI Network/Transport Layer, \textbf{H/PAY} - Header/Payload Bytes, \textbf{Fields} - Header Fields; $\dag$ - Private Data; \emph{Downstream Tasks}: \textbf{Encaps.} - Traffic Encapsulation Identification, \textbf{Serv.} - Traffic Service Classification, \textbf{App.} - Application Classification, \textbf{Query Met.} - Query Metrics Classification, \textbf{Device} - Device Classification;
\item\textbf{Note}: * \fmtTT{TrafficGPT} by~\citet{qu2024trafficgpt} has been naively applied to a security dataset (\ie USTC-TFC$2016$) without mentioning the NID use case. Accordingly, we treat this work as NTG and NTC, without including it in the NID use case.
\end{tablenotes}
\end{threeparttable}
}

\end{table*}

\noindent
\textbf{Definition:} \gls{traffclass} involves categorizing network traffic based on various attributes such as protocols, services, and application types.

Table~\ref{tab:traffic_classification} summarizes the approaches employing \glspl{llm} to address \gls{traffclass}. 
It details the model type, traffic input, classification tasks, and training and evaluation datasets.

All the reviewed works leverage \glspl{llm}, such as \fmtTT{Transformer}, \fmtTT{BERT}, \fmtTT{GPT-2}, and \fmtTT{Mamba},
adapted from other domains (\eg \gls{nlp}) to the traffic context (column ``\textbf{Architecture}'') and optionally modify some architectural elements.
For instance,~\cite{guthula2023netfound} propose a hierarchical \fmtTT{Transformer} model to process data at multiple levels of granularity (\eg intra- and inter-bursts).
The adaptation involves transforming network traffic data into a text-based representation, followed by tokenization, allowing models to learn the complex characteristics of network traffic directly.
Notably, for the tokenization process, considered works leverage \emph{Datagram2Token}~\cite{lin2022, guthula2023netfound, wang2024netmamba, tao2024lambert}, \emph{SentencePiece}~\cite{meng2023netgpt, wang2024lens, qu2024trafficgpt}, \emph{WordPiece}~\cite{wang2024lens}, or \emph{Byte-Pair Encoding}~\cite{sarabi2023}.

As shown, most of the approaches consider the flow as the \gls{to}, 
while~\citet{sarabi2023} leverage 
the network service in terms of destination IP and port (column ``\textbf{TO}'').
The input data typically includes the header and payload of the network layer~\cite{meng2023netgpt, wang2024lens, qu2024trafficgpt, wang2024netmamba,li2024albert}, or only the network-layer~\cite{lin2022} and transport-layer payload~\cite{tao2024lambert}.
Conversely,~\citet{guthula2023netfound} leverage fields from both the transport and application layers along with metadata at both packet and burst levels, while~\cite{sarabi2023} focus on HTTP messages (column ``\textbf{Traffic Input Data}'').

Furthermore, all the reviewed works include a pre-training stage for the \gls{llm} architecture (column ``\textbf{NetPT}''), through a self-supervised learning approach typically involving two tasks:
\begin{enumerate*}[label=(\emph{\alph*})]
    \item \emph{Masked Burst Model} to capture the relationships between different datagram bytes within the same burst, and
    \item \emph{Same-origin Burst Prediction} to model the transmission relationships between preceding and subsequent bursts. 
\end{enumerate*}
Then, the resulting model is fine-tuned by adapting the pre-trained model to various traffic classification tasks and adjusting its parameters to optimize performance on the labeled data.
These processes are performed through the use of different datasets. Primarily, most of the works leverage ISCXVPN-2016~\cite{lin2022, meng2023netgpt,wang2024lens, qu2024trafficgpt, wang2024netmamba, tao2024lambert}, USTCTFC-2016~\cite{meng2023netgpt, qu2024trafficgpt,wang2024netmamba}, CIC-DoHBrw-2020~\cite{meng2023netgpt, wang2024lens, qu2024trafficgpt}, or ISCXTor-2016~\cite{wang2024lens, qu2024trafficgpt, wang2024netmamba}.
Additionally, also traffic from \gls{nids} (\eg CIC-IDS-2017, CIC-IDS-2018) and \gls{iot} devices (\eg CIC-IoT-2022, CIC-IoT-2023) is included to pre-train \glspl{llm}~\cite{lin2022, wang2024lens, qu2024trafficgpt, wang2024netmamba, tao2024lambert} (column ``\textbf{Pre-Training/Fine-Tuning Datasets}'').

The reviewed works address different \gls{traffclass} tasks, differing in granularity and classification types.
Specifically, most approaches focus on classifying the 
service~\cite{lin2022,guthula2023netfound,wang2024lens,wang2024netmamba,li2024albert,tao2024lambert} (ref. \textbf{Serv.}) or application~\cite{lin2022,meng2023netgpt,wang2024lens,qu2024trafficgpt,wang2024netmamba,tao2024lambert,li2024albert} (ref. \textbf{App.}) generating the traffic.
Other works focus on detecting VPN-/Tor-encapsulated traffic~\cite{meng2023netgpt,wang2024lens,li2024albert} 
(ref. \textbf{Encaps.}). 
To this end, these approaches primarily use datasets incorporating diversified traffic on multiple levels (\eg ISCXVPN-2016 and ISCXTor-2016) during both the pre-training and fine-tuning stages.
Conversely, only a few works focus on classifying DNS queries using the DoH protocol~\cite{wang2024lens} (ref. \textbf{Query Met.}).
Noteworthy, some studies introduce datasets specifically for evaluation purposes, which are not used during the pre-training and fine-tuning phases. 
Examples include Cross Platform~\cite{lin2022, wang2024lens, qu2024trafficgpt,li2024albert} and CSTNET-TLS1.3~\cite{lin2022}.

\subsection{Network Intrusion Detection}
\label{sec:intrusion_detection}

\begin{table*}[ht!]
\centering
\caption{
Works dealing with NID through GenAI models (in chronological order).
}
\label{tab:rw-id}
\resizebox{\textwidth}{!}{
\begin{threeparttable}
\begin{tabular}{rcccccccccc}
\toprule

\multirow{2}{*}{\textbf{Paper}} &
\multirow{2}{*}{\textbf{Year}} &
\multicolumn{3}{c}{\textbf{GenAI Model}} & 
\multicolumn{2}{c}{\textbf{Traffic Input}} &
\multicolumn{2}{c}{\textbf{Datasets}} &
\multirow{2}{*}{\textbf{Downstream Task}} \\
\cmidrule(lr){3-5}
\cmidrule(lr){6-7}
\cmidrule(lr){8-9}
 &  & \textbf{Name} & \textbf{Architecture} 
 & \textbf{NetPT}
 & \textbf{TO} & \textbf{Data} 
 & \textbf{Pre-Training} & \textbf{Fine-Tuning}
 \\
\midrule

\gr
\citet{nam2021intrusion} 
& 2021 
& --
& \fmtTT{GPT-1}
& \faTimesCircle[regular]
& B
& CAN ID sequences 
& -- 
& Hyundai Avante CN$7$ 
& NAD 
\\

\citet{yu2021securing} 
& 2021 
& --
& \fmtTT{BERT} 
& \faTimesCircle[regular]
& AS
& APT characteristics 
& --  
& Power Grid Data 
& mMD 
\\

\gr
\citet{li2022extreme} 
& 2022 
& \fmtTT{ESeT}
& \fmtTT{Transformer} 
& \faTimesCircle[regular]
& F 
& \begin{tabular}[c]{@{}c@{}} L4-PAY, \\[\extraarrayvspace] Packet-level byte \\[\extraarrayvspace] encoded features, \\[\extraarrayvspace] Flow-level frequency \\[\extraarrayvspace] domain features \end{tabular} 
& -- 
& \begin{tabular}[c]{@{}c@{}} CIC-IDS$2017$ \\[\extraarrayvspace] CSE-CIC-IDS$2018$ \end{tabular} 
& bMD 
\\

\citet{seyyar2022attack} 
& 2022 
& --
& \fmtTT{BERT} 
& \faTimesCircle[regular]
& B
& HTTP requests
& -- 
& \begin{tabular}[c]{@{}c@{}} CSIC $2010$ \\[\extraarrayvspace] FWAF \\[\extraarrayvspace] HttpParams \end{tabular} 
& bMD 
\\

\gr
\citet{ho2022network} 
& 2022 
& --
& \fmtTT{ViT}
& \faTimesCircle[regular]
& B 
& Stats
& --
& \begin{tabular}[c]{@{}c@{}} CIC-IDS$2017$ \\[\extraarrayvspace] UNSW-NB$15$ \end{tabular} 
& bMD, mMD 
\\ 

\citet{wu2022} 
& 2022 
& \fmtTT{RTIDS}
& \fmtTT{Transformer}
& \faTimesCircle[regular]
& B 
& Stats 
& -- 
& \begin{tabular}[c]{@{}c@{}} CIC-IDS$2017$ \\[\extraarrayvspace] CIC-DDoS$2019$ \end{tabular} 
& mMD  
\\

\gr
\citet{ghourabi2022security} 
& 2022 
& --
& \fmtTT{BERT} 
& \faTimesCircle[regular]
& B 
& Stats 
& -- 
& \begin{tabular}[c]{@{}c@{}} ECU-IoTH \\[\extraarrayvspace] TON\_IoT \\[\extraarrayvspace] Edge-IIoTset \end{tabular} 
& bMD 
\\

\citet{lin2022} 
& 2022 
& \fmtTT{ET-BERT}
& \fmtTT{Transformer} 
& \faCheckCircle[regular]
& F 
& L3-PAY 
& \begin{tabular}[c]{@{}c@{}} ISCXVPN$2016$ \\[\extraarrayvspace] CIC-IDS$2017$ \\[\extraarrayvspace] CSTNET$\dag$ \end{tabular}
& USTC-TFC$2016$
& mMD
\\

\gr
\citet{lai2023} 
& 2023 
& --
& \fmtTT{BERT} 
& \faTimesCircle[regular]
& F 
& Stats 
& -- 
& ISCX NSL-KDD 
& bMD 
\\

Ali et al.~\cite{ali2023} 
& 2023 
& \fmtTT{HuntGPT}
& \fmtTT{GPT-3.5 turbo} 
& \faTimesCircle[regular]
& F 
& Stats 
& -- 
& KDD'$99$ 
& mMD 
\\

\gr
\citet{ullah2023tnn} 
& 2023 
& \fmtTT{TNN-IDS}
& \fmtTT{Transformer}
& \faCheckCircle[regular]
& P, F, B 
& Stats 
& MQTT-IoT-IDS$2020$
& MQTT-IoT-IDS$2020$
& bMD
\\

\citet{wang2023robust} 
& 2023 
& \fmtTT{RUIDS}
& \fmtTT{Transformer} 
& \faTimesCircle[regular]
& BoF 
& \begin{tabular}[c]{@{}c@{}} Stats \end{tabular} 
& -- 
& \begin{tabular}[c]{@{}c@{}} KDD'$99$ \\[\extraarrayvspace] UNSW-NB$15$ \\[\extraarrayvspace] CIC-IDS$2017$ (Friday) \\[\extraarrayvspace] CIC-IDS$2017$ (Wednesday) \end{tabular} 
& bMD 
\\

\gr
\citet{meng2023netgpt} 
& 2023 
& \fmtTT{NetGPT}
& \fmtTT{GPT-2} 
& \faCheckCircle[regular] 
& F
& L3-H+PAY
& \begin{tabular}[c]{@{}c@{}} ISCXVPN$2016$ \\[\extraarrayvspace] USTC-TFC$2016$ \\[\extraarrayvspace] CIRA-CIC-DoHBrw$2020$ \\[\extraarrayvspace] PrivII 2021$\dag$ \end{tabular} 
& \begin{tabular}[c]{@{}c@{}} USTC-TFC$2016$ \\[\extraarrayvspace] CIRA-CIC-DoHBrw$2020$ \\[\extraarrayvspace] Cybermining-$2023$$\dag$ \end{tabular} 
& bMD 
\\

\citet{guthula2023netfound} 
& 2023 
& \fmtTT{netFound}
& \fmtTT{Transformer} 
& \faCheckCircle[regular]
& F
& \begin{tabular}[c]{@{}c@{}} L4/L5 Fields  \\[\extraarrayvspace]
+  \\[\extraarrayvspace]
Metadata  \\[\extraarrayvspace]
\end{tabular}
& Private$\dag$
& CIC-IDS$2017$
& mMD 
\\

\gr
\citet{manocchio2024flowtransformer} 
& 2024 
& \fmtTT{FlowTransformer}
& \begin{tabular}[c]{@{}c@{}} \fmtTT{GPT-3} \\[\extraarrayvspace] \fmtTT{BERT}
\end{tabular} 
& \faTimesCircle[regular]
& F 
& Stats 
& -- 
& \begin{tabular}[c]{@{}c@{}} ISCX NSL-KDD \\[\extraarrayvspace] UNSW-NB$15$ \\[\extraarrayvspace] CIC-IDS$2017$ \\[\extraarrayvspace] CSE-CIC-IDS$2018$ \\[\extraarrayvspace] MQTT-IoT-IDS$2020$\\[\extraarrayvspace] TON\_IoT \end{tabular} 
& bMD 
\\

\citet{wang2024lens} 
& 2024 
& \fmtTT{LENS}
& \fmtTT{T5 1.1}
& \faCheckCircle[regular]
& F 
& L3-H+PAY 
& \begin{tabular}[c]{@{}c@{}} ISCXVPN$2016$ \\[\extraarrayvspace] ISCXTor-$2016$ \\[\extraarrayvspace] USTC-TFC$2016$ \\[\extraarrayvspace] CIRA-CIC-DoHBrw$2020$ \\[\extraarrayvspace] CIC-IoT-$2023$ \end{tabular}  
& \begin{tabular}[c]{@{}c@{}} USTC-TFC$2016$ \\[\extraarrayvspace] CIRA-CIC-DoHBrw$2020$ \\[\extraarrayvspace] CIC-IoT-$2023$ \end{tabular}
& bMD, mMD 
\\

\gr
\citet{wang2024netmamba}
& 2024 
& \fmtTT{NetMamba}
& \fmtTT{Mamba}  
& \faCheckCircle[regular]
& F 
& L4-PAY 
& \begin{tabular}[c]{@{}c@{}} ISCXVPN$2016$ \\[\extraarrayvspace] ISCXTor-$2016$ \\[\extraarrayvspace] USTC-TFC$2016$ \\[\extraarrayvspace] Cross-Platform \\[\extraarrayvspace] CIC-IoT-$2022$ \end{tabular} 
& \begin{tabular}[c]{@{}c@{}} USTC-TFC$2016$ \\[\extraarrayvspace] CIC-IoT-$2022$ \end{tabular} 
& mMD 
\\

\citet{wang2024lightweight} 
& 2024 
& \fmtTT{BT-TPF}
& \fmtTT{BERT-of-Theseus} 
& \faCheckCircle[regular]
& N.D. 
& \begin{tabular}[c]{@{}c@{}} Stats
\end{tabular} 
& CIC-IDS$2017$ 
& \begin{tabular}[c]{@{}c@{}} CIC-IDS$2017$ \\[\extraarrayvspace] TON\_IoT \end{tabular} 
& mMD 
\\

\gr
\citet{melicias2024gpt} 
& 2024 
& --
& \fmtTT{GPT-1} 
& \faTimesCircle[regular]
& B 
& Stats 
& --
& Edge-IIoTset 
& mMD 
\\

\citet{ferrag2024} 
& 2024 
& \fmtTT{SecurityBERT}
& \fmtTT{BERT}  
& \faTimesCircle[regular]
& F 
& L3/L4/L5 Fields 
& -- 
& Edge-IIoTset 
& mMD 
\\

\bottomrule
\end{tabular}
\begin{tablenotes}[flushleft]
\footnotesize
\item\emph{GenAI Model Architecture}: \textbf{NetPT} - Networking Pre-Train, \faCheckCircle[regular] - Present, \faTimesCircle[regular] - Absent; \emph{Traffic-Input}: \textbf{TO} - Traffic Object: \textbf{AS} - Attack Sequence, \textbf{B} - Bidirectional Flow, \textbf{BoF} - Bag of Flows, \textbf{F} - Flow, \textbf{P} - Packet, \textbf{L3/L4/L5} - ISO/OSI Network/Transport Layer, \textbf{H/PAY} - Header/Payload Bytes, \textbf{Fields} - Header Fields; $\dag$ - Private Data; \emph{Downstream Tasks}: \textbf{NAD} - Network Anomaly Detection, \textbf{mMD} - multiclass Misuse Detection,\textbf{ bMD} - binary Misuse Detection.
\end{tablenotes}
\end{threeparttable}
}
\end{table*}

\noindent
\textbf{Definition:} \gls{nid}
is an umbrella term that covers network security tasks that aim to identify and recognize malicious behavior from network traffic and collectively contribute to the design of so-called \gls{nids}.

Accordingly, in this section, we review works that propose \gls{genai} solutions ending up under the \gls{nid}
use-case umbrella. 
Such works are described in Tab.~\ref{tab:rw-id} alongside four main views, namely
($i$) \textbf{``Downstream Task''} details,
($ii$) the leveraged \textbf{``GenAI Model''} characteristics,
($iii$) the \textbf{``Traffic Input''} fed, and
($iv$) \textbf{``Pre-Training''}  and \textbf{``Fine-Tuning Datasets''}.

First, we categorize the literature based on the \textbf{``Downstream Task''}.
Indeed \gls{nids} can be taxonomized based on their modeling objective, namely 
\gls{md} or \gls{nad}.
\gls{md} relies on recognizing both normal (benign) and anomalous (malicious) behaviors in a supervised fashion, 
whereas 
\gls{nad} focuses only on normal network traffic and identifies anomalies as deviations from legitimate behavior.
Within \gls{md}, we also distinguish between \gls{bmd} and \gls{mmd}.
The main difference is that \gls{mmd} can identify specific attack types (and thus enable attack-tailored countermeasures), whereas \gls{bmd} simply distinguishes between legitimate and malicious traffic.
By looking at the reviewed literature, only the work by \citet{nam2021intrusion} performs \gls{nad}, while the remaining studies are divided between \gls{bmd}~\cite{li2022extreme, seyyar2022attack, ghourabi2022security, lai2023, ullah2023tnn, wang2023robust, meng2023netgpt, manocchio2024flowtransformer, wang2024lens} and \gls{mmd}~\cite{yu2021securing, wu2022, lin2022, ali2023, guthula2023netfound, wang2024lens, wang2024netmamba, wang2024lightweight, melicias2024gpt, ferrag2024},
with~\cite{ho2022network} performing both.

Secondly, regarding the \textbf{``GenAI Model''}, 
the most common choice falls in basic \fmtTT{Transformers} or \fmtTT{ViT} (\ie full encoder-decoder category)~\cite{li2022extreme, wu2022, ullah2023tnn, wang2023robust, lin2022, guthula2023netfound, ho2022network},
then \fmtTT{BERT} or variants follow (\ie encoder-only category)~\cite{yu2021securing, seyyar2022attack, ghourabi2022security, lai2023, ferrag2024, wang2024lightweight, manocchio2024flowtransformer},
and minor attention is posed on \fmtTT{GPT}-based solutions (\ie decoder-only category)~\cite{nam2021intrusion, ali2023, melicias2024gpt, meng2023netgpt}.
Other models, such as Google's \fmtTT{T5}~\cite{wang2024lens} and \fmtTT{Mamba}~\cite{wang2024netmamba} are also explored.
In detail, 
\fmtTT{Transformers} and \fmtTT{ViT} are commonly used ``as-is'' in many studies~\cite{li2022extreme, wu2022, ullah2023tnn}. 
\citet{wang2023robust} enhance the transformer encoder's training phase by incorporating a contrastive loss term, and the encoder output is subsequently fed into a transformer decoder for sample reconstruction.
Similarly, \fmtTT{BERT} is often employed with minimal modifications~\cite{yu2021securing, seyyar2022attack, lai2023, ferrag2024}. 
\citet{ghourabi2022security} extends \fmtTT{BERT} by proposing a framework that utilizes it alongside \emph{LightGBM} (Light Gradient Boosting Machine, an open-source distributed gradient boosting framework).
\citet{manocchio2024flowtransformer} introduce \fmtTT{FlowTransformer}, which integrates \fmtTT{BERT} and \fmtTT{GPT-3}. Their architecture includes a shallow encoder and decoder, \fmtTT{GPT} as a deep decoder, and \fmtTT{BERT} as a deep encoder, with a \gls{mlp} for classification.
Regarding studies based on \fmtTT{GPT}, 
\citet{nam2021intrusion} combine two \fmtTT{GPT-1} networks in a bi-directional manner, employing a forward and a backward \fmtTT{GPT} followed by a dense layer and softmax.
\citet{ali2023} use \fmtTT{GPT-3.5 turbo} for explainability purposes alongside various \gls{xai} techniques.
Lastly, \citet{melicias2024gpt} leverage \fmtTT{GPT-1} for data augmentation. They demonstrate that \fmtTT{GPT}-based methods can generate invalid data, leading to performance degradation in \gls{mmd}.

Concerning the particular data format fed to models (see \textbf{``Traffic Input''} column),
the majority of studies~\cite{ho2022network, wu2022, ghourabi2022security, lai2023, ali2023, ullah2023tnn, wang2023robust, manocchio2024flowtransformer, wang2024lightweight, melicias2024gpt} use pre-processed features, such as flow-based statistics typically derived from pre-processed datasets.
Four works~\cite{nam2021intrusion, yu2021securing, seyyar2022attack} utilize inputs specific to their particular domain, such as
HTTP requests
(for anomalous HTTP requests detection), \gls{can} ID sequences (for \gls{can} intrusion detection) or \gls{apt} attack sequence (for \gls{iot} \gls{apt} attack detection).
The remaining studies use fields extracted from packet headers~\cite{guthula2023netfound, ferrag2024}, payload bytes~\cite{li2022extreme, lin2022, wang2024netmamba}, or both~\cite{meng2023netgpt, wang2024lens}.

One third of the works%
~\cite{lin2022, ullah2023tnn, meng2023netgpt, guthula2023netfound, wang2024lens, wang2024netmamba, wang2024lightweight} pre-trains the models.
The \textbf{``Pre-Training Dataset''} may belong to various domains 
such as: VPN traffic (\ie~ISCXVPN-2016~\cite{lin2022, meng2023netgpt, wang2024lens, wang2024netmamba}), Tor traffic (\ie~ISCXTor-2016~\cite{wang2024lens, wang2024netmamba}), and malicious traffic in \gls{iot} and non-\gls{iot} contexts (\ie~CIC-IDS2017~\cite{wang2024lightweight, lin2022}, USTC-TFC2016~\cite{meng2023netgpt, wang2024lens, wang2024netmamba}, 
CIRA-CIC-DoHBrw-2020~\cite{wang2024lens}, 
and CIC-IoT-2022~\cite{wang2024lens, wang2024netmamba}).

Differently from the pre-training phase, for the downstream task, the datasets (see column \textbf{``Fine-Tuning Dataset''}) are exclusively related to the cybersecurity domain.
The most common dataset is CIC-IDS2017~\cite{li2022extreme, ho2022network, wu2022, wang2023robust, manocchio2024flowtransformer, guthula2023netfound, wang2024lightweight}.
Other frequently used datasets include those for \gls{iot}-domain attacks (\eg Edge-IIoTset~\cite{ghourabi2022security, ferrag2024, melicias2024gpt}, MQTT-IoT-IDS2020~\cite{ullah2023tnn, manocchio2024flowtransformer}, TON\_IoT~\cite{ghourabi2022security, manocchio2024flowtransformer, wang2024lightweight}, and CIC-IoT-2022~\cite{wang2024netmamba}). 
Some studies rely on outdated datasets (\ie~KDD'99~\cite{ali2023, wang2023robust} and its improved version NSL-KDD~\cite{manocchio2024flowtransformer, lai2023}) collected over $20$ years ago, which no longer accurately reflect contemporary network traffic.

\subsection{Networked System Log Analysis}
\label{sec:log_analysis}

\setlength{\extraarrayvspace}{-0.45ex}
\begin{table*}[ht!]
\renewcommand{\arraystretch}{1.2}
\centering
\footnotesize
\caption{
Works dealing with NSLA through GenAI models (in chronological order).
}
\label{tab:log_analysis}
\resizebox{\textwidth}{!}
{
\begin{threeparttable}
\begin{tabular}{rccccccccm{4.5cm}}
\toprule
\multirow{2}[2]{*}{\textbf{Paper}} & \multirow{2}[2]{*}{\textbf{Year}} & 
\multirow{2}[2]{*}{\textbf{Input Type}}&
\multicolumn{5}{c}{\textbf{GenAI Model}} &
\multirow{2}[2]{*}{\begin{tabular}{c}
     \textbf{Evaluation/Fine-Tuning} \\
     \textbf{Datasets} 
\end{tabular}} &
\multirow{2}[2]{*}{\textbf{Downstream Task}} 
\\ 
\cmidrule(lr){4-8}
& & & \textbf{Name} &\textbf{Architecture} & 
\textbf{Fine-Tuned} & \textbf{Released} & \textbf{Goal} \\
\midrule

\gr
\citet{setianto2021gpt} & $2021$ & Honeypot logs & \fmtTT{GPT-2C} & \fmtTT{GPT-2} & 
\faCheckCircle[regular] & \faTimesCircle[regular] & 
\begin{tabular}{c}
Q\&A\\[\extraarrayvspace]
Parsing
\end{tabular} & \begin{tabular}{c}
CyberLab 
Honeynet
\end{tabular} & 
Parsing Unix commands
in real-time
\\ 
\citet{ott2021robust} & $2021$ & \begin{tabular}{c}
Cloud environment logs
\end{tabular} & -- & \begin{tabular}{c}
     \fmtTT{BERT}, \fmtTT{GPT-2},\\[\extraarrayvspace]
     \fmtTT{XLNet}
\end{tabular} & 
\faTimesCircle[regular] & \faCheckCircle[regular] & \begin{tabular}{c}Log\\[\extraarrayvspace]
Vectorization\end{tabular} & \begin{tabular}{c}
Loghub
(OpenStack)
\end{tabular} & 
     LAD from preprocessed
     log performed via Bi-LSTM
\\

\gr
\citet{pan2023raglog} & $2023$ & Supercomputer logs & 
\fmtTT{RAGLog} & \fmtTT{GPT-3.5} & 
\faTimesCircle[regular] & \faCheckCircle[regular] & \begin{tabular}{c}
LAD
\end{tabular} & 
\begin{tabular}{c}
Loghub \\[\extraarrayvspace]
(BGL, Thunderbird) 
\end{tabular} & 
Detecting anomalies by leveraging
a Retrieval Augmented Generation 
\\

\citet{qi2023loggpt} & $2023$ & Supercomputer logs & 
\fmtTT{LogGPT} & \fmtTT{GPT-1} & 
\faTimesCircle[regular] & \faCheckCircle[regular] & 
\begin{tabular}{c}
Q\&A LAD
\end{tabular}
& \begin{tabular}{c}
Loghub (BGL),
\\[\extraarrayvspace] 
Spirit 
\end{tabular} & 
     Prompt-based LAD\\

\gr
\citet{jiang2023lilac} & $2023$ & Various log types & 
\fmtTT{LILAC} & \fmtTT{GPT-3.5 turbo} & 
\faTimesCircle[regular] & \faCheckCircle[regular] & \begin{tabular}{c}
    Log \\ [\extraarrayvspace]
    Parsing
\end{tabular} & Loghub
& 
Parse logs using in-context learning
and adaptive cache 
\\

\citet{ji2023log} & $2023$ & \begin{tabular}{c}
Deployment framework and\\[\extraarrayvspace]
hardware network logs
\end{tabular} & -- & \fmtTT{GPT-2} & 
\faTimesCircle[regular] & \faCheckCircle[regular] & \begin{tabular}{c}
LAD
\end{tabular}&
\begin{tabular}{c}
Ada and Bob
\end{tabular} & 
Detect anomalies using semantic and
sequential features with alarm strategy
\\

\gr
Mudgal et al.~\cite{mudgal2023assessment}& $2023$ & 
\begin{tabular}{c}
Cloud environment logs,\\[\extraarrayvspace]
supercomputer logs\\[\extraarrayvspace]
and prompts
\end{tabular} & -
& \fmtTT{GPT-3.5 turbo} & 
\faTimesCircle[regular] & \faCheckCircle[regular] & 
\begin{tabular}{c}
     Q\&A Parsing,\\[\extraarrayvspace]
     Analytics,\\[\extraarrayvspace]
    Summarization
\end{tabular}
& Loghub & 
Log parsing and summarization,
API and LAD with ChatGPT
\\

\citet{sun2023design} & $2023$ & \begin{tabular}{c}
     Spring Boot apps,\\[\extraarrayvspace]
     Kubernetes clusters,\\[\extraarrayvspace]
     HTTP services logs
\end{tabular}& -- & Generic LLMs & 
\faTimesCircle[regular] & \faCheckCircle[regular] & \begin{tabular}{c}
LAD
\end{tabular} &
\begin{tabular}{c}
Spring Boot, Kubernetes,\\[\extraarrayvspace]
and HTTP services logs
\end{tabular}
& 
Log management in cloud-native
environments with real-time monitoring
and LAD.
\\

\gr
\citet{han2023loggpt} & $2023$ & \begin{tabular}{c}
     Supercomputers and\\[\extraarrayvspace]
     distributed file system logs
\end{tabular} & \fmtTT{LogGPT} & \fmtTT{GPT-2} & \faCheckCircle[regular] & \faCheckCircle[regular] & \begin{tabular}{c}
LAD
\end{tabular}& \begin{tabular}{c}
Loghub \\[\extraarrayvspace]
(HDFS,
BGL,
Thunderbird)
\end{tabular} & LAD with GPT model fine-tuned using reinforcement learning\\
\citet{voros2023web} & 
$2023$ &
URLs &
-- &
\begin{tabular}{c}
\fmtTT{BERT},\\[\extraarrayvspace] \fmtTT{BERTiny}, \fmtTT{T5},\\[\extraarrayvspace] \fmtTT{GPT-3}
\end{tabular} &
\faCheckCircle[regular] & \faTimesCircle[regular] &   
URL Classification
&
\begin{tabular}{c}
Private security \\[\extraarrayvspace]vendor dataset\dag
\end{tabular}
& 
Web content filtering for 
maintaining network security 
and regulatory compliance 
\\

\gr
\citet{boffa2024logprecis} & $2024$ & \begin{tabular}{c}
     Unix shell and\\[\extraarrayvspace]
     honeypot logs
\end{tabular}& \fmtTT{LogPrécis} &
\begin{tabular}{c}
    \fmtTT{BERT}, \fmtTT{CodeBERT},\\[\extraarrayvspace]
    \fmtTT{CodeBERTa}, \fmtTT{GPT-3}
\end{tabular}
& 
\faCheckCircle[regular] & \faTimesCircle[regular] & \begin{tabular}{c}
Parsing and\\[\extraarrayvspace] Analysis
\end{tabular} &
\begin{tabular}{c}
NLP2Bash, HaaS\dag, \\[\extraarrayvspace]
CyberLab Honeynet, PoliTO\dag
\end{tabular}
& 
Create an attack fingerprint assigning attacker tactics
to each portion of a session
\\

\citet{balasubramanian2024cygent} 
& $2024$ & 
\begin{tabular}{c}
Web server logs\\[\extraarrayvspace] and prompt
\end{tabular} & \fmtTT{CYGENT} &
\begin{tabular}{c}
    \fmtTT{GPT-3.5 turbo},\\[\extraarrayvspace]
    \fmtTT{CodeT5}
\end{tabular}
& 
\faCheckCircle[regular] & \faTimesCircle[regular] & 
\begin{tabular}{c}
Log \\[\extraarrayvspace] 
Summarization
\end{tabular}
&
\begin{tabular}{c}
Web Server Access Logs,
\\[\extraarrayvspace]
Self-generated (Access Logs)\dag
\end{tabular}
& 
Analyze and summarize logs,
detect events and deliver
cybersecurity informations
\\

\gr
\citet{karlsen2024large} & $2024$ & \begin{tabular}{c}
    Supercomputer and\\[\extraarrayvspace]
    web server logs
\end{tabular} & -- &
\begin{tabular}{c}
    \fmtTT{BERT}, \fmtTT{RoBERTa}, \\[\extraarrayvspace] 
    \fmtTT{DistillRoBERTa},\\[\extraarrayvspace] 
    \fmtTT{GPT-2}, \fmtTT{GPT-Neo}
\end{tabular}
& 
\faCheckCircle[regular] & \faTimesCircle[regular] & \begin{tabular}{c}
LAD
\end{tabular} & 
\begin{tabular}{c}
Apache Web Server, CSIC 2010,\\[\extraarrayvspace]
ECML/PKDD 2007, Spirit \\[\extraarrayvspace]
Loghub (Thunderbird, BGL)\\[\extraarrayvspace] 
\end{tabular}
& 
Unsupervised LAD with embedding compression
via autoencoders and self-organizing map
\\

\citet{meyuhas2024} & $2024$ & Network traffic logs & -- &
\begin{tabular}{c}
     \fmtTT{RoBERTa}, %
     \fmtTT{GPT-1}
\end{tabular}& 
\faTimesCircle[regular] & \faCheckCircle[regular] & \begin{tabular}{c}
     Device Function\\[\extraarrayvspace] Labeling
\end{tabular}& \begin{tabular}{c}
IoT Sentinel, Censys, MUDIS,\\[\extraarrayvspace]
UNSW-IoT-Analytics,
IoTFinder
\end{tabular} &
     AI-automated IoT labeling
     with vendor and device function
\\

\gr
Tian et al.~\cite{tian2024dom} 
& 2024 
& DNS logs
& \fmtTT{Dom-BERT}
& \fmtTT{BERT}
& \faCheckCircle[regular]
& \faTimesCircle[regular]
& \begin{tabular}{c}DNS Logs\\[\extraarrayvspace]Reconstruction\end{tabular}
& \textit{self-built}\dag
& Malicious DNS entry detection on logs
\\

\citet{almodovar2024logfit} & $2024$ & \begin{tabular}{c}
     Supercomputer and\\[\extraarrayvspace]
     distributed file system logs
\end{tabular} & LogFiT & 
\begin{tabular}{c}
    RoBERTa\\[\extraarrayvspace]
    Longformer
\end{tabular}
 & \faCheckCircle[regular] & \faTimesCircle[regular] & LAD & 
\begin{tabular}{c}
Loghub \\[\extraarrayvspace](HDFS, BGL, Thunderbird)
\end{tabular}
& Self-supervised LAD with fine-tuned LLM \\

\bottomrule
\end{tabular}
\begin{tablenotes}[flushleft]
\footnotesize
\item\emph{Fine-Tuned/Released GenAI Model Architecture}: \faCheckCircle[regular] - Present, \faTimesCircle[regular] - Absent; \emph{Evaluation/Fine-Tuning Datasets}: $\bm{\dag}$ - Private Data.
\end{tablenotes}
\end{threeparttable}
}
\end{table*}

\noindent
\textbf{Definition:}
\gls{nla}
tasks involve parsing logs to extract meaningful insights and information critical for maintaining and optimizing network operations.

Table~\ref{tab:log_analysis} provides a comprehensive overview of various research proposals employing \gls{genai} for log analysis within the networking domain during $2021$--$24$. 

As the input of \gls{genai} models (column ``\textbf{Input Type}''), considered works 
employ logs related to events and activities of operating system and applications in honeypots~\cite{setianto2021gpt},
cloud environments~\cite{ott2021robust,mudgal2023assessment}, DNS~\cite{tian2024dom},
web servers~\cite{balasubramanian2024cygent, karlsen2024large,sun2023design}, 
Kubernetes clusters~\cite{sun2023design},
supercomputers~\cite{pan2023raglog,mudgal2023assessment, almodovar2024logfit}, 
and
Unix shell~\cite{boffa2024logprecis}.
Differently,~\citet{meyuhas2024} leverage network traffic log, which captures data from the flow of packets exchanged between \gls{iot} devices, while~\cite{voros2023web} employ URLs collected on firewalls and endpoints.
Additional details on the dataset used for testing and optionally fine-tuning the \gls{genai} model are provided in the ``\textbf{Evaluation/Fine-Tuning Dataset}'' column,
with the majority of reviewed works leveraging Loghub datasets~\cite{ott2021robust,pan2023raglog,qi2023loggpt,jiang2023lilac,mudgal2023assessment,han2023loggpt,karlsen2024large, almodovar2024logfit}.
For details about datasets, please refer to Tab.~\ref{tab:dataset-overview}.

Furthermore, some works take as input also prompts to interact with \gls{genai} models~\cite{setianto2021gpt, qi2023loggpt, mudgal2023assessment, balasubramanian2024cygent}.
Listing~\ref{lst:prompt_example}, reported in~\cite{mudgal2023assessment}, provides an example of a possible prompt interaction
for error and root cause identification within system logs.

\vspace{5pt}
\begin{lstlisting}[style=customlog, caption={Example of prompt.}, label={lst:prompt_example}]
Summarize the errors and warnings from these log messages and identify the root cause.
[Sun Dec 04 04:52:49 2005] [notice] workerEnv.init() ok 
 /etc/httpd/conf/workers2.properties
[Sun Dec 04 04:52:49 2005] [notice] workerEnv.init() ok 
 /etc/httpd/conf/workers2.properties
[Sun Dec 04 04:52:52 2005] [error] mod_jk child workerEnv in error state 7
\end{lstlisting}

All the considered works employ transformer-based models from different categories (see ``\textbf{\gls{genai} Architecture}'' column and \cf  Sec.~\ref{sec:background}).
As a \emph{decoder-only} architecture, various \fmtTT{GPT}-based models are employed, such as 
\fmtTT{GPT-2}~\cite{setianto2021gpt,ott2021robust,ji2023log,han2023loggpt,karlsen2024large}, \fmtTT{GPT-3}~\cite{boffa2024logprecis}, \fmtTT{GPT-3.5}~\cite{pan2023raglog} or \fmtTT{GPT-3.5 turbo}~\cite{jiang2023lilac,mudgal2023assessment,balasubramanian2024cygent}, \fmtTT{GPT-Neo}~\cite{karlsen2024large}.
In contrast, as \emph{encoder-only} architectures, \fmtTT{BERT}~\cite{ott2021robust,boffa2024logprecis,meyuhas2024,voros2023web,tian2024dom} is used, as well as its derived versions, such as
\fmtTT{RoBERTa}~\cite{karlsen2024large, meyuhas2024, almodovar2024logfit}, 
\fmtTT{DistilRoBERTa}~\cite{karlsen2024large}, \fmtTT{BERTiny}~\cite{voros2023web},
\fmtTT{CodeBERT}~\cite{boffa2024logprecis}, and \fmtTT{CodeBERTa}~\cite{boffa2024logprecis}.
Additionally,~\citet{balasubramanian2024cygent},~\citet{ott2021robust}, and~\cite{voros2023web} leverage \emph{full encoder-decoder} models, \ie \fmtTT{T5}, \fmtTT{CodeT5}, and \fmtTT{XLNet} which also include an auto-regressive module.
Furthermore,~\citet{almodovar2024logfit} employ Longformer, which overcomes BERT limitations in handling sequences exceeding $512$ tokens.
Most studies leverage publicly-available \gls{genai} models~\cite{ott2021robust,pan2023raglog,qi2023loggpt,jiang2023lilac,ji2023log,mudgal2023assessment,sun2023design, meyuhas2024}, while others fine-tune them~\cite{setianto2021gpt, han2023loggpt, voros2023web, boffa2024logprecis, balasubramanian2024cygent,karlsen2024large, tian2024dom, almodovar2024logfit}
(see ``\textbf{Fine-Tuned}'' column).
Only some studies release the updated version of the model (see the ``\textbf{Released}'' column).
Notably,~\citet{han2023loggpt} leverage reinforcement learning strategy to fine-tune and adapt the \gls{genai} model for \gls{lad},
that is, identifying unusual (\viz anomalous) patterns or behavior in system logs that deviate from the normal ones.

The ``\textbf{Downstream Task}'' column details the specific task each study addresses, while the ``\textbf{\gls{genai} Goal}'' column describes the functions that \gls{genai} models perform to achieve these tasks.

On the one hand, \citet{setianto2021gpt} and \citet{jiang2023lilac} perform accurate real-time log-parsing.
Although both use \gls{genai} models to perform log parsing---%
which involves extracting structured information from unstructured log data---%
the former~\cite{setianto2021gpt} employs a Q\&A interaction with \gls{genai} model, while the latter~\cite{jiang2023lilac} leverages in-context learning and adaptive parsing cache.
In-context learning optimizes the creation of diverse prompts, while adaptive parsing cache stores and updates parsed log templates to avoid redundant queries and ensure accuracy.

On the other hand, a significant number of works focus on \gls{lad}~\cite{qi2023loggpt, pan2023raglog,ji2023log, karlsen2024large, sun2023design, han2023loggpt, boffa2024logprecis, mudgal2023assessment, ott2021robust, balasubramanian2024cygent, almodovar2024logfit}.
Interestingly, four of these do not use \gls{genai} models directly for \gls{lad} but as preprocessing components.
Specifically, \citet{mudgal2023assessment} and \citet{boffa2024logprecis} perform log parsing for \gls{lad}.
The former work uses prompt-based interactions with ChatGPT using pre-defined prompts, while the latter work creates an attack fingerprint and assigns attacker tactics within the \emph{MITRE ATT\&CK tactics}%
\footnote{\url{https://attack.mitre.org/tactics/ics/}} 
to each session portion to reveal the attacker’s goals.
\citet{ott2021robust} focus on vectorization, converting log data into numerical vectors and performing \gls{lad} using nearest template matching to manage incomplete prior knowledge of log templates.
\citet{balasubramanian2024cygent} perform summarization, condensing log files into concise, human-readable formats for \gls{lad}.
Among the works explicitly using \gls{genai} models for \gls{lad},
\citet{qi2023loggpt}, similarly to~\cite{mudgal2023assessment},
use prompt-based interactions with ChatGPT,
leveraging prompt-construction strategies.
Similarly,~\citet{pan2023raglog} employ a Q\&A strategy with log entries and best-matched retrieved entries from a database to determine whether a queried log entry is normal or not.
Furthermore,~\citet{ji2023log} encode normal patterns and define an alarm strategy to filter out false positives based on statistical log data characteristics.
Then,~\citet{almodovar2024logfit} leverage a self-supervised training strategy on normal log data to learn its linguistic and sequential patterns, thereby distinguishing it from malicious logs.

Lastly,
\citet{meyuhas2024} use \gls{genai} models to analyze network traffic and automatically classify \gls{iot} devices by vendor and function, offering insights into the traffic they generate.
\citet{tian2024dom} examine DNS logs to discover malicious DNS entry---%
using \gls{genai} models for log reconstruction---%
while in~\cite{voros2023web} web content filtering based on URLs is accomplished with \gls{genai} models directly tackling URL classification.

\subsection{Network Digital Assistance for Documentation \& Operation}
\label{sec:network_management}
\newcommand{\faBot}{{\includesvg[height=1em]{figs/icons/pc}}}
\newcommand{\faHuman}{{\includesvg[height=1em]{figs/icons/human}}}

\setlength{\extraarrayvspace}{-0.45ex}

\begin{table*}[t]
\renewcommand{\arraystretch}{1.2}
\centering
\footnotesize
\caption{
Works dealing with NDA through GenAI models (in chronological order).
}
\label{tab:net_mng}

\resizebox{\textwidth}{!}
{
\begin{threeparttable}
\begin{tabular}{rccccccccm{3.5cm}}
\toprule
\multirow{2}[2]{*}{\textbf{Paper}} & \multirow{2}[2]{*}{\textbf{Year}} & 
\multirow{2}[2]{*}{\textbf{Input Type}} &
\multicolumn{4}{c}{\textbf{GenAI Model}} &
\multirow{2}[2]{*}{
    \begin{tabular}{c}
    \textbf{Evaluation/Fine-Tuning} \\[\extraarrayvspace]
    \textbf{Datasets} 
\end{tabular}} &
\multicolumn{2}{c}{\textbf{Downstream Task}}
\\ 
\cmidrule(lr){4-7}
\cmidrule(lr){9-10}
& & &  \textbf{Name} & \textbf{Architecture} & \textbf{Fine-Tuned} & \textbf{Released} 
& & \textbf{Target} & \textbf{Description} \\
\midrule

\gr \citet{soman2023observations} &
$2023$ &
Textual prompts &
-- &
\begin{tabular}{c}
\fmtTT{GPT-4}, \fmtTT{GPT-3.5}, \fmtTT{Bard},\\[\extraarrayvspace] \fmtTT{OpenAssistant-LLaMa} 
\end{tabular} &
\faTimesCircle[regular] & \faCheckCircle[regular] & 
Cradlepoint\dag & \faHuman &
Digital assistants for 
telecom domain as user support 
\\

\citet{wang2023making} & 
$2023$ &
Textual prompts &
\fmtTT{NetBuddy} &
 \fmtTT{GPT-4} &
\faTimesCircle[regular] & \faCheckCircle[regular] &   
\emph{self-built}\dag & \faBot &
Automate network configuration by translating requirements in natural language in low-level network configurations  
\\

\gr
\citet{mani2023enhancing} & 
$2023$ &
\begin{tabular}{c}
Graphs and\\[\extraarrayvspace] textual prompts 
\end{tabular} &
-- &
\begin{tabular}{c}
\fmtTT{GPT-4}, \fmtTT{GPT-3},\\[\extraarrayvspace] \fmtTT{GPT-3.5}, \fmtTT{Bard}
\end{tabular} &
\faTimesCircle[regular] & \faCheckCircle[regular] &  
NeMoEval & \faBot &
Create task-specific code for
graph analysis and manipulation 
\\

\citet{mondal2023llms} & 
$2023$ &
Textual prompts &
-- &
 \fmtTT{GPT-4} &
\faTimesCircle[regular] & \faCheckCircle[regular] &   
\emph{self-built}\dag & \faBot &
Automate router configuration by translating prompts in natural language in router configurations 
\\

\gr
\citet{roychowdhury2024unlocking} &
$2024$ &
Textual prompts &
-- &
\fmtTT{LLaMa 2.0} &
\faCheckCircle[regular] & \faTimesCircle[regular] &  
TeleQnA & \faHuman &
Digital assistants for 
question answering standard documents 
\\

\citet{shen2024large} & 
$2024$ &
Textual prompts &
-- &
\begin{tabular}{c}
\fmtTT{GPT-3}, \fmtTT{GPT-4}
\end{tabular} &
\faTimesCircle[regular] & \faCheckCircle[regular] &  
\begin{tabular}{c}
\emph{self-built}\dag 
\end{tabular} & \faBot &
Coordination of existing edge AI models
to cater to the user’s needs 
and enables automatic AI training 
\\

\gr
\citet{duclos2024utilizing} &
$2024$ &
Text documents &
-- &
\fmtTT{CodeLLaMa} & 
\faCheckCircle[regular] & \faTimesCircle[regular] &  
\emph{self-built}\dag & \faBot &
Translation of protocol specifications 
into structured models suitable for
Cryptographic Protocol Shapes Analyzer
\\

\citet{ahmed2024linguistic} &
$2024$ &
Textual prompts &
-- & 
\begin{tabular}{c}
\fmtTT{LLaMa 2.0}, \fmtTT{Falcon}, \\[\extraarrayvspace]\fmtTT{Mistral 7B}, \fmtTT{Zephyr 7B-}$\beta$
\end{tabular} &
\faTimesCircle[regular] & \faCheckCircle[regular] & 
\begin{tabular}{c}
SPEC5GClassification, \\[\extraarrayvspace]SPEC5GSummarization, \\[\extraarrayvspace]TeleQnA
\end{tabular} & \faHuman &
Digital assistant for
text classification, summarization, 
question answering 
\\

\gr
\citet{piovesan2024telecom} &
$2024$ &
Textual prompts &
-- &
\begin{tabular}{c}
\fmtTT{Phi-2}, \fmtTT{GPT-3.5}, \\[\extraarrayvspace]\fmtTT{GPT-4} 
\end{tabular} &
\faTimesCircle[regular] & \faCheckCircle[regular] & 
TeleQnA & \faHuman &
Digital assistants for 
question answering standard documents 
\\

\citet{karapantelakis2024using} &
$2024$ &
Textual prompts &
-- & 
\begin{tabular}{c}
\fmtTT{GPT-3.5 turbo}, \fmtTT{GPT-4},\\[\extraarrayvspace] \fmtTT{LLaMA 2.0}, \fmtTT{Falcon},\\[\extraarrayvspace] \fmtTT{TeleRoBERTa} 
\end{tabular} &
\faCheckCircle[regular] & \faTimesCircle[regular] & 
TeleQuAD\dag & \faHuman &
Digital assistants for 
Third Generation Partnership Projects 
\\

\gr
\citet{ghasemirahni2024deploying} & 
$2024$ &
\begin{tabular}{c}
Code and\\[\extraarrayvspace] textual prompts 
\end{tabular} &
\fmtTT{FlowMage} & 
\begin{tabular}{c}
\fmtTT{GPT-3.5 turbo},\\[\extraarrayvspace] \fmtTT{GPT-4o},\\[\extraarrayvspace] \fmtTT{CodeLLaMa},\\[\extraarrayvspace] \fmtTT{Gemini 1.0 Pro} 
\end{tabular} &
\faTimesCircle[regular] & \faCheckCircle[regular]&  
\emph{self-built}\dag & \faBot &
Software infrastructure optimization,
deployment configuration,
and execution pipeline 
\\

\citet{erak2024leveraging} & 
$2024$ &
Textual prompts &
-- & 
\begin{tabular}{c}
\fmtTT{Phi-2}, \fmtTT{GPT-4o mini},\\[\extraarrayvspace] \fmtTT{GPT-4o}
\end{tabular} &
\faCheckCircle[regular] & \faTimesCircle[regular] & 
TeleQnA & 
\faHuman &
Digital assistants for Third Generation Partnership Projects 
\\

\gr
\citet{ayed2024hermes} & 
$2024$ &
Textual prompts &
\fmtTT{Hermes} & 
\fmtTT{GPT-4o}, \fmtTT{LLaMa 3.1} &
\faTimesCircle[regular] & \faCheckCircle[regular] & 
\emph{self-built}\dag & 
\faBot &
Create logical blocks accompanied by code to execute specific networking intents 
\\

\bottomrule

\end{tabular}
\begin{tablenotes}[flushleft]
\footnotesize
\item\emph{Fine-Tuned/Released GenAI Model Architecture}: \faCheckCircle[regular] - Present, \faTimesCircle[regular] - Absent; \emph{Evaluation/Fine-Tuning Datasets}: $\bm{\dag}$ - Private Data; \emph{Downstream Task}: \faHuman - The Output of the Model is Intended for a Human, \faBot - The Output of the Model is Intended for a Device.
\end{tablenotes}
\end{threeparttable}
 }
\end{table*}

\noindent
\textbf{Definition:}
\gls{nda} refers to the process of administering, controlling, and optimizing network operations to ensure efficient functionality, performance, and security.
\gls{nda} is pivotal due to the heterogeneous nature of networks, which often consist of diverse hardware and software components from different vendors. 
This diversity introduces complexity, making it challenging to navigate through various network configurations, protocols, and standards effectively~\cite{wang2023network}.

Table~\ref{tab:net_mng} provides a summary of the papers dealing with \gls{nda}, emphasizing the related key aspects.
The publication dates of the works highlight the recent interest of the scientific community in this topic ($2023-24$).
This interest has primarily converged on two downstream tasks, as seen from the \textbf{``Downstream Task Description''} column,
which can be identified as follows:
\begin{itemize}
    \item Using \gls{genai} as virtual assistants to query standard documents in the networking/telecommunication domain, thereby providing support to users (\viz~\emph{Network Digital Assistance for Documentation}).
    \item Employing \gls{genai} as assistants for the network operative phase, \eg~handling network topologies, setup of device configurations, and infrastructure management (\viz~\emph{Network Digital Assistance for Operation}).
\end{itemize}

Detailing, $8$ works~\cite{ahmed2024linguistic, piovesan2024telecom, karapantelakis2024using, roychowdhury2024unlocking, shen2024large, soman2023observations, duclos2024utilizing, erak2024leveraging} focus on the \emph{design of a virtual assistant specific to the telecommunication domain}.
The authors of \cite{roychowdhury2024unlocking, piovesan2024telecom} highlight the difficulty in analyzing and extracting information from standard documents in the telecommunication domain, as it involves identifying sources from multiple documents and related references.
Thus, the authors investigate whether \glspl{llm} can be used as digital assistants for Q\&A on standard documents.
\citet{ahmed2024linguistic} extend the functionalities proposed in~\cite{roychowdhury2024unlocking} to enrich the digital assistant's functionalities.
In particular, they also include tasks related to text classification and summarization, as well as Q\&A.
Unlike the previous ones, in ~\cite{karapantelakis2024using, erak2024leveraging}, a digital assistant specifically designed for \glspl{gpp} standards is introduced. 
Unlike previous research focused on Q\&A for standardized documents, the assistant proposed in~\cite{soman2023observations} specifically targets user-support tasks.
These include finding information about products and services, initial assistance with installation and configuration, and operational tasks like troubleshooting and performance monitoring.
\citet{duclos2024utilizing} develop an \gls{llm}-based assistant designed to translate \glspl{rfc} into a format compatible with \gls{cpsa}.

The works in~\cite{mani2023enhancing, ghasemirahni2024deploying, duclos2024utilizing, wang2023making, mondal2023llms, ayed2024hermes} address the \emph{design of a network operation assistant.}
Contrary to the previous studies, these papers exhibit greater heterogeneity due to the inherent task diversity associated with the operational phase.
\citet{mani2023enhancing} tackle the complexities of network topology and communication graph analysis by exploiting \glspl{llm}. 
They demonstrate how these models can be used to generate task-specific code for graph manipulation, thereby enabling more intuitive network management through natural language interactions.
In~\cite{ghasemirahni2024deploying}, \glspl{llm} are employed to enhance the scalability of 
network functions
as traffic volume increases.
The authors introduce a system that utilizes \glspl{llm} to perform code analysis and extract crucial information about software behavior, semantics, and system-level performance. 
The extracted information is then used to optimize the infrastructure, deployment configuration, and execution pipeline.
\citet{shen2024large} propose to integrate \glspl{llm} into an \gls{ai}-enabled network with two distinct tasks: 
($i$) serving as a user interface to intercept and understand user requests; and 
($ii$) automating the training of \gls{ai} nodes in the network---%
for instance, iteratively finding the best learning rate scheduler.
In~\cite{wang2023making} and~\cite{mondal2023llms}, \glspl{llm} are proposed as tools to facilitate the creation of network configurations from natural language descriptions.
Both studies highlight the limitations of \glspl{llm} in this specific task, highlighting a high number of errors in the generated configurations. 
Consequently, both papers incorporate a verification module to validate and correct the output of the model.
\citet{ayed2024hermes} proposes a framework based on \glspl{llm} to generate logical blocks related to a specific user intent (e.g.,~the deployment of a new base station in a network). Each logical block is accompanied by the corresponding code necessary for its implementation.

The \textbf{``Input Type''} column displays the specific data fed to \gls{genai} models.
Predominantly, these inputs are textual prompts where a query is submitted to the model.
The majority of works employing such inputs fall under the category of downstream tasks for telecommunication documentation and support functions~\cite{soman2023observations, roychowdhury2024unlocking, shen2024large, duclos2024utilizing, ahmed2024linguistic, piovesan2024telecom, karapantelakis2024using, erak2024leveraging}.
Concerning the second downstream task---\ie \gls{genai} as assistant for the network operative phase---inputs vary depending on the specific operation. 
Specifically, 
\citet{mani2023enhancing} combine textual descriptions with network topology graphs, while \citet{ghasemirahni2024deploying} integrate code snippets with textual prompts for analysis purposes.
In~\cite{wang2023making, mondal2023llms} a description of the configurations in natural language is used as input.
Similarly, \citet{ayed2024hermes} use a description of the available data together with a network modeling task.
Looking at the \textbf{``Evaluation/Fine-Tuning Datasets''} column, the datasets utilized in the literature vary according to the specific downstream task. 
TeleQnA is the most frequently used dataset \cite{roychowdhury2024unlocking, ahmed2021ecu, piovesan2024telecom, erak2024leveraging}. 
This dataset is designed to evaluate the knowledge of \glspl{llm} within the Telecom domain, featuring multiple-choice questions categorized into various categories. 
TaleQnAD \cite{karapantelakis2024using} is similar to TeleQnA but specialized in \gls{gpp} standards.
On the other hand, NeMoEval \cite{mani2023enhancing} is used as benchmark for \glspl{llm} for two different applications:
traffic analysis using communication graphs and
network lifecycle management (\eg capacity planning, network topology design, deployment planning, and diagnostic operations).

The \textbf{``GenAI Model''} column highlights that most studies,
except for a few \cite{roychowdhury2024unlocking, duclos2024utilizing, wang2023making, mondal2023llms},
conduct comparative analyses among various currently-available generative models.
The literature considers several \glspl{llm}, such as those from OpenAI (\eg \fmtTT{GPT-4o}, \fmtTT{GPT-4o mini} \fmtTT{GPT-4}, \fmtTT{GPT-3.5}, \fmtTT{GPT-3}), Google (\eg \fmtTT{Bard}, \fmtTT{Gemini}), Meta (\eg \fmtTT{CodeLLaMA}, \fmtTT{LLaMA 2.0}, \fmtTT{LLaMA 3.1}), and Microsoft (\eg \fmtTT{Phi-2}).
Regarding the use of open models, \citet{mani2023enhancing} test open \glspl{llm} (\ie~\fmtTT{StarCoder} and \fmtTT{InCoder}) but omitted their results due to inconsistency.
Delving deeper, the majority of works~\cite{soman2023observations, mani2023enhancing, shen2024large, ahmed2024linguistic, piovesan2024telecom, ghasemirahni2024deploying, wang2023making, mondal2023llms, ayed2024hermes} do not fine-tune the models but use them off-the-shelf, as indicated by the \textbf{``Fine-Tuned''} column. 
On the other hand, the remaining works \cite{roychowdhury2024unlocking, duclos2024utilizing, karapantelakis2024using, erak2024leveraging} refine the models using domain-specific datasets. 
The \textbf{``Released''} column shows that none of the works that perform fine-tuning also release the fine-tuned models;
only the original models are available in those cases.
To reduce the computational burden of \glspl{llm}, the authors of~\cite{piovesan2024telecom, erak2024leveraging} investigate the use of \glspl{slm} leveraging \fmtTT{Phi-2}.
In both works, such a model is compared with larger ones (\ie \fmtTT{GPT-3.5} and \fmtTT{GPT-4}, \fmtTT{GPT-4o}, \fmtTT{GPT-4o mini}).
In \cite{piovesan2024telecom}, the authors propose equipping the model with \gls{rag} to incorporate authoritative knowledge external to the model's initial training data. 
The combination of \fmtTT{Phi-2} and \gls{rag} achieves results comparable to those of \fmtTT{GPT-3.5}.
Similarly, in \cite{erak2024leveraging}, a fine-tuned \fmtTT{Phi-2} in tandem with \gls{rag} and \emph{SelfExtend}~\cite{jin2024llm} (used to extend the model's the context window
during inference) surpasses the performance achieved by the larger \fmtTT{GPT-4o}.
\citet{wang2023making} and \citet{mondal2023llms} introduce a verification module to address the shortcomings of \fmtTT{GPT-4}'s device configuration generation.
Their findings indicate that the \gls{llm} output frequently contains errors,
necessitating a verification component to ensure accuracy and provide corrective feedback.

\section{A Model-Centric Overview on GenAI for Network Monitoring and Management}
\label{sec:genai_model_overview}
\newcommand{\faHuggingFace}{{\includesvg[height=1em]{figs/icons/hf-logo.svg}}}
\newcommand{\faGoogleCloud}{{\includesvg[height=1em]{figs/icons/google-cloud.svg}}}
\newcommand{\faEleutherAI}{{\includesvg[height=1em]{figs/icons/libre.svg}}}
\newcommand{\faOpenAI}{{\includesvg[height=1em]{figs/icons/openai.svg}}}
\renewcommand{\faMicrosoft}{{\includesvg[height=1em]{figs/icons/microsoft.svg}}}
\renewcommand{\faGoogle}{{\includesvg[height=1em]{figs/icons/google.svg}}}
\newcommand{\faMeta}{{\includesvg[height=1em]{figs/icons/meta.svg}}}
\newcommand{\faEricsson}{{\includesvg[height=1em]{figs/icons/ericsson-notext.svg}}}
\newcommand{\faStabilityAI}{{\includesvg[height=1em]{figs/icons/stability-ai.svg}}}
\newcommand{\faTii}{{\includesvg[height=1em]{figs/icons/tii-notext.svg}}}
\newcommand{\faDeepmind}{{\includesvg[height=1em]{figs/icons/deepmind.svg}}}
\newcommand{\faMistralAI}{{\includesvg[height=1em]{figs/icons/mistral-ai.svg}}}
\renewcommand{\faSalesforce}{{\includesvg[height=1em]{figs/icons/salesforce.svg}}}
\newcommand{\faCMU}{{\includesvg[height=1em]{figs/icons/carnegie-mellon-university.svg}}}
\newcommand{\faUT}{{\includesvg[height=1em]{figs/icons/university-of-toronto.svg}}}
\newcommand{\faPU}{{\includesvg[height=1em]{figs/icons/princeton-university.svg}}}
\newcommand{\faOU}{{\includesvg[height=1em]{figs/icons/oldenburg-university.svg}}}
\newcommand{\faSU}{{\includesvg[height=1em]{figs/icons/stanford-university.svg}}}
\newcommand{\faUCB}{{\includesvg[height=1em]{figs/icons/university-of-california-berkeley.svg}}}
\newcommand{\faUCSD}{{\includesvg[height=1em]{figs/icons/university-of-california-san-diego.svg}}}
\newcommand{\faBU}{{\includesvg[height=1em]{figs/icons/beihang-university.svg}}}
\newcommand{\faOpenAss}{{\includesvg[height=1em]{figs/icons/open-assistant.svg}}}
\newcommand{\faTimeSeries}{{\includesvg[height=1em]{figs/icons/time-series.svg}}}
\newcommand{\faTTIC}{{\includesvg[height=1em]{figs/icons/ttic.svg}}}

\begin{table*}[ht]
\centering
\caption{GenAI models leveraged in the NMM use cases considered.
GenAI models are grouped by category and ordered by release year within each category. 
}
\label{table:genai_architectures}

\footnotesize

\resizebox{\textwidth}{!}{
\begin{tabular}{clccccccccc}

\toprule

\multicolumn{6}{c}{\textbf{GenAI Architecture}} & \multicolumn{5}{c}{\textbf{NMM Use Case}} \\
\cmidrule(lr){1-6} \cmidrule(lr){7-11}
\textbf{Cat.} & \textbf{Name} & \textbf{Input} & \textbf{Year} & \textbf{Res. Org.} & \textbf{Lic.} & \textbf{NTG} & \textbf{NTC} & \textbf{NID} & \textbf{NSLA} & \textbf{NDA} \\
\midrule

\rc
&\gc \multirow{-1}{*}{$\mathtt{Transformer}\:\ddag$} &\gc \multirow{-1}{*}{\faFont} &\gc \multirow{-1}{*}{$2017$} &\gc \multirow{-1}{*}{\faGoogle\,\,\faDeepmind\,\,\faUT} &\gc \multirow{-1}{*}{\CIRCLE} &\gc \cite{qu2024trafficgpt} &\gc \cite{qu2024trafficgpt}\cite{guthula2023netfound} &\gc \cite{guthula2023netfound}\cite{li2022extreme}\cite{wu2022}\cite{ullah2023tnn}\cite{wang2023robust} &\gc -- &\gc -- \\

\rc
& $\mathtt{XLNet}$ \textbf{*} & \faFont & $2019$ & \multirow{-1}{*}{\faGoogle\,\,\faCMU} & \CIRCLE & -- & -- & -- & \cite{ott2021robust} & -- \\

\rc
&\gc $\mathtt{T5}$ &\gc \faFont &\gc $2019$ &\gc \multirow{-1}{*}{\faGoogle} &\gc \CIRCLE &\gc -- &\gc -- &\gc -- &\gc \cite{voros2023web} &\gc -- \\

\rc
& \multirow{-1}{*}{$\mathtt{CodeT5}$} & \multirow{-1}{*}{\faCode} & \multirow{-1}{*}{$2020$} & \multirow{-1}{*}{\faSalesforce} & \multirow{-1}{*}{\CIRCLE} & -- & -- & -- & \multirow{-1}{*}{\cite{balasubramanian2024cygent}} & -- \\

\rc
&\gc \multirow{-1}{*}{$\mathtt{T5}\,\mathtt{1.1}$} &\gc \multirow{-1}{*}{\faFont} &\gc \multirow{-1}{*}{$2021$} &\gc \multirow{-1}{*}{\faGoogle} &\gc \multirow{-1}{*}{\CIRCLE} &\gc \cite{wang2024lens} &\gc \cite{wang2024lens} &\gc \cite{wang2024lens} &\gc -- &\gc -- \\

\rc
& $\mathtt{Bard}$ & \faFont & $2023$ & \multirow{-1}{*}{\faGoogle} & \Circle & -- & -- & -- & -- & \cite{soman2023observations}\cite{mani2023enhancing} \\

\rc
&\gc $\mathtt{Gemini}\,\mathtt{1.0}\,\mathtt{Pro}$ &\gc \faFont &\gc $2023$ &\gc \multirow{-1}{*}{\faDeepmind} &\gc \Circle &\gc -- &\gc -- &\gc -- &\gc -- &\gc \cite{ghasemirahni2024deploying} \\

\rc
& $\mathtt{Mistral}\,\mathtt{7B}$ & \faFont & $2023$ & \multirow{-1}{*}{\faMistralAI} & \CIRCLE & -- & -- & -- & -- & \cite{ahmed2024linguistic} \\

\multirow{-9}{*}{\rc FED} 
&\gc $\mathtt{Zephyr}\,\mathtt{7B}\text{-}\beta$ &\gc \faFont &\gc $2023$ &\gc \multirow{-1}{*}{\faHuggingFace} &\gc \CIRCLE &\gc -- &\gc -- &\gc -- &\gc -- &\gc \cite{ahmed2024linguistic} \\

\midrule %

\ec
& \multirow{-1}{*}{$\mathtt{BERT}\:\ddag$} & \multirow{-1}{*}{\faFont} & \multirow{-1}{*}{$2018$} & \multirow{-1}{*}{\faGoogle} & \multirow{-1}{*}{\CIRCLE} & -- & \cite{lin2022}\cite{tao2024lambert} & \cite{lin2022}\cite{manocchio2024flowtransformer}\cite{ferrag2024}\cite{yu2021securing}\cite{seyyar2022attack}\cite{ghourabi2022security}\cite{lai2023} & \cite{boffa2024logprecis}\cite{ott2021robust}\cite{voros2023web}\cite{karlsen2024large}\cite{tian2024dom} & --\\

\ec
&\gc $\mathtt{BERTiny}$ &\gc \faFont &\gc $2019$ &\gc \multirow{-1}{*}{\faHuggingFace} &\gc \CIRCLE &\gc -- &\gc -- &\gc -- &\gc \cite{voros2023web} &\gc --\\

\ec
& $\mathtt{DistilRoBERTa}$ & \faFont & $2019$ & \multirow{-1}{*}{\faHuggingFace} & \CIRCLE & -- & -- & -- & \cite{karlsen2024large} & -- \\

\ec
&\gc \multirow{-1}{*}{$\mathtt{RoBERTa}$} &\gc \multirow{-1}{*}{\faFont} &\gc \multirow{-1}{*}{$2019$} &\gc \multirow{-1}{*}{\faMeta} &\gc \multirow{-1}{*}{\CIRCLE} &\gc -- &\gc \cite{sarabi2023} &\gc -- &\gc \cite{karlsen2024large}\cite{meyuhas2024} &\gc -- \\

\ec
& $\mathtt{BERT}\text{-}\mathtt{of}\text{-}\mathtt{Theseus}$ & \faFont & $2020$ & \multirow{-1}{*}{\faUCSD\,\,\faBU\,\,\faMicrosoft} & \CIRCLE & -- & -- & \cite{wang2024lightweight} & -- & -- \\

\ec
&\gc $\mathtt{CodeBERT}$ &\gc \faCode &\gc $2020$ &\gc \multirow{-1}{*}{\faMicrosoft} &\gc \CIRCLE &\gc -- &\gc -- &\gc -- &\gc \cite{boffa2024logprecis} &\gc -- \\

\ec
& $\mathtt{CodeBERTa}$ & \faCode & $2020$ & \multirow{-1}{*}{\faHuggingFace} & \CIRCLE & -- & -- & -- & \cite{boffa2024logprecis} & -- \\

\ec
&\gc $\mathtt{ALBERT}$ &\gc \faFont &\gc $2020$ &\gc \multirow{-1}{*}{\faGoogle \,\,\faTTIC} &\gc \CIRCLE &\gc -- &\gc \cite{li2024albert} &\gc -- &\gc -- &\gc -- \\

\multirow{-9}{*}{\ec EO} 
&$\mathtt{ViT}$ &\faImage &$2021$ &\multirow{-1}{*}{\faGoogle} &\CIRCLE & -- & -- & \cite{ho2022network} & -- & -- \\

\midrule %

\cc
& \gc\multirow{-1}{*}{$\mathtt{GPT}\text{-}\mathtt{1}\:\ddag$} & \gc\multirow{-1}{*}{\faFont} & \gc\multirow{-1}{*}{$2018$} & \gc\multirow{-1}{*}{\faOpenAI} & \gc\multirow{-1}{*}{\CIRCLE} & \gc-- & \gc-- & \gc\cite{nam2021intrusion}\cite{melicias2024gpt} & \gc\cite{qi2023loggpt}\cite{meyuhas2024} & \gc-- \\

\cc
& \multirow{-1}{*}{$\mathtt{GPT}\text{-}\mathtt{2}$} & \multirow{-1}{*}{\faFont} & \multirow{-1}{*}{$2019$} & \multirow{-1}{*}{\faOpenAI} & \multirow{-1}{*}{\CIRCLE} & 
\cite{meng2023netgpt} \cite{bikmukhamedov2021} & \cite{meng2023netgpt} & \cite{meng2023netgpt} & \cite{han2023loggpt}\cite{setianto2021gpt}\cite{ott2021robust}\cite{ji2023log}\cite{karlsen2024large} & -- \\

\cc
& \gc\multirow{-1}{*}{$\mathtt{GPT}\text{-}\mathtt{3}$} & \gc\multirow{-1}{*}{\faFont} & \gc\multirow{-1}{*}{$2020$} & \gc\multirow{-1}{*}{\faOpenAI} & \gc\multirow{-1}{*}{\Circle} & \gc\cite{Kholgh2023PACGPT} & \gc-- & \gc\multirow{-1}{*}{\cite{manocchio2024flowtransformer}} & \gc\multirow{-1}{*}{\cite{boffa2024logprecis}\cite{voros2023web}} & \gc\multirow{-1}{*}{\cite{mani2023enhancing}\cite{shen2024large}} \\

\cc
& $\mathtt{GPT}\text{-}\mathtt{Neo}$ & \faFont & $2021$ & \multirow{-1}{*}{\faEleutherAI} & \CIRCLE & -- & -- & -- & \cite{karlsen2024large} & -- \\

\cc
& \gc\multirow{-1}{*}{$\mathtt{GPT}\text{-}\mathtt{3.5}$} & \gc\multirow{-1}{*}{\faFont} & \gc\multirow{-1}{*}{$2022$} & \gc\multirow{-1}{*}{\faOpenAI} & \gc\multirow{-1}{*}{\Circle} & \gc-- & \gc-- & \gc-- & \gc\cite{pan2023raglog} & \gc\cite{soman2023observations}\cite{mani2023enhancing}\cite{piovesan2024telecom} \\

\cc
& \multirow{-1}{*}{$\mathtt{GPT}\text{-}\mathtt{3.5}\,\mathtt{turbo}$} & \multirow{-1}{*}{\faFont} & \multirow{-1}{*}{$2022$} & \multirow{-1}{*}{\faOpenAI} & \multirow{-1}{*}{\Circle} & -- & -- & \cite{ali2023} & \cite{jiang2023lilac}\cite{mudgal2023assessment}\cite{balasubramanian2024cygent} & \cite{karapantelakis2024using}\cite{ghasemirahni2024deploying} \\

\cc
& \gc$\mathtt{GPT}\text{-}\mathtt{4}$ & \gc\faFont & \gc$2023$ & \gc\multirow{-1}{*}{\faOpenAI} & \gc\Circle & \gc-- & \gc-- & \gc-- & \gc-- & \gc
\cite{karapantelakis2024using}\cite{soman2023observations}\cite{wang2023making}\cite{mani2023enhancing}\cite{mondal2023llms}\cite{shen2024large}\cite{piovesan2024telecom} \\

\cc
& $\mathtt{Falcon}$ & \faFont & $2023$ & \multirow{-1}{*}{\faTii} & \CIRCLE & -- & -- & -- & -- & \cite{karapantelakis2024using} \\

\cc
& \gc\multirow{-1}{*}{$\mathtt{OASST}\,\mathtt{LLaMA}$} & \gc\multirow{-1}{*}{\faFont} & \gc\multirow{-1}{*}{$2023$} & \gc\multirow{-1}{*}{\faOpenAss} & \gc\multirow{-1}{*}{\LEFTcircle} & \gc-- & \gc-- & \gc-- & \gc-- & \gc\cite{soman2023observations} \\

\cc
& \multirow{-1}{*}{$\mathtt{LLaMA}\,\mathtt{2.0}$} & \multirow{-1}{*}{\faFont} & \multirow{-1}{*}{$2023$} & \multirow{-1}{*}{\faMeta} & \multirow{-1}{*}{\LEFTcircle} & -- & -- & -- & -- & \cite{karapantelakis2024using}\cite{roychowdhury2024unlocking}\cite{ahmed2024linguistic} \\

\cc
& \gc\multirow{-1}{*}{$\mathtt{CodeLLaMA}$} & \gc\multirow{-1}{*}{\faCode} & \gc\multirow{-1}{*}{$2023$} & \gc\multirow{-1}{*}{\faMeta} & \gc\multirow{-1}{*}{\LEFTcircle} & \gc-- & \gc-- & \gc-- & \gc-- & \gc\multirow{-1}{*}{\cite{duclos2024utilizing}\cite{ghasemirahni2024deploying}} \\

\cc
& $\mathtt{Phi}\,\mathtt{2.0}$ & \faFont & $2023$ & \multirow{-1}{*}{\faMicrosoft} & \CIRCLE & -- & -- & -- & -- & \cite{piovesan2024telecom}\cite{erak2024leveraging} \\

\cc
& \gc$\mathtt{GPT}\text{-}\mathtt{4o}\,\mathtt{mini}$ & \gc\faFont & \gc$2024$ & \gc\multirow{-1}{*}{\faOpenAI} & \gc\Circle & \gc-- & \gc-- & \gc-- & \gc-- & \gc\cite{erak2024leveraging} \\

\multirow{-13}{*}{\cc DO} 
& $\mathtt{GPT}\text{-}\mathtt{4o}$ & \faFont & $2024$ & \multirow{-1}{*}{\faOpenAI} & \Circle & -- & -- & -- & -- & 
\cite{ghasemirahni2024deploying}\cite{erak2024leveraging}\cite{ayed2024hermes} \\

\cc
& \gc$\mathtt{LLaMa\, 3.1}$ & \gc\faFont & \gc$2024$ & \gc\multirow{-1}{*}{\faMeta} & \gc\LEFTcircle & \gc-- & \gc-- & \gc-- & \gc-- & \gc\cite{ayed2024hermes} \\

\midrule %

\oc
& $\mathtt{Diffusion}\,\mathtt{Model}\:\ddag$ & \faImage & $2020$ &  \multirow{-1}{*}{\faSU\,\,\faUCB} & \CIRCLE & \cite{sivaroopan2023netdiffus} &-- &-- &-- &-- \\

\multirow{-2}{*}{\oc SDP}
& \gc$\mathtt{Stable}\,\mathtt{Diffusion}\,\mathtt{1.5}$ & \gc\faImage & \gc$2022$ & \gc\multirow{-1}{*}{\faStabilityAI} & \gc\LEFTcircle & \gc\cite{jiang2024netdiffusion} & \gc-- & \gc-- & \gc-- & \gc-- \\

\midrule %

\multirow{-1}{*}{\yc SSM}
& \multirow{-1}{*}{$\mathtt{Mamba}\:\ddag$} & \multirow{-1}{*}{\faFont} & \multirow{-1}{*}{$2023$} & \multirow{-1}{*}{\faCMU\,\,\faPU} & \multirow{-1}{*}{\CIRCLE} & \cite{chu2024mamba} & \cite{wang2024netmamba} & \cite{wang2024netmamba} & -- & -- \\

\end{tabular}
}

\resizebox{\textwidth}{!}{
\begin{tabular}{m{\textwidth}}

\toprule

\gr
\makecell[c]{\textbf{\textsc{Legend}}}\\

\midrule

\textbf{GenAI Model}:
Category~(\textbf{Cat.}),
Research Organization~(\textbf{Res. Org.}),
Licensing~(\textbf{Lic.}).\\

\textbf{NMM Use Case}:
Network Traffic Generation~(\textbf{NTG}),
Network Traffic Classification~(\textbf{NTC}),
Network Intrusion Detection~(\textbf{NID}),
Networked System Log Analysis~(\textbf{NSLA}),
Network Digital Assistance~(\textbf{NDA}).\\

\midrule

\textbf{Category}:
Fully Encoder-Decoder~(\rb{FED}),
Encoder-Only~(\eb{EO}),
Decoder-Only~(\cb{DO}),
Sequential Denoising Process~(\ob{SDP}),
State-Space Model~(\yb{SSM}).\\

\textbf{Input}:
Natural Language~(\faFont),
Code~(\faCode),
Image~(\faImage).
\\

\textbf{Research Organization}:
Beihang University~(\multirow{-1}{*}{\faBU}),
Carnegie Mellon University~(\multirow{-1}{*}{\faCMU}),
Deepmind~(\multirow{-1}{*}{\faDeepmind}),
EleutherAI~(\multirow{-1}{*}{\faEleutherAI}),
Google Research~(\multirow{-1}{*}{\faGoogle}),
Meta~(\multirow{-1}{*}{\faMeta}),
Microsoft~(\multirow{-1}{*}{\faMicrosoft}),
MistralAI~(\multirow{-1}{*}{\faMistralAI}),
OpenAI~(\multirow{-1}{*}{\faOpenAI}),
OpenAssistant~(\multirow{-1}{*}{\faOpenAss}),
Princeton University~(\multirow{-1}{*}{\faPU}),
Salesforce~(\multirow{-1}{*}{\faSalesforce}),
StabilityAI~(\multirow{-1}{*}{\faStabilityAI}),
Stanford University~(\multirow{-1}{*}{\faSU}),
Technological Innovation Institute~(\multirow{-1}{*}{\faTii}),
Toyota Technological Institute at Chicago~(\multirow{-1}{*}{\faTTIC}),
University of California Berkeley~(\multirow{-1}{*}{\faUCB}),
University of California San Diego~(\multirow{-1}{*}{\faUCSD}),
University of Toronto~(\multirow{-1}{*}{\faUT}).\\

\textbf{Licensing}:
open-source~(\CIRCLE),
non-commercial~(\LEFTcircle),
proprietary~(\Circle). \\

\midrule

\makecell[r]{$\ddag$ The first row of each category contains the progenitor architecture.} \\

\makecell[r]{\textbf{*} $\mathtt{XLNet}$ is based on a Transformer architecture integrated with an AutoRegressive component~\cite{yang2019xlnet}.} \\

\bottomrule

\end{tabular}
}

\end{table*}

\newcommand{\faStarHalfStroke}{{\includesvg[height=1.05em]{figs/icons/faStarHalfStroke.svg}}}
\newcommand{\faCircleHalfStroke}{{\includesvg[height=1em]{figs/icons/faCircleHalfStroke.svg}}}

\begin{table*}[ht!]
\centering
\footnotesize
\caption{
Naming adopted, base GenAI architecture, and repositories by revised works.
When the proposal is a tool/framework, we indicate its name with a $\star$ and only report the best-performing base GenAI architecture.
}
\label{tab:openess}

\resizebox{\textwidth}{!}{
\begin{threeparttable}
\begin{tabular}{rrcccccccccccccc}

\toprule

\multirow{2}{*}{\textbf{
Base GenAI Architecture}
} & 
\multirow{2}{*}{\textbf{Proposal Name}} &
\multirow{2}{*}{\textbf{V}}&
\multirow{2}{*}{\textbf{F}}&
\multirow{2}{*}{\textbf{Modifications}} & 
\multicolumn{2}{c}{\textbf{Components}} &
\multicolumn{5}{c}{\textbf{NMM Use Case}} &
\multirow{2}{*}{\textbf{Paper}} &
\multirow{2}{*}{\textbf{Year}} &
\multirow{2}{*}{\begin{tabular}{c}
\textbf{Code}\\\textbf{Repo}
\end{tabular}}\\
\cmidrule(lr){6-7}
\cmidrule(lr){8-12}

 & & & & & \textbf{Pre GenAI} & \textbf{Post GenAI} &
 \textbf{NTG} & \textbf{NTC} & \textbf{NID} & \textbf{NSLA} & \textbf{NDA} & & \\

\toprule
\rc
 & $\mathtt{ESeT}$ & 
\faCheckCircle[regular] &
\faCheckCircle[regular] & 
-- & 
\begin{tabular}{c}
Multi-level Feature\\[\extraarrayvspace]
Extractor
\end{tabular}
& 
\begin{tabular}{c}
Credibility Selector,\\[\extraarrayvspace]
Feature Augmentor
\end{tabular}
& \Circle & \Circle & \CIRCLE & \Circle & \Circle & \cite{li2022extreme} & $2022$ & --\\
\rc
& \gc$\mathtt{RTIDS}$ & 
\gc\faCheckCircle[regular] &
\gc\faCheckCircle[regular] & 
\gc -- & \gc 
SMOTE, Feature Selection
& \gc --
& \gc \Circle & \gc \Circle & \gc \CIRCLE & \gc \Circle & \gc \Circle & \gc \cite{wu2022} & \gc $2022$ & \gc --\\
\rc
& $\mathtt{netFound}$ & 
\faTimesCircle[regular] &
\faCheckCircle[regular] & 
\begin{tabular}{c}
Hierarchical\\[\extraarrayvspace]Attention\\[\extraarrayvspace]Transformer 
\end{tabular} &
-- & -- &
\Circle & \CIRCLE & \CIRCLE & \Circle & \Circle & \cite{guthula2023netfound} & $2023$ & --\\
\rc
& \gc $\mathtt{TNN\text{-}IDS}$ & 
\gc\faCheckCircle[regular] & 
\gc\faTimesCircle[regular] &
\gc -- & \gc Feature Extractor & \gc --
& \gc \Circle & \gc \Circle & \gc \CIRCLE & \gc \Circle & \gc \Circle & \gc \cite{ullah2023tnn} & \gc $2023$ & \gc --\\
\rc
& $\mathtt{RUIDS}$ & 
\faTimesCircle[regular] &
\faCheckCircle[regular] & 
Contrastive Loss & \begin{tabular}{c}Sampling and\\[\extraarrayvspace]Masking Module\end{tabular} & 
\begin{tabular}{c}
Masked Context\\[\extraarrayvspace]
Reconstructor 
\end{tabular}
& 
\Circle & \Circle & \CIRCLE & \Circle & \Circle & \cite{wang2023robust} & $2023$ & --\\
\rc
\multirow{-9}{*}{$\mathtt{Transformer}$} & \gc $\mathtt{TrafficGPT}$ & 
\gc\faTimesCircle[regular] &
\gc\faCheckCircle[regular] & 
\gc Tokenizer & 
\gc --
& \gc --&
\gc \CIRCLE & \gc \CIRCLE & \gc \Circle & \gc \Circle & \gc \Circle & \gc \cite{qu2024trafficgpt} & \gc $2024$ & \gc --\\
\midrule
\rc
\multirow{1}{*}{$\mathtt{T5}$ $1.1$} & $\mathtt{LENS}$ & 
\faTimesCircle[regular] &
\faCheckCircle[regular] & 
\begin{tabular}{c}
Tokenizer,\\[\extraarrayvspace]Pre-Training and\\[\extraarrayvspace]Fine-Tuning Stages
\end{tabular} & 
--
& -- 
& \CIRCLE & \CIRCLE & \CIRCLE & \Circle & \Circle & \cite{wang2024lens} & $2024$ & --\\
\midrule
\ec
 & \gc $\mathtt{ET\text{-}BERT}$ & 
\gc \faTimesCircle[regular] &
\gc \faCheckCircle[regular] & 
\gc \begin{tabular}{c}
Tokenizer,\\[\extraarrayvspace]Pre-Training and\\[\extraarrayvspace]Fine-Tuning Stages
\end{tabular} &
\gc -- & \gc -- & 
\gc \Circle & \gc \CIRCLE & \gc \CIRCLE & \gc \Circle & \gc \Circle & \gc \cite{lin2022} & \gc $2022$ & \gc \href{https://github.com/linwhitehat/ET-BERT}{\faGithub[regular]}\\
\ec
& $\mathtt{LAMBERT}$ & 
\faCheckCircle[regular] &
\faCheckCircle[regular] & 
 -- &  
\begin{tabular}{c}
Pre-Processing,
\end{tabular}
& 
& \Circle & \CIRCLE & \Circle & \Circle & \Circle & \cite{tao2024lambert} & $2024$ & \href{https://github.com/MysteryObstacle/Lambert}{\faGithub[regular]}\\
\ec
& \gc $\mathtt{Dom\text{-}BERT}$ & 
\gc \faCheckCircle[regular] &
\gc \faCheckCircle[regular] & 
\gc -- & \gc Neighbor Sampling & \gc -- &
\gc \Circle & \gc \Circle & \gc \Circle & \gc \CIRCLE & \gc \Circle & \gc \cite{tian2024dom} & \gc $2024$ & \gc --\\
\ec
\multirow{-5}{*}{$\mathtt{BERT}$} & $\mathtt{SecurityBERT}$ & 
\faTimesCircle[regular] &
\faCheckCircle[regular] & 
 Tokenizer & 
\begin{tabular}{c}
Pre-Processing
\end{tabular}
&  --
& \Circle & \Circle & \CIRCLE & \Circle & \Circle & \cite{ferrag2024} & $2024$ & --\\
\midrule
\ec
\multirow{1}{*}{$\mathtt{BERT\text{-}of\text{-}Theseus}$} & \gc $\mathtt{BT\text{-}TPF}$ & 
\gc \faTimesCircle[regular] &
\gc \faCheckCircle[regular] & 
\gc KD Loss & \gc Siamese Network & \gc -- & 
\gc \Circle & \gc \Circle & \gc \CIRCLE & \gc \Circle & \gc \Circle & \gc \cite{wang2024lightweight} & \gc $2024$ & \gc --\\
\midrule
\ec
\multirow{1}{*}{$\mathtt{CodeBERT}$} & $\mathtt{LogPrecis}\:\star$ & 
\faCheckCircle[regular] &
\faCheckCircle[regular] & 
-- & 
--
& -- &
\Circle & \Circle & \Circle & \CIRCLE & \Circle & \cite{boffa2024logprecis} & $2024$ & --\\
\midrule
\cc
\multirow{1}{*}{$\mathtt{GPT\text{-}1}$} & \gc $\mathtt{LogGPT}$ & 
\gc \faCheckCircle[regular] & 
\gc \faTimesCircle[regular] & 
\gc -- & 
\gc \begin{tabular}{c}
Log parser,\\[\extraarrayvspace]
Prompt Constructor
\end{tabular}
& 
\gc \begin{tabular}{c}
Response Parser
\end{tabular}
& \gc \Circle & \gc \Circle & \gc \Circle & \gc \CIRCLE & \gc \Circle & \gc \cite{qi2023loggpt} & \gc $2023$ & \gc --\\
\midrule
\cc
& $\mathtt{GPT\text{-}2C}$ & 
\faCheckCircle[regular] &
\faCheckCircle[regular] & 
-- & -- & -- 
& \Circle & \Circle & \Circle & \CIRCLE & \Circle  & \cite{setianto2021gpt} & $2021$& --\\
\cc
& \gc $\mathtt{LogGPT}$ & 
\gc\faCheckCircle[regular] &
\gc\faCheckCircle[regular] & 
\gc-- & \gc Prompt Generation & \gc Reward Function
& \gc \Circle & \gc \Circle & \gc \Circle & \gc \CIRCLE & \gc \Circle  & \gc \cite{han2023loggpt} & \gc $2023$& \gc \href{https://github.com/nokia/LogGPT}{\faGithub[regular]}\\
\cc
\multirow{-3}{*}{$\mathtt{GPT\text{-}2}$} & $\mathtt{NetGPT}$ & 
\faCheckCircle[regular] &
\faCheckCircle[regular] & 
-- & 
-- & --
& \CIRCLE & \CIRCLE & \CIRCLE & \Circle & \Circle & \cite{meng2023netgpt} & $2023$ & --\\
\midrule
\cc
 & \gc $\mathtt{PAC\text{-}GPT}$ & \gc \faCheckCircle[regular] & 
\gc \faTimesCircle[regular] &
\gc -- & \gc --
& \gc Scapy Generator & \gc \CIRCLE & \gc \Circle & \gc \Circle & \gc \Circle & \gc \Circle  & \gc \cite{Kholgh2023PACGPT} & \gc $2023$ & \gc --\\
\cc
\multirow{-2}{*}{$\mathtt{GPT\text{-}3}$} & $\mathtt{FlowTransformer}\:\star$ &\faCheckCircle[regular] & \faTimesCircle[regular] &
-- & Input Encoder & Classifier 
& \Circle & \Circle & \CIRCLE & \Circle & \Circle & \cite{manocchio2024flowtransformer} & $2024$ & \href{https://github.com/liamdm/FlowTransformer}{\faGithub[regular]}\\
\midrule
\cc
\multirow{1}{*}{$\mathtt{GPT\text{-}3.5}$} & \gc $\mathtt{RAGLog}$ & 
\gc \faCheckCircle[regular] & 
\gc \faTimesCircle[regular] &
\gc -- & 
\gc \begin{tabular}{c}
Log Database,\\[\extraarrayvspace]
Retriever
\end{tabular}
& 
\gc --
& \gc \Circle & \gc \Circle & \gc \Circle & \gc \CIRCLE & \gc \Circle & \gc \cite{pan2023raglog} & \gc $2023$ & \gc -- \\
\midrule

\cc
 & $\mathtt{HuntGPT}$ & 
\faCheckCircle[regular] & 
\faTimesCircle[regular] &
-- & Explainer Module & -- 
& \Circle & \Circle & \CIRCLE & \Circle & \Circle & \cite{ali2023} & $2023$ & --\\
\cc
& \gc $\mathtt{LILAC}$ & 
\gc\faCheckCircle[regular] &
\gc\faCheckCircle[regular] & 
\gc -- & \gc
\begin{tabular}{c}
     Adaptive Parsing Cache\\[\extraarrayvspace]
     ICL-Enhanced Parser\\[\extraarrayvspace]
     Prompt Constructor
\end{tabular}
& \gc --&
\gc \Circle & \gc \Circle & \gc \Circle & \gc \CIRCLE & \gc \Circle & \gc \cite{jiang2023lilac} & \gc $2023$ & \gc --\\
\cc
\multirow{-4}{*}{$\mathtt{GPT\text{-}3.5\text{-}turbo}$} & $\mathtt{CYGENT}\:\star$ & 
\faCheckCircle[regular] &
\faCheckCircle[regular] & 
-- & Prompt Generator & -- 
& \Circle & \Circle & \Circle & \CIRCLE & \Circle & \cite{balasubramanian2024cygent} & $2024$ & --\\
\midrule

\cc
\multirow{1}{*}{$\mathtt{GPT\text{-}4}$} & \gc $\mathtt{NetBuddy}$ & 
\gc \faCheckCircle[regular] & 
\gc \faTimesCircle[regular] &
\gc -- & \gc -- & \gc Output Verifier 
& \gc \Circle & \gc \Circle & \gc \Circle & \gc \Circle & \gc \CIRCLE & \gc \cite{wang2023making} & \gc $2023$ & \gc --\\
\midrule

\cc
 & $\mathtt{FlowMage}\:\star$ & 
\faCheckCircle[regular] & 
\faTimesCircle[regular] &
-- & Prompt Generator & -- 
& \Circle & \Circle & \Circle & \Circle & \CIRCLE & \cite{ghasemirahni2024deploying} & $2024$ & --\\
\cc
\multirow{-2}{*}{$\mathtt{GPT\text{-}4o}$} & \gc $\mathtt{Hermes}\:\star$ & 
\gc\faCheckCircle[regular] & 
\gc\faTimesCircle[regular] &
\gc Chain of LLMs & \gc-- & \gc-- 
& \gc\Circle & \gc\Circle & \gc\Circle & \gc\Circle & \gc\CIRCLE & \gc\cite{ayed2024hermes} & \gc$2024$ & \gc--\\

\midrule
\oc
\multirow{1}{*}{$\mathtt{Diffusion}$ $\mathtt{Model}$} & $\mathtt{NetDiffus}$ & 
 \faCheckCircle[regular] &
 \faCheckCircle[regular] & 
-- &  Txt2Img Module &  -- & 
 \CIRCLE &  \Circle & \Circle & \Circle & \Circle &  \cite{sivaroopan2023netdiffus} &  $2023$ & --\\
\midrule
\oc
\multirow{1}{*}{$\mathtt{Stable}$ $\mathtt{Diffusion}$ $1.5$} & \gc$\mathtt{NetDiffusion}$ & 
\gc\faCheckCircle[regular] &
\gc\faCheckCircle[regular] & 
\gc -- & \gc Txt2Img Module & \gc ControlNet &
\gc\CIRCLE & \gc\Circle & \gc\Circle & \gc\Circle & \gc\Circle & \gc\cite{jiang2024netdiffusion} & \gc $2024$ & \gc\href{https://github.com/noise-lab/NetDiffusion_Generator}{\faGithub[regular]}\\
\midrule
\yc
\multirow{1}{*}{$\mathtt{Mamba}$} &  $\mathtt{NetMamba}$ & 
 \faCheckCircle[regular] &
 \faCheckCircle[regular] & 
 -- & 
 \begin{tabular}{c}
Traffic Representation\\[\extraarrayvspace]
Module
\end{tabular} &  -- &  \Circle & \CIRCLE & \CIRCLE & \Circle & \Circle &  \cite{wang2024netmamba} &  $2024$ & \href{https://github.com/wangtz19/NetMamba}{\faGithub[regular]}\\
\bottomrule

\end{tabular}
\begin{tablenotes}[flushleft]
    \footnotesize
    \item \textbf{V:}~Vanilla Version; \textbf{F:}~Fine-Tuned Version. \textbf{Architecture Category}: Fully Encoder-Decoder (\rb{FED}), Encoder-Only~(\eb{EO}), Decoder-Only~(\cb{DO}), Sequential Denoising Process~(\ob{SDP}), State-Space Model~(\yb{SSM}).
    \item \textbf{NMM Use Case}: Network Traffic Generation~(\textbf{NTG}), Network Traffic Classification~(\textbf{NTC}), Network Intrusion Detection~(\textbf{NID}), Networked System Log Analysis~(\textbf{NSLA}), Network Digital Assistance~(\textbf{NDA}).
\end{tablenotes}
\end{threeparttable}
}
\end{table*}

In this section, we present a model-centric overview of the works categorized in the previous sections.
This section is divided into two parts: ($i$) the first part reports details about the base \gls{genai} architecture leveraged by each work (Sec.~\ref{subsec:genai_models_overview}), while ($ii$) the second part focuses on modifications to \gls{genai} architecture the authors performed in their proposals (Sec.~\ref{subsec:adhoc_genai_solutions}).

\subsection{Overview of GenAI Models\label{subsec:genai_models_overview}
}

This section provides a broad view of the use of (foundation) \gls{genai} models in \gls{nmm}.
Accordingly, Tab.~\ref{table:genai_architectures} is centered around each \textbf{``GenAI Architecture''} and its application in the considered \textbf{``\gls{nmm} Use Cases''}.
First, we want to emphasize the persistent trend towards increasingly complex \gls{genai} models. 
However, such complexity is not fully justified when considering the effectiveness of simpler \gls{ml}/\gls{dl} models in accomplishing the \gls{nmm} tasks discussed in the present survey~\cite{yao2024survey}.

\glsreset{fed}
\glsreset{eo}
\glsreset{do}
\glsreset{sdp}
\glsreset{ssm}

The \gls{genai} architectures leveraged for network tasks defined in Sec.~\ref{subsec:dowstream_tasks} can be broadly divided---%
according to the nature of the underlying layers---%
into $5$ categories (\cf~Sec.~\ref{sec:background}), namely:
\begin{enumerate*}[label=(\textit{\roman*})]
  \item \gls{fed},
  commonly leveraged for tasks where both the input and output are sequences, such as sequence-to-sequence tasks;
  \item \gls{eo},
  intended for language comprehension, specifically for interpreting and encoding input text for various subsequent applications;
  \item \gls{do},
  frequently used for autoregressive text generation, meaning it generates text tokens based on the preceding token;
  \item \gls{sdp},
  applied for producing high-quality data by means of successive denoising;
  \item Selective and Structured \gls{ssm},
 representing a sophisticated means created for effectively handling and modeling long sequential data.
\end{enumerate*}

\gls{genai} architectures
like \fmtTT{Transformer}, \fmtTT{T5}, \fmtTT{Gemini}, \fmtTT{Mistral}, \fmtTT{Zephyr}, and \fmtTT{XLNet}%
\footnote{\fmtTT{XLNet}, proposed by Google and Carnegie Mellon University in~\cite{yang2019xlnet}, combines a Transformer-based model with an AutoRegressive component.}
belong to the \gls{fed}
category.
The 
\gls{eo} category encompasses \fmtTT{BERT} along with its enhancements (\eg \fmtTT{DistilRoBERTa}) and variations (\eg \fmtTT{CodeBERT}), as well as \fmtTT{ViT}. The 
\gls{do}
category features \fmtTT{GPT}, \fmtTT{Falcon}, \fmtTT{LLaMA}, and \fmtTT{Phi}. The 
\gls{sdp}
category includes \fmtTT{Diffusion} and \fmtTT{Stable Diffusion} architectures, while the 
\gls{ssm}
category consists of the sole \fmtTT{Mamba} architecture.

Table~\ref{table:genai_architectures} clearly shows that certain categories are more frequently utilized than others (column \textbf{``Cat.''}).
The \gls{do} category accounts for the majority of works, followed by the \gls{fed} and \gls{eo} categories.
When examining the models employed (\textbf{``Name''}), \fmtTT{Transformer} and \fmtTT{BERT} (by Google), the \fmtTT{GPT} family (by OpenAI), and \fmtTT{LLaMA} (by Meta) are the most widely used.
Equally important, the reviewed works use most of the architectures designed for language-processing tasks. Accordingly, a large portion of them use plain-text-arranged information as model input (\textbf{``Input''}).
Occasionally, this input is formatted as code, especially when the model is fine-tuned for code-generation tasks, like \fmtTT{CodeBERT}, \fmtTT{CodeLLaMA}, or \fmtTT{CodeT5}.
Only a small fraction of the works~\cite{sivaroopan2023netdiffus,jiang2024netdiffusion,ho2022network} considers input traffic shaped as images.

A different perspective on \gls{genai} models utilized for \gls{nmm} use cases focuses on the research organization (\textbf{``Res.~Org.''}) that introduced them and the associated licensing framework (\textbf{``Lic.''}).
From this viewpoint, \emph{three key points} emerge:
\begin{itemize}
    \item \textit{Non-academic organizations dominate the development of these architectures}. 
    Google has been prolific, especially in the \gls{fed} and \gls{eo} categories, OpenAI has primarily developed \gls{do} solutions, and Meta and Microsoft have contributed to both the \gls{eo} and \gls{do} \gls{genai} categories.

    \item \textit{The open-source paradigm is also embraced by non-academic entities}. 
    Google, except for its private \fmtTT{Bard} and \fmtTT{Gemini} models, and Microsoft have released many models as open-source. In contrast, OpenAI's recent products, from \fmtTT{GPT-3} onward, are closed-source. Meta, except for the open-source \fmtTT{RoBERTa} model, and StabilityAI employ non-commercial licenses.

    \item  \textit{Both academic and non-academic entities have pioneered each category of models}. These progenitor architectures are reported with a  ``$\ddag$'' in Tab.~\ref{table:genai_architectures}.
    Google developed both \fmtTT{Transformer} and \fmtTT{BERT} models (the first in collaboration with the University of Toronto). OpenAI designed \fmtTT{GPT-1}. \fmtTT{Diffusion} and \fmtTT{Mamba} models are completely proposed by academic entities, namely, the former from Stanford and California universities and the latter from Carnegie Mellon and Princeton universities.

\end{itemize}

Concerning the specific use cases of the models in \gls{nmm} (\textbf{``NMM Use Case''} column),
\fmtTT{BERT} and \fmtTT{GPT}-like models address a wide range of applications, with \fmtTT{BERT} predominantly used for \gls{nid} and \fmtTT{GPT}-like models for \gls{nda}. The \fmtTT{Transformer} model is applied to \gls{traffgen}, \gls{traffclass}, and \gls{nid}.
The \fmtTT{LLaMA} family has been exclusively used for \gls{nda}.
Notably, the \gls{traffgen} use case is addressed by various \gls{genai} model categories,
including \gls{fed} models like \fmtTT{ViT},
\gls{do} models such as  \fmtTT{GPT-2} and \fmtTT{GPT-3}, 
and diffusion-based models.

\subsection{
Ad-Hoc GenAI Solutions\label{subsec:adhoc_genai_solutions}
}

Table~\ref{tab:openess} outlines the naming conventions used by state-of-the-art solutions.
It specifies the base architecture and whether it is used ``as-is'' or fine-tuned, details the modifications made (if any), lists the components included in the pipeline before and after the \gls{genai} model (if any), and indicates the availability of repositories for each proposed framework.

On the one hand, the majority of the works reported in Tab.~\ref{tab:openess} ($21$ out of $28$) utilize the base architecture in its vanilla version (see \textbf{``V''} column), typically adding pre-\gls{genai} components like feature selectors/extractors and traffic-to-image modules, or post-\gls{genai} components such as output refinement modules, exemplified by the ControlNet used in \fmtTT{NetDiffusion}.
Among these, $8$ works simply leverage the \gls{genai} model without modifications or fine-tuning (see \textbf{``F''} column) but adding at least one pre-\gls{genai} or post-\gls{genai} component.
Conversely, only $8$ works propose modifications to the \gls{genai} model and also perform fine-tuning for the targeted use cases.

For the \gls{traffgen} use case, notable examples include \fmtTT{TrafficGPT}~\cite{qu2024trafficgpt} and \fmtTT{LENS}~\cite{wang2024lens}, both of which modify the tokenizer to handle network traffic. \fmtTT{LENS} also changes the pre-training and fine-tuning phases.
In the \gls{traffclass} and \gls{nid} use cases,  \fmtTT{LENS}~\cite{wang2024lens} is again notable, along with \fmtTT{netFound}~\cite{guthula2023netfound}, which uses a Hierarchical Attention Transformer architecture, and \fmtTT{ET-BERT}~\cite{lin2022}, which redefines the tokenizer and the pre-training/fine-tuning procedures.
Notably, none of the works addressing \gls{nla} and \gls{nda} modify and fine-tune the base \gls{genai} architecture, as these use cases align closely with the core philosophy of \gls{genai}, \eg document summarization and question-answering.

Finally, we also note whether a paper provides access to a related public repository 
(\textbf{``Code Repo''} column). 
\textit{Only $6$ works make their framework code available}~\cite{lin2022, han2023loggpt, wang2024netmamba, liu2024large, manocchio2024flowtransformer, jiang2024netdiffusion}.
This lack of shared code significantly hampers reproducibility and hinders further development and verification by the research community.

\section{Datasets and Platforms}
\label{sec:dataset_code_platforms}

\newcommand{\ntccolor}{\rowcolor{lightgreen}}
\newcommand{\nidcolor}{\rowcolor{lightyellow}}
\newcommand{\nlacolor}{\rowcolor{lightpink}}
\newcommand{\ndacolor}{\rowcolor{lightviolet}}
\newcommand{\ntgcolor}{\rowcolor{lightcyan}}

\newcommand{\ntccolorcell}{\cellcolor{lightgreen}}
\newcommand{\nidcolorcell}{\cellcolor{lightyellow}}
\newcommand{\nlacolorcell}{\cellcolor{lightpink}}
\newcommand{\ndacolorcell}{\cellcolor{lightviolet}}
\newcommand{\ntgcolorcell}{\cellcolor{lightcyan}}

\newcommand{\ntccolorbox}[1]{\colorbox{lightgreen}{#1}}
\newcommand{\nidcolorbox}[1]{\colorbox{lightyellow}{#1}}
\newcommand{\nlacolorbox}[1]{\colorbox{lightpink}{#1}}
\newcommand{\ndacolorbox}[1]{\colorbox{lightviolet}{#1}}
\newcommand{\ntgcolorbox}[1]{\colorbox{lightcyan}{#1}}

\definecolor{lightgreen}{RGB}{216, 248, 201}
\definecolor{lightyellow}{RGB}{255, 238, 194}
\definecolor{lightpink}{RGB}{255, 214, 231} 
\definecolor{lightviolet}{RGB}{217, 204, 222}
\definecolor{lightcyan}{RGB}{204, 245, 255}

\setlength{\extraarrayvspace}{-0.4ex}

\begin{table*}[tp]
\centering
\caption{Overview of datasets used in the pre-training, tine-tuning, and evaluation phases of GenAI models in the NMM use cases considered. Datasets are reported in chronological order.
}
\label{tab:dataset-overview}
\resizebox{0.99\textwidth}{!}{
\begin{threeparttable}
    \begin{tabular}{l c l c c c}
    \toprule
        \textbf{Dataset Name} & \textbf{Year} & \textbf{Network Data} & \textbf{Pre-Training} & \textbf{Fine-Tuning} & \textbf{Evaluation} \\
        \midrule
        
        \gr KDD'99~\cite{tavallaee2009detailed} & 1999 & 
        Per-flow traffic features
        & \nidcolorcell -- & \nidcolorcell \cite{ali2023}\cite{wang2023robust} & \nidcolorcell \cite{ali2023}\cite{wang2023robust} \\
        
        \midrule
        
        ECML/PKDD 2007~\cite{raissi2007web} & 2007 & Web server logs & \nlacolorcell -- & \nlacolorcell \cite{karlsen2024large} & \nlacolorcell \cite{karlsen2024large} \\
        
        \midrule

        \gr ISCX NSL-KDD~\cite{tavallaee2009detailed} & 2009 & Per-flow traffic features & \nidcolorcell -- & \nidcolorcell \cite{manocchio2024flowtransformer}\cite{lai2023} & \nidcolorcell \cite{manocchio2024flowtransformer}\cite{lai2023} \\
        
        \midrule

        & & & \nidcolorcell -- & \nidcolorcell \cite{seyyar2022attack} & \nidcolorcell \cite{seyyar2022attack} \\
        \multirow{-2}{*}{CSIC 2010~\cite{csic_2010_web_application_attacks}} & \multirow{-2}{*}{2010} & \multirow{-2}{*}{HTTP requests} & \nlacolorcell -- & \nlacolorcell \cite{karlsen2024large} & \nlacolorcell \cite{karlsen2024large} \\
        
        \midrule
        
        \gr HttpParams~\cite{httpparamsdataset} & 2015 & HTTP requests & \nidcolorcell -- & \nidcolorcell \cite{seyyar2022attack} & \nidcolorcell \cite{seyyar2022attack} \\
        
        \midrule
        
        \multirow{2}{*}{UNSW-NB15~\cite{moustafa2015unsw}} & 
        \multirow{2}{*}{2015} & 
        \multirow{2}{*}{%
        \begin{tabular}[c]{@{}l@{}}
        Raw traffic data\\[\extraarrayvspace]Per-flow traffic features
        \end{tabular}
        }
        & \ntgcolorcell \cite{bikmukhamedov2021} & \ntgcolorcell \cite{bikmukhamedov2021} & \ntgcolorcell \cite{bikmukhamedov2021} \\
        & & & \nidcolorcell -- & \nidcolorcell \cite{manocchio2024flowtransformer}\cite{ho2022network}\cite{wang2023robust} & \nidcolorcell \cite{manocchio2024flowtransformer}\cite{ho2022network}\cite{wang2023robust} \\
        
        \midrule
        
        \gr IoT Sentinel~\cite{miettinen2017iot} & 2016 & Traffic logs & \nlacolorcell -- & \nlacolorcell -- & \nlacolorcell \cite{meyuhas2024} \\
        
        \midrule

        & & & \ntgcolorcell \cite{meng2023netgpt}\cite{bikmukhamedov2021}\cite{wang2024lens}\cite{qu2024trafficgpt} & \ntgcolorcell \cite{meng2023netgpt}\cite{bikmukhamedov2021}\cite{wang2024lens}\cite{qu2024trafficgpt} & \ntgcolorcell \cite{meng2023netgpt}\cite{bikmukhamedov2021}\cite{wang2024lens}\cite{qu2024trafficgpt} \\
        \multirow{-2}{*}{ISCXVPN2016~\cite{draper2016characterization}} & \multirow{-2}{*}{2016} & 
        \multirow{-2}{*}{%
        \begin{tabular}[c]{@{}l@{}}
        Raw traffic data\\[\extraarrayvspace]Per-flow traffic features
        \end{tabular}
        } 
        & \ntccolorcell \cite{meng2023netgpt}\cite{lin2022}\cite{wang2024netmamba}\cite{wang2024lens}\cite{qu2024trafficgpt}\cite{tao2024lambert} & \ntccolorcell \cite{meng2023netgpt}\cite{lin2022}\cite{wang2024netmamba}\cite{wang2024lens}\cite{qu2024trafficgpt}\cite{tao2024lambert} & \ntccolorcell \cite{meng2023netgpt}\cite{lin2022}\cite{wang2024netmamba}\cite{wang2024lens}\cite{qu2024trafficgpt}\cite{tao2024lambert} \\
        
        \midrule
        
        \gr & & & \ntgcolorcell \cite{wang2024lens}\cite{qu2024trafficgpt} & \ntgcolorcell \cite{wang2024lens}\cite{qu2024trafficgpt} & \ntgcolorcell \cite{wang2024lens}\cite{qu2024trafficgpt} \\
        \gr \multirow{-2}{*}{ISCXTor2016~\cite{lashkari2017characterization}} & \multirow{-2}{*}{2016} & 
        \multirow{-2}{*}{%
        \begin{tabular}[c]{@{}l@{}}
        Raw traffic data\\[\extraarrayvspace]Per-flow traffic features
        \end{tabular}
        }
        & \ntccolorcell \cite{wang2024lens}\cite{qu2024trafficgpt}\cite{wang2024netmamba} & \ntccolorcell \cite{wang2024lens}\cite{lin2022}\cite{wang2024netmamba} & \ntccolorcell \cite{wang2024lens}\cite{lin2022}\cite{wang2024netmamba} \\
        
        \midrule
        
        & & & \ntgcolorcell \cite{bikmukhamedov2021} & \ntgcolorcell \cite{bikmukhamedov2021} & \ntgcolorcell \cite{bikmukhamedov2021} \\
        \multirow{-2}{*}{UNSW-IoT-Analytics~\cite{sivanathan2018classifying}} & \multirow{-2}{*}{2016} & 
        \multirow{-2}{*}{%
        \begin{tabular}[c]{@{}l@{}}
        Raw traffic data \& logs\\[\extraarrayvspace]Per-flow traffic features
        \end{tabular} 
        }
        & \nlacolorcell -- & \nlacolorcell -- & \nlacolorcell \cite{meyuhas2024} \\
        
        \midrule
        
        \gr & & & \ntgcolorcell \cite{meng2023netgpt}\cite{qu2024trafficgpt}\cite{wang2024lens} & \ntgcolorcell \cite{meng2023netgpt}\cite{wang2024lens}\cite{qu2024trafficgpt} & \ntgcolorcell \cite{meng2023netgpt}\cite{wang2024lens}\cite{qu2024trafficgpt} \\
        \gr & & & \ntccolorcell \cite{meng2023netgpt}\cite{wang2024netmamba}\cite{wang2024lens}\cite{qu2024trafficgpt} & \ntccolorcell \cite{qu2024trafficgpt} & \ntccolorcell \cite{qu2024trafficgpt} \\
        \gr \multirow{-3}{*}{USTC-TFC2016~\cite{ustc-tfc2016}} & \multirow{-3}{*}{2016} & \multirow{-3}{*}{Raw traffic data} & \nidcolorcell \cite{meng2023netgpt}\cite{wang2024netmamba}\cite{wang2024lens} & \nidcolorcell \cite{meng2023netgpt}\cite{lin2022}\cite{wang2024netmamba}\cite{wang2024lens} & \nidcolorcell \cite{meng2023netgpt}\cite{lin2022}\cite{wang2024netmamba}\cite{wang2024lens} \\
        
        \midrule
        
        \multirow{2}{*}{CIC-IDS2017~\cite{sharafaldin2018toward}} & \multirow{2}{*}{2017} & 
        \multirow{2}{*}{%
        \begin{tabular}[c]{@{}l@{}}
        Raw traffic data\\[\extraarrayvspace]Per-flow traffic features
        \end{tabular}
        }
        & \ntccolorcell \cite{lin2022}\cite{tao2024lambert} & \ntccolorcell -- & \ntccolorcell -- \\
        
        & & & \nidcolorcell \cite{wang2024lightweight} & \nidcolorcell \cite{manocchio2024flowtransformer}\cite{guthula2023netfound}\cite{li2022extreme}\cite{ho2022network}\cite{wu2022}\cite{wang2023robust}\cite{wang2024lightweight} & \nidcolorcell \cite{manocchio2024flowtransformer}\cite{guthula2023netfound}\cite{li2022extreme}\cite{ho2022network}\cite{wu2022}\cite{wang2023robust}\cite{wang2024lightweight} \\

        \midrule

        \gr Deep Fingerprinting~\cite{sirinam2018deep} & 2017 & Per-flow traffic features & \ntgcolorcell \cite{sivaroopan2023netdiffus} & \ntgcolorcell \cite{sivaroopan2023netdiffus} & \ntgcolorcell \cite{sivaroopan2023netdiffus} \\
        
        \midrule

        LabeledFlows~\cite{labeledflows17} & 2017 & Per-flow traffic features & \ntgcolorcell\cite{zhang2024netdiff} & \ntgcolorcell- & \ntgcolorcell\cite{zhang2024netdiff}\\
        \midrule
        
        & & & \ntgcolorcell \cite{qu2024trafficgpt} & \ntgcolorcell \cite{qu2024trafficgpt} & \ntgcolorcell \cite{qu2024trafficgpt} \\
        \multirow{-2}{*}{Cross-Platform~\cite{van2020flowprint}} & \multirow{-2}{*}{2018} & \multirow{-2}{*}{Raw traffic data} & \ntccolorcell \cite{wang2024netmamba} & \ntccolorcell \cite{lin2022}\cite{wang2024netmamba}\cite{qu2024trafficgpt} & \ntccolorcell \cite{lin2022}\cite{wang2024netmamba}\cite{qu2024trafficgpt} \\
        
        \midrule
        
        \gr & & & \ntccolorcell \cite{tao2024lambert} & \ntccolorcell -- & \ntccolorcell -- \\
        \gr \multirow{-2}{*}{CSE-CIC-IDS2018~\cite{sharafaldin2018toward}} & \multirow{-2}{*}{2018} & 
        \multirow{-2}{*}{%
        \begin{tabular}[c]{@{}l@{}}
        Raw traffic data\\[\extraarrayvspace]Per-flow traffic features
        \end{tabular}
        }
        & \nidcolorcell -- & \nidcolorcell \cite{manocchio2024flowtransformer}\cite{li2022extreme} & \nidcolorcell \cite{manocchio2024flowtransformer}\cite{li2022extreme} \\
        
        \midrule
        
        NLP2Bash~\cite{lin2018nl2bash} & 2018 & Unix shell logs & \nlacolorcell -- & \nlacolorcell \cite{boffa2024logprecis} & \nlacolorcell \cite{boffa2024logprecis} \\
        
        \midrule
        
        \gr CIC-DDoS2019~\cite{sharafaldin2019developing} & 2019 & 
        Per-flow traffic features
        & \nidcolorcell -- & \nidcolorcell \cite{wu2022} & \nidcolorcell \cite{wu2022} \\
        
        \midrule
        
        IoTFinder~\cite{perdisci2020iotfinder} & 2019 & Traffic logs & \nlacolorcell -- & \nlacolorcell -- & \nlacolorcell \cite{meyuhas2024} \\
        
        \midrule
        
        \gr Web Server Access Logs~\cite{zaker_2019} & 2019 & Web server logs \& prompts & \nlacolorcell -- & \nlacolorcell \cite{balasubramanian2024cygent} & \nlacolorcell \cite{balasubramanian2024cygent} \\
        
        \midrule
        
        LabeledFlows~\cite{labeledflows19} & 2019 & Per-flow traffic features & \ntgcolorcell\cite{zhang2024netdiff} & \ntgcolorcell- & \ntgcolorcell\cite{zhang2024netdiff}\\
        \midrule
        
        Apache Web Server~\cite{hilmi2020} & 2020 & Web server logs & \nlacolorcell -- & \nlacolorcell \cite{karlsen2024large} & \nlacolorcell \cite{karlsen2024large} \\
        
        \midrule
        
        \gr & & & \ntgcolorcell \cite{meng2023netgpt}\cite{wang2024lens}\cite{qu2024trafficgpt} & \ntgcolorcell \cite{meng2023netgpt}\cite{wang2024lens}\cite{qu2024trafficgpt} & \ntgcolorcell \cite{meng2023netgpt}\cite{wang2024lens}\cite{qu2024trafficgpt} \\
        \gr \multirow{-2}{*}{CIRA-CIC-DoHBrw2020~\cite{montazerishatoori2020detection}} & \multirow{-2}{*}{2020} & 
        \multirow{-2}{*}{%
        Raw traffic data
        }
        & \ntccolorcell \cite{meng2023netgpt}\cite{wang2024lens}\cite{qu2024trafficgpt} & \ntccolorcell \cite{wang2024lens} & \ntccolorcell \cite{wang2024lens} \\
        
        \midrule
        
        CyberLab Honeynet~\cite{sedlar_2020_3687527} & 2020 & Traffic logs & \nlacolorcell -- & \nlacolorcell \cite{boffa2024logprecis}\cite{setianto2021gpt} & \nlacolorcell \cite{boffa2024logprecis}\cite{setianto2021gpt} \\
        
        \midrule
        
        \gr Loghub~\cite{zhu2023loghub} & 2020 & System logs \& prompts & \nlacolorcell -- & \nlacolorcell \cite{han2023loggpt}\cite{karlsen2024large} & \nlacolorcell 
        \cite{han2023loggpt}\cite{ott2021robust}\cite{pan2023raglog}\cite{qi2023loggpt}\cite{jiang2023lilac}\cite{mudgal2023assessment}\cite{karlsen2024large} \\
        
        \midrule
        
        MUDIS~\cite{bremler2022one} & 2020 & Traffic logs & \nlacolorcell -- & \nlacolorcell -- & \nlacolorcell \cite{meyuhas2024} \\
        
        \midrule
        
        \gr MQTT-IoT-IDS2020~\cite{hindy2020machine} & 2020 & 
        \begin{tabular}[c]{@{}l@{}}
        Raw traffic data\\[\extraarrayvspace]Per-flow traffic features
        \end{tabular}
        & \nidcolorcell \cite{ullah2023tnn} & \nidcolorcell \cite{manocchio2024flowtransformer}\cite{ullah2023tnn} & \nidcolorcell \cite{manocchio2024flowtransformer}\cite{ullah2023tnn} \\
        
        \midrule
        
        & & & \ntgcolorcell -- & \ntgcolorcell \cite{Kholgh2023PACGPT} & \ntgcolorcell \cite{Kholgh2023PACGPT} \\
        \multirow{-2}{*}{TON\_IoT~\cite{moustafa2021new}} & \multirow{-2}{*}{2020} & 
        \multirow{-2}{*}{%
        \begin{tabular}[c]{@{}l@{}}
        Raw traffic data\\[\extraarrayvspace]Per-flow traffic features
        \end{tabular} 
        }
        & \nidcolorcell -- & \nidcolorcell \cite{manocchio2024flowtransformer}\cite{ghourabi2022security}\cite{wang2024lightweight} & \nidcolorcell \cite{manocchio2024flowtransformer}\cite{ghourabi2022security}\cite{wang2024lightweight} \\
        
        \midrule
        
        \gr CSTNET-TLS1.3~\cite{lin2022} & 2021 & Raw traffic data & \ntccolorcell \cite{tao2024lambert} & \ntccolorcell \cite{lin2022}\cite{tao2024lambert} & \ntccolorcell \cite{lin2022}\cite{tao2024lambert} \\
        
        \midrule
        
        FWAF~\cite{fwaf-dataset} & 2021 & HTTP requests & \nidcolorcell -- & \nidcolorcell \cite{seyyar2022attack} & \nidcolorcell \cite{seyyar2022attack} \\
        
        \midrule
        
        \gr Spirit~\cite{eu_spirit_consortium_2021_4767861} & 2021 & 
        System logs 
        & \nlacolorcell -- & \nlacolorcell 
        \cite{qi2023loggpt}\cite{karlsen2024large} & \nlacolorcell \cite{karlsen2024large} \\
        
        \midrule
        
        & & & \ntgcolorcell \cite{qu2024trafficgpt} & \ntgcolorcell \cite{qu2024trafficgpt} & \ntgcolorcell \cite{qu2024trafficgpt} \\
        & & & \ntccolorcell
        \cite{qu2024trafficgpt}\cite{wang2024netmamba} & \ntccolorcell -- & \ntccolorcell -- \\
        \multirow{-3}{*}{CIC-IoT-2022~\cite{dadkhah2022towards}} & \multirow{-3}{*}{2022} & 
        \multirow{-3}{*}{%
        \begin{tabular}[c]{@{}l@{}}
        Raw traffic data\\[\extraarrayvspace]Per-flow traffic features
        \end{tabular}
        }
        & \nidcolorcell
        \cite{wang2024netmamba} & \nidcolorcell \cite{wang2024netmamba} & \nidcolorcell \cite{wang2024netmamba} \\
        
        \midrule
        
        \gr Edge-IIoTset~\cite{ferrag2022edge} & 2022 & 
        \begin{tabular}[c]{@{}l@{}}
        Raw traffic data\\[\extraarrayvspace]Per-flow traffic features
        \end{tabular}
        & \nidcolorcell -- & \nidcolorcell \cite{ferrag2024}\cite{melicias2024gpt} & \nidcolorcell \cite{ferrag2024}\cite{melicias2024gpt} \\
        
        \midrule

        & & & \ntccolorcell \cite{wang2024lens} & \ntccolorcell -- & \ntccolorcell -- \\
        \multirow{-2}{*}{CIC-IoT-2023~\cite{neto2023ciciot2023}} & \multirow{-2}{*}{2023} & 
        \multirow{-2}{*}{\begin{tabular}[c]{@{}l@{}}
        Raw traffic data\\[\extraarrayvspace]Per-flow traffic features
        \end{tabular}} & \nidcolorcell \cite{wang2024lens} & \nidcolorcell -- & \nidcolorcell -- \\
        
        \midrule
        
        \gr NeMoEval~\cite{mani2023enhancing} & 2023 & Network graphs \& prompts & \ndacolorcell -- & \ndacolorcell -- & \ndacolorcell \cite{mani2023enhancing} \\
        
        \midrule
        
        SPEC5G~\cite{karim2023spec5g} & 2023 & Prompts & \ndacolorcell -- & \ndacolorcell -- & \ndacolorcell \cite{ahmed2024linguistic} \\
        
        \midrule
        
        \gr TeleQnA~\cite{maatouk2023teleqna} & 2023 & Prompts & \ndacolorcell -- & \ndacolorcell \cite{roychowdhury2024unlocking}\cite{ahmed2024linguistic}\cite{piovesan2024telecom}\cite{erak2024leveraging} & \ndacolorcell \cite{roychowdhury2024unlocking}\cite{ahmed2024linguistic}\cite{piovesan2024telecom}\cite{erak2024leveraging} \\
        
        \midrule
        
        \multirow{2}{*}{Censys~\cite{durumeric2015search}} & \multirow{2}{*}{2024} & \multirow{2}{*}{\begin{tabular}[c]{@{}l@{}}
        Traffic logs\\[\extraarrayvspace]HTTP requests
        \end{tabular}} &
        \ntccolorcell \cite{sarabi2023} & \ntccolorcell \cite{sarabi2023} & \ntccolorcell \cite{sarabi2023} \\
        & & &
        \nlacolorcell -- & \nlacolorcell -- & \nlacolorcell \cite{meyuhas2024} \\
        
        \bottomrule
    \end{tabular}
    \begin{tablenotes}[flushleft]
    \footnotesize
    \item \textbf{Use Cases:} \ntgcolorbox{Network Traffic Generation (NTG)}--\ntccolorbox{Network Traffic Classification (NTC)}--\nidcolorbox{Network Intrusion Detection (NID)}--\nlacolorbox{Networked System Log Analysis (NSLA)}--\ndacolorbox{Network Digital Assistance (NDA)}. 
    \item
    The datasets highlighted with more than one color are used for multiple use cases.
    \end{tablenotes}
\end{threeparttable}
}
\end{table*}

In the ever-changing networking landscape, \gls{genai} emerges as a revolutionary tool that fosters innovative solutions and new perspectives.
In this section, we explore two principal aspects of the \gls{genai} tools that are used from a networking perspective, inspecting (\textit{i}) the leveraged datasets for pre-training or fine-tuning/evaluation phases (Sec.~\ref{subsec:datasets}) and (\textit{ii}) the available platforms (Sec.~\ref{subsec:platforms}) that facilitate the development of \gls{genai} solutions for \gls{nmm}.

\subsection{Datasets}
\label{subsec:datasets}

Data play a critical role in applying \gls{genai} in networking since they directly affect the development of \gls{ai} models and their performance.
In this section, we discuss the adoption of various datasets in the \textbf{``Pre-Training''}, \textbf{``Fine-Tuning''}, and \textbf{``Evaluation''} phases of a \gls{genai} model life-cycle, when dealing with the \gls{nmm} use cases defined in Sec.~\ref{subsec:dowstream_tasks}. 
Table~\ref{tab:dataset-overview} presents such an overview by associating \emph{different colors} to the \gls{nmm} use cases, along with general information about the datasets, namely their \textbf{``Year''} (according to the collection time-span) and the \textbf{``Network Data''} each dataset provides.%
\footnote{As presented in Sec.~\ref{sec:genai_landscape}, different works could apply different pre-processing operations and extract different information from considered datasets to feed \gls{genai} models.}
It is worth noticing that the vast amount of unlabeled data collected (and now available) thanks to programmable devices and Software-Defined Networking (SDN) can be effectively utilized to refine the pre-training process of GenAI models, similarly as in other domains~\cite{zhu2022zoonet, yu2019network, d2019survey}.
Still,
we remark that many researchers do not release the data they used in their work, making it difficult to find publicly available datasets that are adequately representative of the specific context.
To contribute valuable resources to the community and foster reproducibility, we focus on datasets that are \textit{publicly available} or obtainable\textit{ upon-request} ($35$ out of $55$).
Notably, \gls{nda} is the \gls{nmm} use case with fewer datasets reported in Tab.~\ref{tab:dataset-overview}; 
this is because most of them (\ie~$7$ out of $10$) are closed source (\cf Sec.~\ref{sec:network_management}).

Looking at the temporal evolution, older datasets tend to be more general and not explicitly collected or formatted for \gls{genai} purposes. 
Conversely, more recent datasets~\cite{zhu2023loghub,mani2023enhancing,karim2023spec5g,maatouk2023teleqna} are designed to optimize the training or evaluation of \gls{genai} models. 
These datasets often include explicit prompts to better facilitate the generation of coherent and contextually relevant content.
This is particularly true for \gls{nla} and \gls{nda} use cases---being more affine to the \gls{nlp} domain---where tailored datasets are used. 
In contrast, the other \gls{nmm} use cases often repurpose traffic datasets originally collected for different objectives to train/evaluate \gls{genai} models.
For example, in~\cite{lin2022} numerous raw-traffic datasets collected in different domains~\cite{lashkari2017characterization,sharafaldin2018toward,van2020flowprint,lin2022} 
are fused to fine-tune the \fmtTT{ET-BERT} model and allow it to deal with multiple 
\gls{traffclass}/\gls{nid}-related tasks (\eg \gls{traffclass} on VPN/TLS/Tor, malware classification).

Other trends can be inferred when considering the phases of a \gls{genai} model life-cycle. 
On the one hand, the datasets mostly used for pre-training are typically more generic than those used in the other phases, since they aim to allow the model learning a foundational understanding of the context.
On the other hand, the datasets used for fine-tuning are vertical to the specific \gls{nmm} use case, as their goal is to specialize the pre-trained model.
Similarly, the evaluation phase requires data specific to the \gls{nmm} use case to be solved. 
Therefore, most fine-tuning datasets are also used for evaluation. 
Additionally, the evaluation datasets are usually meticulously labeled to facilitate accurate quantitative assessment of the related models in a supervised manner.
Unfortunately, regarding this latter aspect, different methods often rely on distinct datasets for performance evaluation. 
This inconsistency results in significant efforts for manual data processing and ultimately in unfair comparisons. 
To address this limitation, initial large-scale and unified benchmark datasets specifically designed for assessing \gls{genai} models are beginning to be proposed~\cite{qian2024netbench}. These datasets---particularly for validating foundation models in \gls{traffclass} and \gls{traffgen} use cases---aim to replace the current reliance on heterogeneous compositions of older datasets.

From Tab.~\ref{tab:dataset-overview}, we can also extract some differences in how data are used for different \gls{nmm} use cases.
Interestingly, all the works falling within the \gls{traffclass} (colored in pale green) train from scratch the \gls{genai} architecture on one or more datasets.
On the other hand, as can be noted from the ``pre-training'' column, all the works that perform \gls{nla} and \gls{nda}, along with the majority addressing intrusion detection, start with a pre-trained model and only fine-tune it.
While it is expected for \gls{nla} and \gls{nda} use cases, whose models are usually fed with textual inputs (and can thus leverage pre-trained \glspl{llm}), this does not apply to works performing intrusion detection.
Indeed, the latter commonly employ models pre-trained for affine tasks,
such as 
\gls{traffclass} (and vice versa). 
Accordingly, for \gls{traffclass}, some datasets~\cite{sharafaldin2018toward,neto2023ciciot2023} are used exclusively for pre-training purposes. 
Finally, datasets 
employed
in \gls{traffgen} are commonly leveraged for other tasks as well, except for \gls{nda}, which, as aforementioned, relies on ad-hoc \gls{genai}-tailored data.

Table~\ref{tab:dataset-overview} highlights the presence of a relatively large number of publicly available datasets.
However, it is crucial for the scientific community to focus on the quality of these datasets.
Many datasets fail to adequately represent real-world scenarios due to various factors, such as the use of synthetically generated data, small-scale testbeds, or heavy anonymization~\cite{jacobs2022ai}.
For instance, CIC-IDS2017, which is the most common dataset for the \gls{nid} task (cf.~Sec.~\ref{sec:intrusion_detection}), suffers from several issues that hinder its utility as a benchmark~\cite{engelen2021troubleshooting}.
Therefore, an important step for the scientific community would be to critically filter public datasets to select only those that provide a reliable representation of real-world scenarios.

\subsection{Dedicated Platforms}
\label{subsec:platforms}

\gls{genai} solutions require huge amounts of resources to be effective---%
during both the pre-training and the fine-tuning phases.
Thus the deployment of these solutions can benefit from simplifications in the management of computing infrastructures, which translates into reduced maintenance costs.
This calls for
\emph{dedicated platforms} that offer integrated environments that streamline the end-to-end \gls{ai} workflow.
These platforms provide essential tools and services for data processing, model training, deployment, and monitoring.

In the context of \gls{genai}, platforms can be categorized based on two main aspects: (\textit{a}) whether they provide only first-party models (\ie~developed by the platform owner) or also include third-party models, and
(\textit{b}) whether they offer open-source models in addition to closed-source ones, which are typically accessible exclusively through \glspl{ui}, \glspl{api}, or \glspl{cli}.

Here we survey some of the most well-known and utilized platforms according to this categorization.

To the best of our knowledge, no platform that offers only \textbf{first-party} models, provides them as \textbf{open source}.
\emph{OpenAI GPT} is a notable case that offers access to a vast amount of \textbf{first-party} 
pre-trained natural language processing models---such as \fmtTT{GPT} and its advancements---which can be fine-tuned using computational resources provided by OpenAI. 
However, the models are \textbf{closed}, and they remain available and can be operated only on OpenAI servers. 
The functionality of OpenAI GPT can be accessed through \gls{ui}/\gls{api} via free/paid plans or by signing a partnership.

On the other hand, most platforms offer both \textbf{first-party and third-party} models, including \textbf{open-source and closed} alternatives. 
In this case, the user leverages \gls{cli}, \gls{ui}, or interactive notebooks for performing pre-training, fine-tuning, and operation of 
models.
Notably, the business models of these platforms allow users to \emph{deploy and execute even open-source models only on their own cloud} 
with different degrees of customization.
In more detail, \emph{Amazon Bedrock} is a cloud computing platform that provides access to several foundation models from various companies (\eg AI$21$ Labs, Anthropic, Cohere, Meta, Mistral AI, Stability AI, and Amazon). 
Similarly, \emph{Microsoft Azure AI} offers a wide range of services for \gls{genai}, including pre-built and customizable \glspl{api} and models. 
The users can train models on the Azure cloud with their data and access a variety of pre-trained models (from OpenAI, Hugging Face, Stability AI, Meta, etc.).
\emph{Google Cloud AI Platform (Vertex AI)} provides services for the development and fine-tuning of pre-trained \glspl{llm} from Google (\ie~\fmtTT{Gemini}, \fmtTT{Gemma}) and other open models (\eg \fmtTT{LLaMa}, \fmtTT{Claude}). Users can customize hardware resources (\eg GPUs, storage, and virtual machines) according to their needs.

Other platforms provide only \textbf{open-source} models, both \textbf{first- and third-party}. 
\emph{Hugging Face}, one of the most popular and active platforms in the \gls{ai} and particularly \gls{genai} community, belongs to this category. 
Hugging Face hosts a vast amount of open-source pre-trained models and also provides remote resources for model training via \gls{api} on a subscription basis. 
\emph{Nvidia NIM} has a more specific focus on provider hardware support and exposes an \gls{api} to access to numerous open pre-trained \glspl{llm} hosted on its infrastructure. These models are optimized for Nvidia architectures and accelerated through the Nvidia software stack.
Differently than these platforms that allow the users to \emph{download, customize, and operate the models outside of them}, \emph{IBM Watson} provides pre-trained models (\eg \fmtTT{Granite}, \fmtTT{LLaMa}, \fmtTT{Mixtral}) for \gls{genai} deployed on its cloud, which can be accessed only via \glspl{api} and notebooks.
Based on a slightly different philosophy, \emph{Cloudflare AI} is based on a network of serverless GPUs specifically designed for deploying and running \gls{ai} models from anywhere. 
Cloudflare AI offers numerous open models for text generation and text-to-image tasks (\eg \fmtTT{LLaMa}, \fmtTT{Gemma}, \fmtTT{Zephyr}, \fmtTT{Stable Diffusion}) whose interaction can be carried out via \gls{api} and \gls{cli}.

Among the frameworks categorized in Tab.~\ref{table:genai_architectures} (\cf Sec.~\ref{sec:genai_model_overview}), all the base architectures are available on (and downloadable from) Hugging Face, except for the (closed) GPT-based models, which are available on OpenAI, and \fmtTT{BERT}, which is accessible through the Google Cloud AI Platform.

\section{Wrap Up and Future Prospects}
\label{sec:future directions}
\begin{figure*}[t]
    \centering
    \subfloat{
    \includegraphics[width=0.35\textwidth]{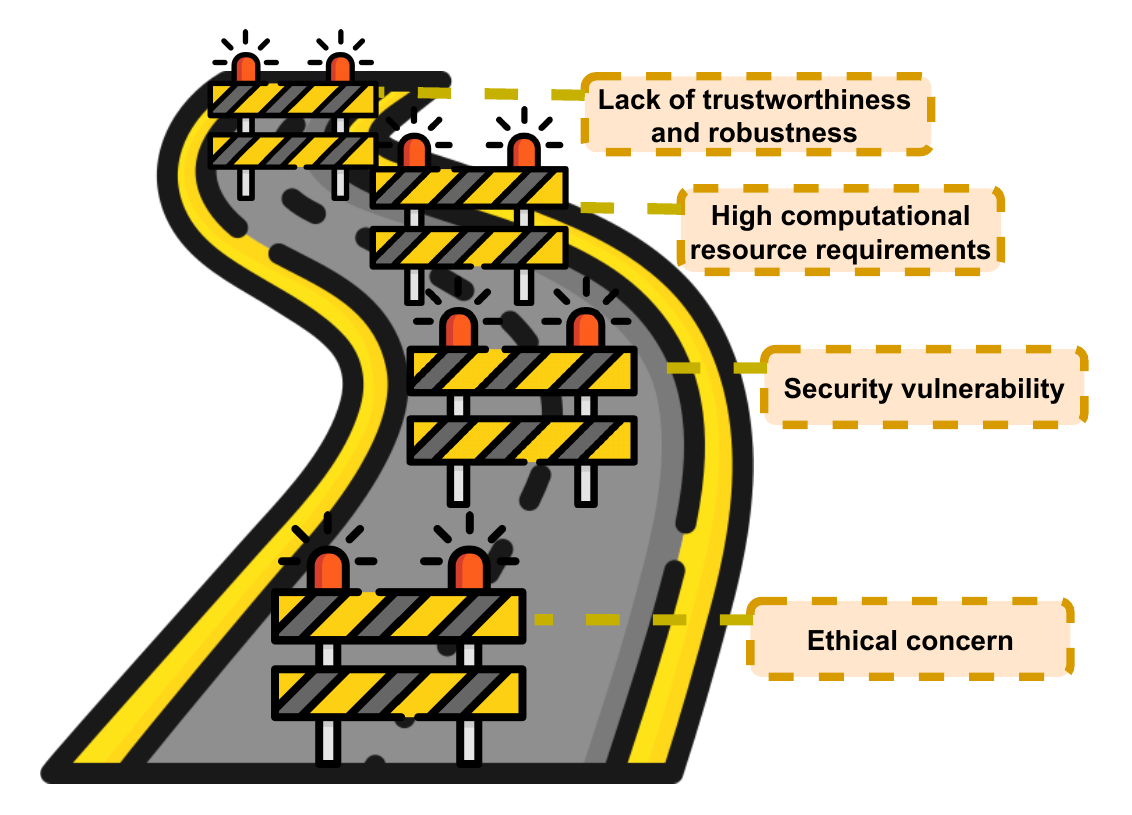}}
    \subfloat{
    \includegraphics[width=0.65\textwidth]{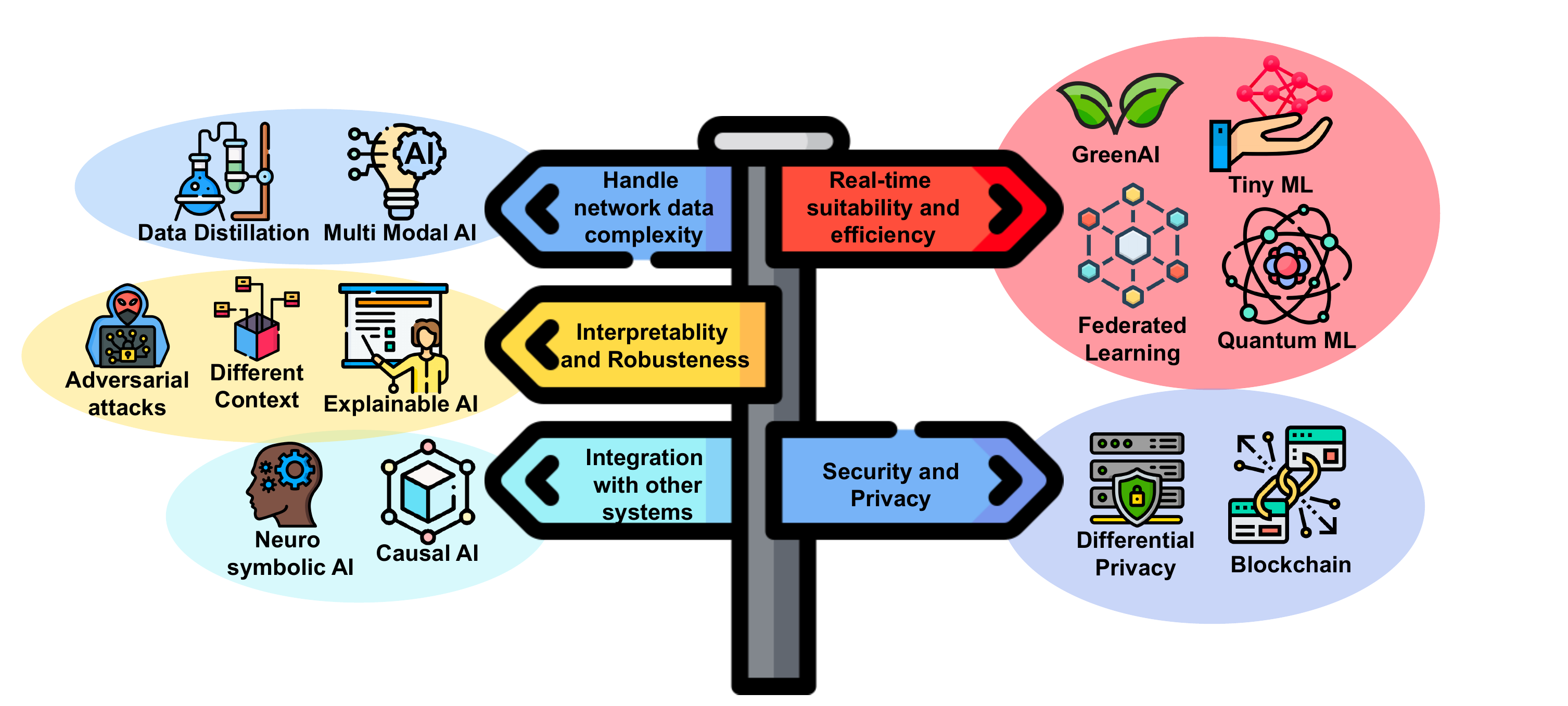}
    }
    \caption{
    Limitations (\textit{left}-side), represented by roadblocks that highlight the current challenges in the adoption and implementation of GenAI for network monitoring and management.
    Future directions (\textit{right}-side), grouped into five categories distinguished by different colors, aimed at addressing these limitations.}
    \label{fig: genai_limitations}
\end{figure*}

The advent of \gls{genai} presents a transformative potential for network monitoring and management tasks. 
By leveraging advanced generative models, 
network systems can 
address the growing complexity and dynamic nature of modern networks with a more predictive and proactive stance, moving beyond the traditional reactive measures.

Despite the positive aspects of this breakthrough approach, \gls{genai} also has several \emph{limitations} summarized in the left side of Figure~\ref{fig: genai_limitations}. 
One of the major concerns is 
(\textit{i}) \emph{the lack of trustworthiness and robustness}~\cite{jacobs2022ai}.
The issue of trustworthiness is 
due to ``closed box'' nature of the \gls{genai}-model architectures.
Hence, users often find it difficult to 
understand the decision-making processes of a \gls{genai} model, thus complicating the debugging, refinement, and efforts to improve model performance and address biases.
Furthermore,
the evaluation of \gls{genai} models focuses only on a few publicly-available datasets, thus not reflecting dynamic real-world scenarios. 
Additionally, network attacks or crafted inputs to evade malware identification can \emph{intentionally} modify the nature of network traffic.
These points raise doubts about the robustness of \gls{genai} solutions and their applicability to network monitoring and management in practice.

Furthermore, (\textit{ii}) \emph{these models require substantial computational resources for pre-training and fine-tuning}. 
Achieving high performance typically involves processing extensive datasets, which demands significant computational power and time. 
Additionally, current \gls{genai} models consist of trillions of parameters and require extensive training periods and significant resources that may not be available in smaller computational environments.
In fact, some studies~\cite{piovesan2024telecom, erak2024leveraging} suggest using more compact models (\eg \gls{slm}) for network monitoring and management tasks, which, with a significantly smaller number of parameters, can deliver performance comparable to larger models.
This high resource requirement represents a major hindrance to the widespread adoption of \gls{genai} and poses challenges for its online application in network environments. 
As a result, integrating such models into operational networks remains complex and costly.
Additionally, the complexity of these models also introduces (\textit{iii}) \emph{potential vulnerabilities to adversarial attacks}, posing security risks. 
Attackers can exploit their architectures, manipulating their behavior or compromising data integrity.
Expressly, adversarial attacks may use subtle perturbations to trick the model into making incorrect predictions, risking data security and system reliability.
Lastly, the use of \gls{genai} raises (\textit{iv}) \emph{ethical concerns} related to data privacy and misuse of generated content.
Personal data are used in training models, potentially causing
significant privacy issues.
Then, the ability to generate misleading content (such as deepfakes) can lead to misinformation and manipulation.

In the following, we identify possible \textbf{future directions} to improve and overcome the drawbacks of \gls{genai} in network monitoring and management. We graphically summarize these perspectives on the right side of Figure~\ref{fig: genai_limitations}.

\noindent
\textbf{Real-time suitability and efficiency:}
for network monitoring and management tasks, timely operation is a strict requirement.
Nevertheless, \gls{genai} models demand substantial computational power and extensive datasets for both training and fine-tuning. 
\emph{Federated Learning}~\cite{zhang2021survey, li2020review} can help overcome these challenges by enabling decentralized training across multiple devices, thereby minimizing the need for centralized data storage and processing~\cite{huang2024federated}.
Additionally, large model sizes can be a significant bottleneck, resulting in longer inference times.
A key future direction is to develop efficient \gls{genai} models that can be effectively trained on smaller datasets.
Furthermore, model compression, efficient architecture design, and hardware acceleration can improve the processing speed and reduce computational overhead.
In this direction, integrating \emph{Green AI}~\cite{schwartz2020green} principles in \gls{genai} development allows
an environmentally-sustainable progress,
reducing the energy consumption and carbon footprint associated with model training and deployment.
Specifically, 
\emph{TinyML}~\cite{ray2022review, dutta2021tinyml} can offer a powerful combination of efficiency and intelligence. 
TinyML enables real-time, low-power data processing on edge devices, while \gls{genai} provides advanced predictive modeling and simulation.
This synergy allows for proactive network management, offering localized insights and responses that enhance network efficiency, even in resource-constrained environments (\eg when running \gls{genai} models directly on handheld devices).
Additionally, 
quick model adaptation to network shifts and anomalies
is crucial for maintaining efficient and responsive network operations.
In this context, \emph{Quantum ML}~\cite{ciliberto2018quantum, dunjko2018machine} may significantly boost this capability in the long term, speeding up model training and enhancing predictive precision in intricate and evolving scenarios.

\vspace{5pt}
\noindent
\textbf{Handle network data complexity:}
integrating \emph{multimodal \gls{genai}} promises to revolutionize network monitoring and management. 
By leveraging the capability of \gls{ai} to process and analyze diverse data types---ranging from textual logs and metrics to visual network topologies---network operators can achieve unprecedented levels of insight and automation.
This holistic approach allows for more accurate anomaly detection, predictive maintenance, and dynamic resource allocation, ultimately leading to more resilient and efficient network infrastructures.
The ability of multimodal \gls{ai} to synthesize information from multiple sources enhances ``on-the-fly'' decision-making while paving the way for adaptive, self-healing networks, aligning with the vision of a fully automated, intelligent network management paradigm.
Furthermore, more advanced techniques,
such as Reinforcement Learning~\cite{du2023beyond, du2024enhancing}, enhance the adaptivity of \glspl{llm}, enabling them to evolve during their operational mode.
Additionally, \textit{data distillation} procedures can be enforced to manage the large scale and redundancy of datasets.
This approach helps in reducing the dataset size by retaining only the most essential samples and discarding unnecessary ones. 
Such a procedure can significantly decrease both the training time and the resources required.
\vspace{5pt}
\noindent
\textbf{Interpretability and robustness:}
the convergence of \emph{\gls{xai}}~\cite{xu2019explainable, dwivedi2023explainable} and \emph{\gls{genai}} heralds a new era in network monitoring and management, where transparency and innovation go hand in hand. 
\gls{xai} provides much-needed clarity in the decision-making processes of \gls{ai} systems, enabling network operators to trust and understand the actions taken by their automated tools~\cite{nascita2024survey}.
When coupled with the \gls{genai}'s ability to simulate and predict network behaviors, this synergy offers a robust framework for proactive management. For instance, network anomalies can be not only detected but also explained leveraging \gls{xai} tools and addressed with \gls{ai}-generated tailored responses. 
This fusion ensures actionable and transparent \gls{ai}-driven insights,
fostering a deeper integration of \gls{ai} in network operations. 
As a result, network management becomes more intelligent, reliable, and user-centric, 
enabling networks to be
both self-optimizing and comprehensible.
Additionally, to improve model robustness, we can leverage diverse and comprehensive datasets to improve the \textit{generalizability of GenAI models across different contexts}, optionally including multi-task learning techniques, and their \textit{resistance to adversarial attacks}, exploiting adversarial training or data augmentation strategies.

\vspace{5pt}
\noindent
\textbf{Integration with other systems:}
integrating \gls{genai} models with existing network systems enhances their utility in network monitoring and management by ensuring interoperability with the current infrastructure, developing robust \glspl{api} for seamless embedding, and enabling data fusion from multiple sources. This integration also ensures scalability and leverages automation to handle routine tasks, providing comprehensive and intelligent network management solutions.
Future directions could focus on integrating \emph{Causal \gls{ai}}~\cite{kaddour2022causal}, which emphasizes understanding cause-and-effect relationships, and \emph{Neuro-symbolic \gls{ai}}~\cite{sarker2021neuro, hitzler2022neuro}, 
which combines the learning capabilities of neural networks with the logical reasoning of symbolic \gls{ai} into \gls{genai} models. 
This integration could improve the ability of these models to handle complex long-term tasks and multi-step decision-making, empowering them to accomplish intelligent planning.

\vspace{5pt}
\noindent
\textbf{Security and privacy:}
as \gls{genai} models are increasingly used in network monitoring and management, ensuring their security and privacy is essential. 
Future directions should focus on
integrating \emph{Blockchain}~\cite{zheng2018blockchain, pilkington2016blockchain} technology can provide a decentralized and immutable framework for secure data sharing and model updates, further enhancing the overall security and transparency of \gls{genai} applications.
Moreover, implementing privacy-preserving techniques, such as \emph{differential
privacy}, can safeguard user data, and ensure secure deployment with robust access control and secure communication channels.
In addition, addressing ethical considerations by ensuring transparent data usage, unbiased model training, and accountability in \gls{ai} decisions will be crucial in building trust and reliability in \gls{genai} applications for network management.

\vspace{5pt}
In conclusion, integrating \gls{genai} into network monitoring and management holds significant promise. However, it is crucial to carefully manage expectations and avoid overly optimistic assumptions about its capabilities. Tackling the associated limitations is essential to ensure a realistic and effective implementation.
Collaboration between academia and industry is vital to ensure that the generative models developed are not only theoretically sound but also practical and scalable in real-world applications. Establishing benchmarks and standardized datasets to evaluate the performance of \gls{genai} in network monitoring and management can provide a foundation for continuous improvement and innovation.
By driving advancements in predictive analytics, anomaly detection, and automation, future research can pave the way for more intelligent, efficient, and secure network systems. As we continue to explore this intersection of \gls{genai} and networking, it is imperative to address the associated challenges and ethical considerations to fully harness the potential of \gls{genai}.

\section*{Acknowledgments}
This work is partially supported by the European Union under the Italian National Recovery and Resilience Plan (NRRP) of NextGenerationEU, partnership on ``Telecommunications of the Future'' (PE00000001 -- program ``RESTART'').
Also, this work is partially carried out within the ``xInternet'' Project supported by the MUR PRIN 2022 program (D.D.104---02/02/2022) funded by the NextGenerationEU.

\begingroup
\footnotesize
\bibliographystyle{IEEEtranN}
\bibliography{genAI_references}

\begin{thebibliography}{189}
\providecommand{\natexlab}[1]{#1}
\providecommand{\url}[1]{#1}
\csname url@samestyle\endcsname
\providecommand{\newblock}{\relax}
\providecommand{\bibinfo}[2]{#2}
\providecommand{\BIBentrySTDinterwordspacing}{\spaceskip=0pt\relax}
\providecommand{\BIBentryALTinterwordstretchfactor}{4}
\providecommand{\BIBentryALTinterwordspacing}{\spaceskip=\fontdimen2\font plus
\BIBentryALTinterwordstretchfactor\fontdimen3\font minus \fontdimen4\font\relax}
\providecommand{\BIBforeignlanguage}[2]{{%
\expandafter\ifx\csname l@#1\endcsname\relax
\typeout{** WARNING: IEEEtranN.bst: No hyphenation pattern has been}%
\typeout{** loaded for the language `#1'. Using the pattern for}%
\typeout{** the default language instead.}%
\else
\language=\csname l@#1\endcsname
\fi
#2}}
\providecommand{\BIBdecl}{\relax}
\BIBdecl

\bibitem[Yang et~al.(2024)Yang, Jin, Tang, Han, Feng, Jiang, Zhong, Yin, and Hu]{yang2024harnessing}
J.~Yang, H.~Jin, R.~Tang, X.~Han, Q.~Feng, H.~Jiang, S.~Zhong, B.~Yin, and X.~Hu, ``Harnessing the power of llms in practice: A survey on chatgpt and beyond,'' \emph{ACM Transactions on Knowledge Discovery from Data}, vol.~18, no.~6, pp. 1--32, 2024.

\bibitem[sta()]{statista1}
``{Generative Artificial Intelligence (AI) Market Size Worldwide from 2020 to 2030},'' \url{https://www.statista.com/forecasts/1449838/generative-ai-market-size-worldwide}.

\bibitem[Huang et~al.(2024{\natexlab{a}})Huang, Yang, Zhou, Shen, and Zhuang]{huang2024digital}
X.~Huang, H.~Yang, C.~Zhou, X.~Shen, and W.~Zhuang, ``{When Digital Twin Meets Generative AI: Intelligent Closed-Loop Network Management},'' \emph{arXiv preprint arXiv:2404.03025}, 2024.

\bibitem[Du et~al.(2024{\natexlab{a}})Du, Niyato, Kang, Xiong, Zhang, Cui, Shen, Mao, Han, Jamalipour, et~al.]{du2024age}
H.~Du, D.~Niyato, J.~Kang, Z.~Xiong, P.~Zhang, S.~Cui, X.~Shen, S.~Mao, Z.~Han, A.~Jamalipour \emph{et~al.}, ``{The Age of Generative AI and AI-generated Everything},'' \emph{IEEE Network}, 2024.

\bibitem[Lavin et~al.(2022)Lavin, Gilligan-Lee, Visnjic, Ganju, Newman, Ganguly, Lange, Baydin, Sharma, Gibson, et~al.]{lavin2022technology}
A.~Lavin, C.~M. Gilligan-Lee, A.~Visnjic, S.~Ganju, D.~Newman, S.~Ganguly, D.~Lange, A.~G. Baydin, A.~Sharma, A.~Gibson \emph{et~al.}, ``Technology readiness levels for machine learning systems,'' \emph{Nature Communications}, vol.~13, no.~1, p. 6039, 2022.

\bibitem[Rossi and Zhang(2022)]{rossi2022landing}
D.~Rossi and L.~Zhang, ``{Landing AI on Networks: An Equipment Vendor Viewpoint on Autonomous Driving Networks},'' \emph{IEEE Transactions on Network and Service Management}, vol.~19, no.~3, pp. 3670--3684, 2022.

\bibitem[net()]{networkingchannel}
``{The Networking Channel},'' \url{https://networkingchannel.eu/library/}.

\bibitem[att()]{att}
``{AT\&T’s new Generative AI Tool Will Help Employees Be More Effective, Creative, and Innovative},'' \url{https://about.att.com/blogs/2023/generative-ai.html}.

\bibitem[cis()]{cisco}
``{Cisco Artificial Intelligence},'' \url{https://www.cisco.com/site/us/en/solutions/artificial-intelligence/ai-assistant/index.html}.

\bibitem[eri()]{ericsson}
``{How to Make Better Use of Network Insights with Generative AI},'' \url{https://www.ericsson.com/en/blog/2024/2/how-to-make-better-use-of-network-insights-with-generative-ai}.

\bibitem[eic()]{eic}
``{European Innovation Council},'' \url{https://eic.ec.europa.eu/index_en}.

\bibitem[hua()]{huawei}
``{Huawei Introduces AI Technologies to Accelerate Network Transformation Towards All Intelligence in the Net5.5G Era},'' \url{https://www.huawei.com/en/news/2024/4/has-net-5-point-5g-ai}.

\bibitem[gen()]{genainet}
``{Welcome to the Large Generative Al Models inTelecom (GenAlNet) Emerging TechnologyInitiative website},'' \url{https://genainet.committees.comsoc.org/}.

\bibitem[iet()]{ietf}
``{IETF Side Meetings},'' \url{https://wiki.ietf.org/en/meeting/119/sidemeetings}.

\bibitem[aif()]{aiforgood}
``{Specializing Large Language Models for Telecom Networks},'' \url{https://aiforgood.itu.int/event/specializing-large-language-models-for-telecom-networks/}.

\bibitem[nok()]{nokia}
``{Generative AI implications for Telco Operations},'' \url{https://www.bell-labs.com/institute/white-papers/generative-ai-implications-for-telco-operations/}.

\bibitem[tel()]{telefonica}
``{Telefónica Partners with Microsoft to Incorporate Generative AI into Kernel},'' \url{https://www.telefonica.com/en/communication-room/press-room/telefonica-partners-with-microsoft-to-incorporate-generative-ai-into-kernel/}.

\bibitem[tim()]{tim}
``{Generative AI: the challenge of TIM for the future of IT},'' \url{https://www.gruppotim.it/it/newsroom/notiziario-tecnico-tim/Anno-2023/n3-2023/Generative_AI_la_sfida_TIM_per_il_futuro_dell_IT.html}.

\bibitem[Sai et~al.(2024)Sai, Kanadia, and Chamola]{sai2024empowering}
S.~Sai, M.~Kanadia, and V.~Chamola, ``{Empowering IoT with Generative AI: Applications, Case Studies, and Limitations},'' \emph{IEEE Internet of Things Magazine}, vol.~7, no.~3, pp. 38--43, 2024.

\bibitem[Hassanin and Moustafa(2024)]{hassanin2024comprehensive}
M.~Hassanin and N.~Moustafa, ``{A Comprehensive Overview of Large Language Models (LLMs) for Cyber Defences: Opportunities and Directions},'' \emph{arXiv preprint arXiv:2405.14487}, 2024.

\bibitem[Alwahedi et~al.(2024)Alwahedi, Aldhaheri, Ferrag, Battah, and Tihanyi]{alwahedi2024ml}
F.~Alwahedi, A.~Aldhaheri, M.~A. Ferrag, A.~Battah, and N.~Tihanyi, ``{Machine Learning Techniques for IoT Security: Current Research and Future Vision with Generative AI and Large Language Models},'' \emph{Internet of Things and Cyber-Physical Systems}, vol.~4, pp. 167--185, 2024.

\bibitem[Halvorsen et~al.(2024)Halvorsen, Izurieta, Cai, and Gebremedhin]{halvorsen2024applying}
J.~Halvorsen, C.~Izurieta, H.~Cai, and A.~Gebremedhin, ``{Applying Generative Machine Learning to Intrusion Detection: A Systematic Mapping Study and Review},'' \emph{ACM Computing Surveys}, vol.~56, no.~10, 2024.

\bibitem[Zhou et~al.(2024)Zhou, Hu, Yuan, Cui, Jin, Chen, Wu, Yuan, Jiang, Wu, et~al.]{zhou2024large}
H.~Zhou, C.~Hu, Y.~Yuan, Y.~Cui, Y.~Jin, C.~Chen, H.~Wu, D.~Yuan, L.~Jiang, D.~Wu \emph{et~al.}, ``{Large Language Model (LLM) for Telecommunications: A Comprehensive Survey on Principles, Key Techniques, and Opportunities},'' \emph{arXiv preprint arXiv:2405.10825}, 2024.

\bibitem[Celik and Eltawil(2024)]{celik2024dawn}
A.~Celik and A.~M. Eltawil, ``At the dawn of generative {AI} era: A tutorial-cum-survey on new frontiers in {6G} wireless intelligence,'' \emph{IEEE Open Journal of the Communications Society}, 2024.

\bibitem[Karapantelakis et~al.(2024{\natexlab{a}})Karapantelakis, Alizadeh, Alabassi, Dey, and Nikou]{karapantelakis2024generative}
A.~Karapantelakis, P.~Alizadeh, A.~Alabassi, K.~Dey, and A.~Nikou, ``{Generative AI in Mobile Networks: a Survey},'' \emph{Annals of Telecommunications}, vol.~79, no.~1, pp. 15--33, 2024.

\bibitem[Liu et~al.(2024{\natexlab{a}})Liu, Xie, Zhang, and Cui]{liu2024large}
C.~Liu, X.~Xie, X.~Zhang, and Y.~Cui, ``{Large Language Models for Networking: Workflow, Advances and Challenges},'' \emph{arXiv preprint arXiv:2404.12901}, 2024.

\bibitem[Huang et~al.(2023)Huang, Du, Zhang, Niyato, Kang, Xiong, Wang, and Huang]{huang2023large}
Y.~Huang, H.~Du, X.~Zhang, D.~Niyato, J.~Kang, Z.~Xiong, S.~Wang, and T.~Huang, ``{Large Language Models for Networking: Applications, Enabling Techniques, and Challenges},'' \emph{arXiv preprint arXiv:2311.17474}, 2023.

\bibitem[Chaccour et~al.(2024)Chaccour, Karapantelakis, Murphy, and Dohler]{chaccour2024telecom}
C.~Chaccour, A.~Karapantelakis, T.~Murphy, and M.~Dohler, ``{Telecom’s Artificial General Intelligence (AGI) Vision: Beyond the GenAI Frontier},'' \emph{IEEE Network}, 2024.

\bibitem[Ferrag et~al.(2024{\natexlab{a}})Ferrag, Alwahedi, Battah, Cherif, Mechri, and Tihanyi]{ferrag2024generative}
M.~A. Ferrag, F.~Alwahedi, A.~Battah, B.~Cherif, A.~Mechri, and N.~Tihanyi, ``{Generative AI and Large Language Models for Cyber Security: All Insights You Need},'' \emph{arXiv preprint arXiv:2405.12750}, 2024.

\bibitem[Xu et~al.(2024)Xu, Du, Niyato, Kang, Xiong, Mao, Han, Jamalipour, Kim, Shen, et~al.]{xu2024unleashing}
M.~Xu, H.~Du, D.~Niyato, J.~Kang, Z.~Xiong, S.~Mao, Z.~Han, A.~Jamalipour, D.~I. Kim, X.~Shen \emph{et~al.}, ``{Unleashing the Power of Edge-Cloud Generative AI in Mobile Networks: A Survey of AIGC Services},'' \emph{IEEE Communications Surveys \& Tutorials}, 2024.

\bibitem[Wang et~al.(2024{\natexlab{a}})Wang, Du, Niyato, Kang, Xiong, Kim, and Letaief]{wang2024toward}
J.~Wang, H.~Du, D.~Niyato, J.~Kang, Z.~Xiong, D.~I. Kim, and K.~B. Letaief, ``{Toward Scalable Generative AI via Mixture of Experts in Mobile Edge Networks},'' \emph{arXiv preprint arXiv:2402.06942}, 2024.

\bibitem[Aceto et~al.(2021)Aceto, Bovenzi, Ciuonzo, Montieri, Persico, and Pescap{\'e}]{aceto2021characterization}
G.~Aceto, G.~Bovenzi, D.~Ciuonzo, A.~Montieri, V.~Persico, and A.~Pescap{\'e}, ``Characterization and prediction of mobile-app traffic using {Markov} modeling,'' \emph{IEEE Transactions on Network and Service Management}, vol.~18, no.~1, pp. 907--925, 2021.

\bibitem[Aceto et~al.(2024)Aceto, Giampaolo, Guida, Izzo, Pescap{\`e}, Piccialli, and Prezioso]{aceto2024synthetic}
G.~Aceto, F.~Giampaolo, C.~Guida, S.~Izzo, A.~Pescap{\`e}, F.~Piccialli, and E.~Prezioso, ``{Synthetic and Privacy-Preserving Traffic Trace Generation using Generative AI Models for Training Network Intrusion Detection Systems},'' \emph{Journal of Network and Computer Applications}, p. 103926, 2024.

\bibitem[Hui et~al.(2022)Hui, Wang, Wang, Yang, Liu, Jin, and Li]{hui2022knowledge}
S.~Hui, H.~Wang, Z.~Wang, X.~Yang, Z.~Liu, D.~Jin, and Y.~Li, ``Knowledge enhanced {GAN} for {IoT} traffic generation,'' in \emph{ACM Web Conference (WWW)}, 2022, pp. 3336--3346.

\bibitem[Gudovskiy et~al.(2022)Gudovskiy, Ishizaka, and Kozuka]{gudovskiy2022cflow}
D.~Gudovskiy, S.~Ishizaka, and K.~Kozuka, ``{CFLOW-AD:} real-time unsupervised anomaly detection with localization via conditional normalizing flows,'' in \emph{IEEE/CVF Winter Conference on Applications of Computer Vision (WACV)}, 2022, pp. 98--107.

\bibitem[Kingma and Welling(2014)]{kingma2013auto}
D.~P. Kingma and M.~Welling, ``Auto-encoding variational {Bayes},'' in \emph{International Conference on Learning Representations (ICLR)}, 2014.

\bibitem[Vaswani et~al.(2017)Vaswani, Shazeer, Parmar, Uszkoreit, Jones, Gomez, Kaiser, and Polosukhin]{vaswani2017attention}
A.~Vaswani, N.~Shazeer, N.~Parmar, J.~Uszkoreit, L.~Jones, A.~N. Gomez, {\L}.~Kaiser, and I.~Polosukhin, ``{Attention is All You Need},'' \emph{Advances in Neural Information Processing Systems (NeurIPS)}, vol.~30, 2017.

\bibitem[Dinh et~al.(2015)Dinh, Krueger, and Bengio]{dinh2014nice}
L.~Dinh, D.~Krueger, and Y.~Bengio, ``{NICE}: Non-linear independent components estimation,'' in \emph{International Conference on Learning Representations (ICLR), Workshop Track}, 2015.

\bibitem[Adeleke et~al.(2022)Adeleke, Bastin, and Gurkan]{adeleke2022network}
O.~A. Adeleke, N.~Bastin, and D.~Gurkan, ``{Network Traffic Generation: A Survey and Methodology},'' \emph{ACM Computing Surveys}, vol.~55, no.~2, pp. 1--23, 2022.

\bibitem[Goodfellow et~al.(2014)Goodfellow, Pouget-Abadie, Mirza, Xu, Warde-Farley, Ozair, Courville, and Bengio]{goodfellow2014generative}
I.~Goodfellow, J.~Pouget-Abadie, M.~Mirza, B.~Xu, D.~Warde-Farley, S.~Ozair, A.~Courville, and Y.~Bengio, ``{Generative Adversarial Nets},'' \emph{Advances in Neural Information Processing Systems (NeurIPS)}, vol.~27, 2014.

\bibitem[Mirza and Osindero(2014)]{mirza2014conditional}
M.~Mirza and S.~Osindero, ``Conditional generative adversarial nets,'' \emph{arXiv preprint arXiv:1411.1784}, 2014.

\bibitem[Radford et~al.(2016)Radford, Metz, and Chintala]{radford2015unsupervised}
A.~Radford, L.~Metz, and S.~Chintala, ``Unsupervised representation learning with deep convolutional generative adversarial networks,'' in \emph{International Conference on Learning Representations (ICLR)}, 2016.

\bibitem[Larsen et~al.(2016)Larsen, S{\o}nderby, Larochelle, and Winther]{larsen2016autoencoding}
A.~B.~L. Larsen, S.~K. S{\o}nderby, H.~Larochelle, and O.~Winther, ``Autoencoding beyond pixels using a learned similarity metric,'' in \emph{International Conference on Machine Learning (ICML)}, 2016, pp. 1558--1566.

\bibitem[Dinh et~al.(2017)Dinh, Sohl-Dickstein, and Bengio]{dinh2016density}
L.~Dinh, J.~Sohl-Dickstein, and S.~Bengio, ``Density estimation using {Real NVP},'' in \emph{International Conference on Learning Representations (ICLR)}, 2017.

\bibitem[Bommasani et~al.(2021)Bommasani, Hudson, Adeli, Altman, Arora, von Arx, Bernstein, Bohg, Bosselut, Brunskill, et~al.]{bommasani2021opportunities}
R.~Bommasani, D.~A. Hudson, E.~Adeli, R.~Altman, S.~Arora, S.~von Arx, M.~S. Bernstein, J.~Bohg, A.~Bosselut, E.~Brunskill \emph{et~al.}, ``On the opportunities and risks of foundation models,'' \emph{arXiv preprint arXiv:2108.07258}, 2021.

\bibitem[Devlin et~al.(2018)Devlin, Chang, Lee, and Toutanova]{devlin2018bert}
J.~Devlin, M.-W. Chang, K.~Lee, and K.~Toutanova, ``{{BERT}: Pre-training of Deep Bidirectional Transformers for Language Understanding},'' \emph{arXiv preprint arXiv:1810.04805}, 2018.

\bibitem[Radford et~al.(2018)Radford, Narasimhan, Salimans, Sutskever, et~al.]{radford2018improving}
A.~Radford, K.~Narasimhan, T.~Salimans, I.~Sutskever \emph{et~al.}, ``{Improving Language Understanding by Generative Pre-Training},'' \emph{OpenAI Tech Report}, 2018.

\bibitem[Ho et~al.(2020)Ho, Jain, and Abbeel]{ho2020denoising}
J.~Ho, A.~Jain, and P.~Abbeel, ``{Denoising Diffusion Probabilistic Models},'' \emph{Advances in Neural Information Processing Systems (NeurIPS)}, vol.~33, pp. 6840--6851, 2020.

\bibitem[Kingma and Gao(2024)]{kingma2024understanding}
D.~Kingma and R.~Gao, ``Understanding diffusion objectives as the elbo with simple data augmentation,'' \emph{Advances in Neural Information Processing Systems (NeurIPS)}, vol.~36, 2024.

\bibitem[Gu and Dao(2023)]{gu2023mamba}
A.~Gu and T.~Dao, ``{Mamba: Linear-Time Sequence Modeling with Selective State Spaces},'' \emph{arXiv preprint arXiv:2312.00752}, 2023.

\bibitem[Kim et~al.(2024)Kim, Lee, Kim, Park, Yoo, Kwon, and Lee]{kim2024memory}
J.~Kim, J.~H. Lee, S.~Kim, J.~Park, K.~M. Yoo, S.~J. Kwon, and D.~Lee, ``Memory-efficient fine-tuning of compressed large language models via sub-4-bit integer quantization,'' \emph{Advances in Neural Information Processing Systems (NeurIPS)}, vol.~36, 2024.

\bibitem[Hu et~al.(2022)Hu, Shen, Wallis, Allen-Zhu, Li, Wang, Wang, and Chen]{hu2021lora}
E.~J. Hu, Y.~Shen, P.~Wallis, Z.~Allen-Zhu, Y.~Li, S.~Wang, L.~Wang, and W.~Chen, ``{LoRa}: Low-rank adaptation of large language models,'' in \emph{International Conference on Learning Representations (ICLR)}, 2022.

\bibitem[Jiang et~al.(2024{\natexlab{a}})Jiang, Liu, Gember-Jacobson, Bhagoji, Schmitt, Bronzino, and Feamster]{jiang2024netdiffusion}
X.~Jiang, S.~Liu, A.~Gember-Jacobson, A.~N. Bhagoji, P.~Schmitt, F.~Bronzino, and N.~Feamster, ``{NetDiffusion: Network Data Augmentation Through Protocol-Constrained Traffic Generation},'' \emph{Proceedings of the ACM on Measurement and Analysis of Computing Systems}, vol.~8, no.~1, pp. 1--32, 2024.

\bibitem[Ding et~al.(2023)Ding, Qin, Yang, Wei, Yang, Su, Hu, Chen, Chan, Chen, et~al.]{ding2023parameter}
N.~Ding, Y.~Qin, G.~Yang, F.~Wei, Z.~Yang, Y.~Su, S.~Hu, Y.~Chen, C.-M. Chan, W.~Chen \emph{et~al.}, ``Parameter-efficient fine-tuning of large-scale pre-trained language models,'' \emph{Nature Machine Intelligence}, vol.~5, no.~3, pp. 220--235, 2023.

\bibitem[ggu()]{gguf-website}
\BIBentryALTinterwordspacing
``{GGUF}.'' [Online]. Available: \url{https: //github.com/ggerganov/ggml/blob/master/docs/gguf.md}
\BIBentrySTDinterwordspacing

\bibitem[Karapantelakis et~al.(2024{\natexlab{b}})Karapantelakis, Shakur, Nikou, Moradi, Orlog, Gaim, Holm, Nimara, and Huang]{karapantelakis2024using}
A.~Karapantelakis, M.~Shakur, A.~Nikou, F.~Moradi, C.~Orlog, F.~Gaim, H.~Holm, D.~D. Nimara, and V.~Huang, ``{Using Large Language Models to Understand Telecom Standards},'' in \emph{IEEE International Conference on Machine Learning for Communication and Networking (ICMLCN)}, 2024.

\bibitem[Meng et~al.(2023)Meng, Lin, Wang, and Zhang]{meng2023netgpt}
X.~Meng, C.~Lin, Y.~Wang, and Y.~Zhang, ``{NetGPT: Generative Pretrained Transformer for Network Traffic},'' \emph{arXiv preprint arXiv:2304.09513}, 2023.

\bibitem[Pacheco et~al.(2018)Pacheco, Exposito, Gineste, Baudoin, and Aguilar]{pacheco2018towards}
F.~Pacheco, E.~Exposito, M.~Gineste, C.~Baudoin, and J.~Aguilar, ``{Towards the Deployment of Machine Learning Solutions in Network Traffic Classification: A Systematic Survey},'' \emph{IEEE Communications Surveys \& Tutorials}, vol.~21, no.~2, pp. 1988--2014, 2018.

\bibitem[Azab et~al.(2022)Azab, Khasawneh, Alrabaee, Choo, and Sarsour]{azab2022network}
A.~Azab, M.~Khasawneh, S.~Alrabaee, K.-K.~R. Choo, and M.~Sarsour, ``{Network Traffic Classification: Techniques, Datasets, and Challenges},'' \emph{Digital Communications and Networks}, 2022.

\bibitem[Aceto et~al.(2023)Aceto, Ciuonzo, Montieri, Persico, and Pescape]{aceto2023aipowered}
G.~Aceto, D.~Ciuonzo, A.~Montieri, V.~Persico, and A.~Pescape, ``{AI-powered Internet Traffic Classification: Past, Present, and Future},'' \emph{IEEE Communications Magazine}, pp. 1--7, 2023.

\bibitem[Lin et~al.(2022)Lin, Xiong, Gou, Li, Shi, and Yu]{lin2022}
X.~Lin, G.~Xiong, G.~Gou, Z.~Li, J.~Shi, and J.~Yu, ``{ET-BERT: A Contextualized Datagram Representation with Pre-training Transformers for Encrypted Traffic Classification},'' in \emph{ACM Web Conference (WWW)}, 2022, p. 633–642.

\bibitem[Wang et~al.(2024{\natexlab{b}})Wang, Xie, Wang, Wang, Zhao, and Cui]{wang2024netmamba}
T.~Wang, X.~Xie, W.~Wang, C.~Wang, Y.~Zhao, and Y.~Cui, ``{NetMamba: Efficient Network Traffic Classification via Pre-training Unidirectional Mamba},'' \emph{arXiv preprint arXiv:2405.11449}, 2024.

\bibitem[Chou and Jiang(2021)]{chou2021survey}
D.~Chou and M.~Jiang, ``{A Survey on Data-driven Network Intrusion Detection},'' \emph{ACM Computing Surveys}, vol.~54, no.~9, pp. 1--36, 2021.

\bibitem[Manocchio et~al.(2024)Manocchio, Layeghy, Lo, Kulatilleke, Sarhan, and Portmann]{manocchio2024flowtransformer}
L.~D. Manocchio, S.~Layeghy, W.~W. Lo, G.~K. Kulatilleke, M.~Sarhan, and M.~Portmann, ``{FlowTransformer: A Transformer Framework for Flow-based Network Intrusion Detection Systems},'' \emph{Expert Systems with Applications}, vol. 241, p. 122564, 2024.

\bibitem[Ferrag et~al.(2024{\natexlab{b}})Ferrag, Ndhlovu, Tihanyi, Cordeiro, Debbah, Lestable, and Thandi]{ferrag2024}
M.~A. Ferrag, M.~Ndhlovu, N.~Tihanyi, L.~C. Cordeiro, M.~Debbah, T.~Lestable, and N.~S. Thandi, ``{Revolutionizing Cyber Threat Detection With Large Language Models: A Privacy-Preserving BERT-based Lightweight Model for IoT/IIoT Devices},'' \emph{IEEE Access}, vol.~12, pp. 23\,733--23\,750, 2024.

\bibitem[He et~al.(2021)He, He, Chen, Yang, Su, and Lyu]{he2021survey}
S.~He, P.~He, Z.~Chen, T.~Yang, Y.~Su, and M.~R. Lyu, ``{A Survey on Automated Log Analysis for Reliability Engineering},'' \emph{ACM computing surveys}, vol.~54, no.~6, pp. 1--37, 2021.

\bibitem[Han et~al.(2023)Han, Yuan, and Trabelsi]{han2023loggpt}
X.~Han, S.~Yuan, and M.~Trabelsi, ``{LogGPT: Log Anomaly Detection via GPT},'' in \emph{IEEE International Conference on Big Data (BigData)}, 2023, pp. 1117--1122.

\bibitem[Boffa et~al.(2024)Boffa, Drago, Mellia, Vassio, Giordano, Valentim, and Houidi]{boffa2024logprecis}
M.~Boffa, I.~Drago, M.~Mellia, L.~Vassio, D.~Giordano, R.~Valentim, and Z.~B. Houidi, ``{LogPr\'ecis: Unleashing Language Models for Automated Malicious Log Analysis: Pr\'ecis: A Concise Summary of Essential Points, Statements, or Facts},'' \emph{Computers \& Security}, vol. 141, p. 103805, 2024.

\bibitem[Martinez et~al.(2014)Martinez, Yannuzzi, López, López, Ramírez, Serral-Gracià, Masip-Bruin, Maciejewski, and Altmann]{martinez2014network}
A.~Martinez, M.~Yannuzzi, V.~López, D.~López, W.~Ramírez, R.~Serral-Gracià, X.~Masip-Bruin, M.~Maciejewski, and J.~Altmann, ``{Network Management Challenges and Trends in Multi-Layer and Multi-Vendor Settings for Carrier-Grade Networks},'' \emph{IEEE Communications Surveys \& Tutorials}, vol.~16, no.~4, pp. 2207--2230, 2014.

\bibitem[Aarikka-Stenroos and Ritala(2017)]{aarikka2017network}
L.~Aarikka-Stenroos and P.~Ritala, ``{Network Management in the Era of Ecosystems: Systematic Review and Management Framework},'' \emph{Industrial Marketing Management}, vol.~67, pp. 23--36, 2017.

\bibitem[Aboubakar et~al.(2022)Aboubakar, Kellil, and Roux]{aboubakar2022review}
M.~Aboubakar, M.~Kellil, and P.~Roux, ``{A Review of IoT Network Management: Current Status and Perspectives},'' \emph{Journal of King Saud University-Computer and Information Sciences}, vol.~34, no.~7, pp. 4163--4176, 2022.

\bibitem[Wang et~al.(2023{\natexlab{a}})Wang, Zhang, Yang, Zhuang, Qi, Sun, Lu, Feng, and Liao]{wang2023network}
J.~Wang, L.~Zhang, Y.~Yang, Z.~Zhuang, Q.~Qi, H.~Sun, L.~Lu, J.~Feng, and J.~Liao, ``{Network Meets ChatGPT: Intent Autonomous Management, Control and Operation},'' \emph{Journal of Communications and Information Networks}, vol.~8, no.~3, pp. 239--255, 2023.

\bibitem[Holland et~al.(2021)Holland, Schmitt, Feamster, and Mittal]{holland2021new}
J.~Holland, P.~Schmitt, N.~Feamster, and P.~Mittal, ``{New Directions in Automated Traffic Analysis},'' in \emph{ACM SIGSAC Conference on Computer and Communications Security (CCS)}, 2021, pp. 3366--3383.

\bibitem[Shapira and Shavitt(2021)]{shapira2021flowpic}
T.~Shapira and Y.~Shavitt, ``Flowpic: A generic representation for encrypted traffic classification and applications identification,'' \emph{IEEE Transactions on Network and Service Management}, vol.~18, no.~2, pp. 1218--1232, 2021.

\bibitem[Sivaroopan et~al.(2024)Sivaroopan, Bandara, Madarasingha, Jourjon, Jayasumana, and Thilakarathna]{sivaroopan2023netdiffus}
N.~Sivaroopan, D.~Bandara, C.~Madarasingha, G.~Jourjon, A.~P. Jayasumana, and K.~Thilakarathna, ``{NetDiffus: Network Traffic Generation by Diffusion Models through Time-Series Imaging},'' \emph{Computer Networks}, vol. 251, p. 110616, 2024.

\bibitem[Bikmukhamedov and Nadeev(2021)]{bikmukhamedov2021}
R.~F. Bikmukhamedov and A.~F. Nadeev, ``{Multi-Class Network Traffic Generators and Classifiers Based on Neural Networks},'' in \emph{Systems of Signals Generating and Processing in the Field of on Board Communications}, 2021, pp. 1--7.

\bibitem[Kholgh and Kostakos(2023)]{Kholgh2023PACGPT}
D.~K. Kholgh and P.~Kostakos, ``{PAC-GPT: A Novel Approach to Generating Synthetic Network Traffic With GPT-3},'' \emph{IEEE Access}, vol.~11, pp. 114\,936--114\,951, 2023.

\bibitem[Wang et~al.(2024{\natexlab{c}})Wang, Qian, Li, Yao, and Shao]{wang2024lens}
Q.~Wang, C.~Qian, X.~Li, Z.~Yao, and H.~Shao, ``{LENS: A Foundation Model for Network Traffic},'' \emph{arXiv preprint arXiv:2402.03646}, 2024.

\bibitem[Qu et~al.(2024)Qu, Ma, and Li]{qu2024trafficgpt}
J.~Qu, X.~Ma, and J.~Li, ``{TrafficGPT: Breaking the Token Barrier for Efficient Long Traffic Analysis and Generation},'' \emph{arXiv preprint arXiv:2403.05822}, 2024.

\bibitem[Chu et~al.(2024)Chu, Jiang, Liu, Bhagoji, Bronzino, Schmitt, and Feamster]{chu2024mamba}
A.~Chu, X.~Jiang, S.~Liu, A.~Bhagoji, F.~Bronzino, P.~Schmitt, and N.~Feamster, ``Feasibility of state space models for network traffic generation,'' \emph{arXiv preprint arXiv:2406.02784}, 2024.

\bibitem[Zhang et~al.(2024)Zhang, Li, Jin, and Li]{zhang2024netdiff}
S.~Zhang, T.~Li, D.~Jin, and Y.~Li, ``{NetDiff: A Service-Guided Hierarchical Diffusion Model for Network Flow Trace Generation},'' \emph{Proc. ACM Netw.}, vol.~2, no. CoNEXT3, Aug. 2024.

\bibitem[Li et~al.(2024{\natexlab{a}})Li, Wu, and Zhang]{li2024lightweight}
F.~Li, H.~Wu, and J.~Zhang, ``Lightweight diffusion model for synthesizing malicious network traffic,'' in \emph{NAECON 2024 - IEEE National Aerospace and Electronics Conference}, 2024, pp. 409--413.

\bibitem[Wolf et~al.(2024)Wolf, Tritscher, Landes, Hotho, and Schlör]{wolf2024}
M.~Wolf, J.~Tritscher, D.~Landes, A.~Hotho, and D.~Schlör, ``{Benchmarking of Synthetic Network Data: Reviewing Challenges and Approaches},'' \emph{Computers \& Security}, vol. 145, p. 103993, 2024.

\bibitem[Wang and Oates(2015)]{wang2015imaging}
Z.~Wang and T.~Oates, ``{Imaging Time-series to Improve Classification and Imputation},'' in \emph{24th International Conference on Artificial Intelligence (IJCAI), ML Track}, 2015, pp. 3939--3945.

\bibitem[Guthula et~al.(2023)Guthula, Battula, Beltiukov, Guo, and Gupta]{guthula2023netfound}
S.~Guthula, N.~Battula, R.~Beltiukov, W.~Guo, and A.~Gupta, ``{netFound: Foundation Model for Network Security},'' \emph{arXiv preprint arXiv:2310.17025}, 2023.

\bibitem[Sarabi et~al.(2023)Sarabi, Yin, and Liu]{sarabi2023}
A.~Sarabi, T.~Yin, and M.~Liu, ``{An LLM-based Framework for Fingerprinting Internet-connected Devices},'' in \emph{ACM on Internet Measurement Conference (IMC)}, 2023, p. 478–484.

\bibitem[Liu et~al.(2024{\natexlab{b}})Liu, Ma, Liu, Liu, Zhao, Hu, and Ghafoor]{tao2024lambert}
T.~Liu, X.~Ma, L.~Liu, X.~Liu, Y.~Zhao, N.~Hu, and K.~Z. Ghafoor, ``{LAMBERT: Leveraging Attention Mechanisms to Improve the BERT Fine-Tuning Model for Encrypted Traffic Classification},'' \emph{Mathematics}, vol.~12, no.~11, 2024.

\bibitem[Li et~al.(2024{\natexlab{b}})Li, Zhang, Gu, Bai, Xiao, and Wang]{li2024albert}
H.~Li, Y.~Zhang, M.~Gu, J.~Bai, Z.~Xiao, and Y.~Wang, ``{Encrypted Traffic Classification Framework Based on Albert},'' in \emph{IGARSS 2024 - 2024 IEEE International Geoscience and Remote Sensing Symposium}, 2024, pp. 10\,019--10\,024.

\bibitem[Nam et~al.(2021)Nam, Park, and Kim]{nam2021intrusion}
M.~Nam, S.~Park, and D.~S. Kim, ``{Intrusion Detection Method using Bi-directional GPT for In-vehicle Controller Area Networks},'' \emph{IEEE Access}, vol.~9, pp. 124\,931--124\,944, 2021.

\bibitem[Yu et~al.(2021)Yu, Tan, Mumtaz, Al-Rubaye, Al-Dulaimi, Bashir, and Khan]{yu2021securing}
K.~Yu, L.~Tan, S.~Mumtaz, S.~Al-Rubaye, A.~Al-Dulaimi, A.~K. Bashir, and F.~A. Khan, ``{Securing Critical Infrastructures: Deep-Learning-Based Threat Detection in IIoT},'' \emph{IEEE Communications Magazine}, vol.~59, no.~10, pp. 76--82, 2021.

\bibitem[Li et~al.(2022)Li, Yuan, and Li]{li2022extreme}
Y.~Li, X.~Yuan, and W.~Li, ``{An Extreme Semi-supervised Framework Based on Transformer for Network Intrusion Detection},'' in \emph{31st ACM International Conference on Information \& Knowledge Management (CIKM)}, 2022, pp. 4204--4208.

\bibitem[Seyyar et~al.(2022)Seyyar, Yavuz, and {\"U}nver]{seyyar2022attack}
Y.~E. Seyyar, A.~G. Yavuz, and H.~M. {\"U}nver, ``{An Attack Detection Framework Based on BERT and Deep Learning},'' \emph{IEEE Access}, vol.~10, pp. 68\,633--68\,644, 2022.

\bibitem[Ho et~al.(2022)Ho, Yow, Zhu, and Aravamuthan]{ho2022network}
C.~M.~K. Ho, K.-C. Yow, Z.~Zhu, and S.~Aravamuthan, ``{Network Intrusion Detection via Flow-to-Image Conversion and Vision Transformer Classification},'' \emph{IEEE Access}, vol.~10, pp. 97\,780--97\,793, 2022.

\bibitem[Wu et~al.(2022)Wu, Zhang, Wang, and Sun]{wu2022}
Z.~Wu, H.~Zhang, P.~Wang, and Z.~Sun, ``{RTIDS: A Robust Transformer-Based Approach for Intrusion Detection System},'' \emph{IEEE Access}, vol.~10, pp. 64\,375--64\,387, 2022.

\bibitem[Ghourabi(2022)]{ghourabi2022security}
A.~Ghourabi, ``{A Security Model Based on LightGBM and Transformer to Protect Healthcare Systems from Cyberattacks},'' \emph{IEEE Access}, vol.~10, pp. 48\,890--48\,903, 2022.

\bibitem[Lai(2023)]{lai2023}
H.~Lai, ``{Intrusion Detection Technology Based on Large Language Models},'' in \emph{IEEE International Conference on Evolutionary Algorithms and Soft Computing Techniques (EASCT)}, 2023, pp. 1--5.

\bibitem[{Ali} and {Kostakos}(2023)]{ali2023}
T.~{Ali} and P.~{Kostakos}, ``{HuntGPT: Integrating Machine Learning-Based Anomaly Detection and Explainable AI with Large Language Models (LLMs)},'' \emph{arXiv e-prints}, p. arXiv:2309.16021, 2023.

\bibitem[Ullah et~al.(2023)Ullah, Ahmad, Khan, Alshehri, Boulila, Koubaa, Jan, and Ch]{ullah2023tnn}
S.~Ullah, J.~Ahmad, M.~A. Khan, M.~S. Alshehri, W.~Boulila, A.~Koubaa, S.~U. Jan, and M.~M.~I. Ch, ``{TNN-IDS: Transformer Neural Network-based Intrusion Detection System for MQTT-enabled IoT Networks},'' \emph{Computer Networks}, vol. 237, p. 110072, 2023.

\bibitem[Wang et~al.(2023{\natexlab{b}})Wang, Jian, Tan, Wu, and Huang]{wang2023robust}
W.~Wang, S.~Jian, Y.~Tan, Q.~Wu, and C.~Huang, ``{Robust Unsupervised Network Intrusion Detection with Self-supervised Masked Context Reconstruction},'' \emph{Computers \& Security}, vol. 128, p. 103131, 2023.

\bibitem[Wang et~al.(2024{\natexlab{d}})Wang, Li, Yang, Luo, Li, and Mahmoodi]{wang2024lightweight}
Z.~Wang, J.~Li, S.~Yang, X.~Luo, D.~Li, and S.~Mahmoodi, ``{A Lightweight IoT Intrusion Detection Model based on Improved BERT-of-Theseus},'' \emph{Expert Systems with Applications}, vol. 238, p. 122045, 2024.

\bibitem[Mel{\'\i}cias et~al.(2024)Mel{\'\i}cias, Ribeiro, Rabad{\~a}o, Santos, and Costa]{melicias2024gpt}
F.~S. Mel{\'\i}cias, T.~F. Ribeiro, C.~Rabad{\~a}o, L.~Santos, and R.~L. d.~C. Costa, ``{GPT and Interpolation-based Data Augmentation for Multiclass Intrusion Detection in IIoT},'' \emph{IEEE Access}, 2024.

\bibitem[Setianto et~al.(2021)Setianto, Tsani, Sadiq, Domalis, Tsakalidis, and Kostakos]{setianto2021gpt}
F.~Setianto, E.~Tsani, F.~Sadiq, G.~Domalis, D.~Tsakalidis, and P.~Kostakos, ``{GPT-2C: A Parser for Honeypot Logs using Large Pre-trained Language Models},'' in \emph{IEEE/ACM International Conference on Advances in Social Networks Analysis and Mining (ASONAM)}, 2021, pp. 649--653.

\bibitem[Ott et~al.(2021)Ott, Bogatinovski, Acker, Nedelkoski, and Kao]{ott2021robust}
H.~Ott, J.~Bogatinovski, A.~Acker, S.~Nedelkoski, and O.~Kao, ``{Robust and Transferable Anomaly Detection in Log Data using Pre-Trained Language Models},'' in \emph{IEEE/ACM International Workshop on Cloud Intelligence (CloudIntelligence)}, 2021, pp. 19--24.

\bibitem[Pan et~al.(2023)Pan, Wong, and Yuan]{pan2023raglog}
J.~Pan, S.~L. Wong, and Y.~Yuan, ``{RAGLog: Log Anomaly Detection using Retrieval Augmented Generation},'' \emph{arXiv preprint arXiv:2311.05261}, 2023.

\bibitem[Qi et~al.(2023)Qi, Huang, Luan, Yang, Fung, Yang, Qian, Shang, Xiao, and Wu]{qi2023loggpt}
J.~Qi, S.~Huang, Z.~Luan, S.~Yang, C.~Fung, H.~Yang, D.~Qian, J.~Shang, Z.~Xiao, and Z.~Wu, ``{LogGPT: Exploring ChatGPT for Log-based Anomaly Detection},'' in \emph{IEEE International Conference on High Performance Computing \& Communications, Data Science \& Systems, Smart City \& Dependability in Sensor, Cloud \& Big Data Systems \& Application (HPCC/DSS/SmartCity/DependSys)}, 2023, pp. 273--280.

\bibitem[Jiang et~al.(2024{\natexlab{b}})Jiang, Liu, Chen, Li, Huang, Huo, He, Ge, and Lyu]{jiang2023lilac}
Z.~Jiang, J.~Liu, Z.~Chen, Y.~Li, J.~Huang, Y.~Huo, P.~He, J.~Ge, and M.~R. Lyu, ``{LILAC: Log Parsing using LLMs with Adaptive Parsing Cache},'' \emph{Proceedings of the ACM on Software Engineering}, vol.~1, pp. 137--160, 2024.

\bibitem[Ji et~al.(2023)Ji, Han, Zhao, Zhang, and Gong]{ji2023log}
Y.~Ji, J.~Han, Y.~Zhao, S.~Zhang, and Z.~Gong, ``{Log Anomaly Detection Through GPT-2 for Large Scale Systems},'' \emph{ZTE Communications}, vol.~21, no.~3, p.~70, 2023.

\bibitem[Mudgal and Wouhaybi(2023)]{mudgal2023assessment}
P.~Mudgal and R.~Wouhaybi, ``{An Assessment of {ChatGPT} on Log Data},'' in \emph{1st International Conference on AI-generated Content (AIGC)}, 2023, pp. 148--169.

\bibitem[Sun et~al.(2023)Sun, Chen, Zhao, and Peng]{sun2023design}
Y.~Sun, Y.~Chen, H.~Zhao, and S.~Peng, ``{Design and Development of a Log Management System Based on Cloud Native Architecture},'' in \emph{9th IEEE International Conference on Systems and Informatics (ICSAI)}, 2023, pp. 1--6.

\bibitem[V{\"o}r{\"o}s et~al.(2023)V{\"o}r{\"o}s, Bergeron, and Berlin]{voros2023web}
T.~V{\"o}r{\"o}s, S.~P. Bergeron, and K.~Berlin, ``{Web Content Filtering through Knowledge Distillation of Large Language Models},'' in \emph{IEEE International Conference on Web Intelligence and Intelligent Agent Technology (WI-IAT)}, 2023, pp. 357--361.

\bibitem[Balasubramanian et~al.(2024)Balasubramanian, Seby, and Kostakos]{balasubramanian2024cygent}
P.~Balasubramanian, J.~Seby, and P.~Kostakos, ``{CYGENT: A Cybersecurity Conversational Agent with Log Summarization Powered by GPT-3},'' \emph{arXiv preprint arXiv:2403.17160}, 2024.

\bibitem[Karlsen et~al.(2024)Karlsen, Luo, Zincir-Heywood, and Heywood]{karlsen2024large}
E.~Karlsen, X.~Luo, N.~Zincir-Heywood, and M.~Heywood, ``{Large Language Models and Unsupervised Feature Learning: Implications for Log Analysis},'' \emph{Annals of Telecommunications}, pp. 1--19, 2024.

\bibitem[Meyuhas et~al.(2024)Meyuhas, Bremler-Barr, and Shapira]{meyuhas2024}
B.~Meyuhas, A.~Bremler-Barr, and T.~Shapira, ``{IoT} device labeling using large language models,'' \emph{arXiv preprint arXiv:2403.01586}, 2024.

\bibitem[Tian and Li(2024)]{tian2024dom}
Y.~Tian and Z.~Li, ``{Dom-BERT: Detecting Malicious Domains with Pre-training Model},'' in \emph{International Conference on Passive and Active Network Measurement (PAM)}.\hskip 1em plus 0.5em minus 0.4em\relax Springer, 2024, pp. 133--158.

\bibitem[Almodovar et~al.(2024)Almodovar, Sabrina, Karimi, and Azad]{almodovar2024logfit}
C.~Almodovar, F.~Sabrina, S.~Karimi, and S.~Azad, ``Logfit: Log anomaly detection using fine-tuned language models,'' \emph{IEEE Transactions on Network and Service Management}, 2024.

\bibitem[Soman and HG(2023)]{soman2023observations}
S.~Soman and R.~HG, ``{Observations on LLMs for Telecom Domain: Capabilities and Limitations},'' in \emph{3rd ACM International Conference on AI-ML Systems (AIMLSystems)}, 2023, pp. 1--5.

\bibitem[Wang et~al.(2023{\natexlab{c}})Wang, Scazzariello, Farshin, Kostic, and Chiesa]{wang2023making}
C.~Wang, M.~Scazzariello, A.~Farshin, D.~Kostic, and M.~Chiesa, ``Making network configuration human friendly,'' \emph{arXiv preprint arXiv:2309.06342}, 2023.

\bibitem[Mani et~al.(2023)Mani, Zhou, Hsieh, Segarra, Eberl, Azulai, Frizler, Chandra, and Kandula]{mani2023enhancing}
S.~K. Mani, Y.~Zhou, K.~Hsieh, S.~Segarra, T.~Eberl, E.~Azulai, I.~Frizler, R.~Chandra, and S.~Kandula, ``{Enhancing Network Management Using Code Generated by Large Language Models},'' in \emph{22nd ACM Workshop on Hot Topics in Networks (HotNets)}, 2023, pp. 196--204.

\bibitem[Mondal et~al.(2023)Mondal, Tang, Beckett, Millstein, and Varghese]{mondal2023llms}
R.~Mondal, A.~Tang, R.~Beckett, T.~Millstein, and G.~Varghese, ``{What Do LLMs Need to Synthesize Correct Router Configurations?}'' in \emph{22nd ACM Workshop on Hot Topics in Networks (HotNets)}, 2023, pp. 189--195.

\bibitem[Roychowdhury et~al.(2024)Roychowdhury, Jain, and Soman]{roychowdhury2024unlocking}
S.~Roychowdhury, N.~Jain, and S.~Soman, ``{Unlocking Telecom Domain Knowledge Using LLMs},'' in \emph{16th IEEE International Conference on COMmunication Systems \& NETworkS (COMSNETS)}, 2024, pp. 267--269.

\bibitem[Shen et~al.(2024)Shen, Shao, Zhang, Lin, Pan, Li, Zhang, and Letaief]{shen2024large}
Y.~Shen, J.~Shao, X.~Zhang, Z.~Lin, H.~Pan, D.~Li, J.~Zhang, and K.~B. Letaief, ``{Large Language Models Empowered Autonomous Edge AI for Connected Intelligence},'' \emph{IEEE Communications Magazine}, 2024.

\bibitem[Duclos et~al.(2024)Duclos, Fernandez, Moore, Mittal, and Zieglar]{duclos2024utilizing}
M.~Duclos, I.~A. Fernandez, K.~Moore, S.~Mittal, and E.~Zieglar, ``{Utilizing Large Language Models to Translate RFC Protocol Specifications to CPSA Definitions},'' \emph{arXiv preprint arXiv:2402.00890}, 2024.

\bibitem[Ahmed et~al.(2024)Ahmed, Piovesan, De~Domenico, and Choudhury]{ahmed2024linguistic}
T.~Ahmed, N.~Piovesan, A.~De~Domenico, and S.~Choudhury, ``{Linguistic Intelligence in Large Language Models for Telecommunications},'' \emph{arXiv preprint arXiv:2402.15818}, 2024.

\bibitem[Piovesan et~al.(2024)Piovesan, De~Domenico, and Ayed]{piovesan2024telecom}
N.~Piovesan, A.~De~Domenico, and F.~Ayed, ``{Telecom Language Models: Must They Be Large?}'' \emph{arXiv preprint arXiv:2403.04666}, 2024.

\bibitem[Ghasemirahni et~al.(2024)Ghasemirahni, Farshin, Scazzariello, Chiesa, and Kosti{\'c}]{ghasemirahni2024deploying}
H.~Ghasemirahni, A.~Farshin, M.~Scazzariello, M.~Chiesa, and D.~Kosti{\'c}, ``{Deploying Stateful Network Functions Efficiently using Large Language Models},'' in \emph{4th ACM-EUROSYS Workshop on Machine Learning and Systems (EuroMLSys)}, 2024, pp. 28--38.

\bibitem[Erak et~al.(2024)Erak, Alabbasi, Alhussein, Lotfi, Hussein, Muhaidat, and Debbah]{erak2024leveraging}
O.~Erak, N.~Alabbasi, O.~Alhussein, I.~Lotfi, A.~Hussein, S.~Muhaidat, and M.~Debbah, ``{Leveraging Fine-Tuned Retrieval-Augmented Generation with Long-Context Support: For 3GPP Standards},'' \emph{arXiv preprint arXiv:2408.11775}, 2024.

\bibitem[Ayed et~al.(2024)Ayed, Maatouk, Piovesan, De~Domenico, Debbah, and Luo]{ayed2024hermes}
F.~Ayed, A.~Maatouk, N.~Piovesan, A.~De~Domenico, M.~Debbah, and Z.-Q. Luo, ``Hermes: A large language model framework on the journey to autonomous networks,'' \emph{arXiv preprint arXiv:2411.06490}, 2024.

\bibitem[Ahmed et~al.(2021)Ahmed, Byreddy, Nutakki, Sikos, and Haskell-Dowland]{ahmed2021ecu}
M.~Ahmed, S.~Byreddy, A.~Nutakki, L.~F. Sikos, and P.~Haskell-Dowland, ``{ECU-IoHT: A Dataset for Analyzing Cyberattacks in Internet of Health Things},'' \emph{Ad Hoc Networks}, vol. 122, p. 102621, 2021.

\bibitem[Jin et~al.(2024)Jin, Han, Yang, Jiang, Liu, Chang, Chen, and Hu]{jin2024llm}
H.~Jin, X.~Han, J.~Yang, Z.~Jiang, Z.~Liu, C.-Y. Chang, H.~Chen, and X.~Hu, ``{LLM Maybe LongLM: Self-Extend LLM Context Window Without Tuning},'' \emph{arXiv preprint arXiv:2401.01325}, 2024.

\bibitem[Yang et~al.(2019)Yang, Dai, Yang, Carbonell, Salakhutdinov, and Le]{yang2019xlnet}
Z.~Yang, Z.~Dai, Y.~Yang, J.~Carbonell, R.~R. Salakhutdinov, and Q.~V. Le, ``{XLNet: Generalized Autoregressive Pretraining for Language Understanding},'' \emph{Advances in Neural Information Processing Systems (NeurIPS)}, vol.~32, 2019.

\bibitem[Yao et~al.(2024)Yao, Duan, Xu, Cai, Sun, and Zhang]{yao2024survey}
Y.~Yao, J.~Duan, K.~Xu, Y.~Cai, Z.~Sun, and Y.~Zhang, ``{A Survey on Large Language Model (LLM) Security and Privacy: The Good, the Bad, and the Ugly},'' \emph{High-Confidence Computing}, p. 100211, 2024.

\bibitem[Tavallaee et~al.(2009)Tavallaee, Bagheri, Lu, and Ghorbani]{tavallaee2009detailed}
M.~Tavallaee, E.~Bagheri, W.~Lu, and A.~A. Ghorbani, ``{A Detailed Analysis of the KDD CUP 99 Data Set},'' in \emph{IEEE Symposium on Computational Intelligence for Security and Defense Applications (CISDA)}, 2009, pp. 1--6.

\bibitem[Ra{\i}ssi et~al.(2007)Ra{\i}ssi, Brissaud, Dray, Poncelet, Roche, and Teisseire]{raissi2007web}
C.~Ra{\i}ssi, J.~Brissaud, G.~Dray, P.~Poncelet, M.~Roche, and M.~Teisseire, ``{Web Analyzing Traffic Challenge: Description and Results},'' in \emph{European Conference on Machine Learning and Principles and Practice of Knowledge Discovery in Databases (ECML/PKDD)}, 2007, pp. 47--52.

\bibitem[csi()]{csic_2010_web_application_attacks}
\BIBentryALTinterwordspacing
``{CSIC 2010 Web Application Attacks}.'' [Online]. Available: \url{https://www.kaggle.com/datasets/ispangler/csic-2010-web-application-attacks}
\BIBentrySTDinterwordspacing

\bibitem[htt()]{httpparamsdataset}
\BIBentryALTinterwordspacing
``{HttpParamsDataset}.'' [Online]. Available: \url{https://www.kaggle.com/datasets/evg3n1j/httpparamsdataset}
\BIBentrySTDinterwordspacing

\bibitem[Moustafa and Slay(2015)]{moustafa2015unsw}
N.~Moustafa and J.~Slay, ``{UNSW-NB15: A Comprehensive Data Set for Network Intrusion Detection Systems (UNSW-NB15 Network Data Set)},'' in \emph{IEEE Military Communications and Information Systems Conference (MilCIS)}, 2015, pp. 1--6.

\bibitem[Miettinen et~al.(2017)Miettinen, Marchal, Hafeez, Asokan, Sadeghi, and Tarkoma]{miettinen2017iot}
M.~Miettinen, S.~Marchal, I.~Hafeez, N.~Asokan, A.-R. Sadeghi, and S.~Tarkoma, ``{IoT SENTINEL: Automated Device-Type Identification for Security Enforcement in IoT},'' in \emph{IEEE 37th International Conference on Distributed Computing Systems (ICDCS)}, 2017, pp. 2177--2184.

\bibitem[Draper-Gil et~al.(2016)Draper-Gil, Lashkari, Mamun, and Ghorbani]{draper2016characterization}
G.~Draper-Gil, A.~H. Lashkari, M.~S.~I. Mamun, and A.~A. Ghorbani, ``{Characterization of Encrypted and VPN Traffic using Time-related Features},'' in \emph{International Conference on Information Systems Security and Privacy (ICISSP)}, 2016, pp. 407--414.

\bibitem[Lashkari et~al.(2017)Lashkari, Gil, Mamun, and Ghorbani]{lashkari2017characterization}
A.~H. Lashkari, G.~D. Gil, M.~S.~I. Mamun, and A.~A. Ghorbani, ``{Characterization of Tor Traffic using Time based Features},'' in \emph{International Conference on Information Systems Security and Privacy (ICISSP)}, vol.~2, 2017, pp. 253--262.

\bibitem[Sivanathan et~al.(2018)Sivanathan, Gharakheili, Loi, Radford, Wijenayake, Vishwanath, and Sivaraman]{sivanathan2018classifying}
A.~Sivanathan, H.~H. Gharakheili, F.~Loi, A.~Radford, C.~Wijenayake, A.~Vishwanath, and V.~Sivaraman, ``{Classifying IoT Devices in Smart Environments Using Network Traffic Characteristics},'' \emph{IEEE Transactions on Mobile Computing}, vol.~18, no.~8, pp. 1745--1759, 2018.

\bibitem[ust()]{ustc-tfc2016}
\BIBentryALTinterwordspacing
``{USTC-TFC2016}.'' [Online]. Available: \url{https://www.kaggle.com/datasets/randasrour/ustctfc2016}
\BIBentrySTDinterwordspacing

\bibitem[Sharafaldin et~al.(2018)Sharafaldin, Lashkari, Ghorbani, et~al.]{sharafaldin2018toward}
I.~Sharafaldin, A.~H. Lashkari, A.~A. Ghorbani \emph{et~al.}, ``{Toward Generating a New Intrusion Detection Dataset and Intrusion Traffic Characterization},'' \emph{ICISSp}, vol.~1, pp. 108--116, 2018.

\bibitem[Sirinam et~al.(2018)Sirinam, Imani, Juarez, and Wright]{sirinam2018deep}
P.~Sirinam, M.~Imani, M.~Juarez, and M.~Wright, ``{Deep Fingerprinting: Undermining Website Fingerprinting Defenses with Deep Learning},'' in \emph{ACM SIGSAC Conference on Computer and Communications Security (CCS)}, 2018, pp. 1928--1943.

\bibitem[lab({\natexlab{a}})]{labeledflows17}
``{LabeledFlows2017 Dataset},'' \url{https://www.kaggle.com/datasets/jsrojas/ip-network-traffic-flows-labeled-with-87-apps}.

\bibitem[Van~Ede et~al.(2020)Van~Ede, Bortolameotti, Continella, Ren, Dubois, Lindorfer, Choffnes, Van~Steen, and Peter]{van2020flowprint}
T.~Van~Ede, R.~Bortolameotti, A.~Continella, J.~Ren, D.~J. Dubois, M.~Lindorfer, D.~Choffnes, M.~Van~Steen, and A.~Peter, ``{FlowPrint: Semi-Supervised Mobile-App Fingerprinting on Encrypted Network Traffic},'' in \emph{Network and Distributed System Security symposium (NDSS)}, vol.~27, 2020.

\bibitem[Lin et~al.(2018)Lin, Wang, Zettlemoyer, and Ernst]{lin2018nl2bash}
X.~V. Lin, C.~Wang, L.~Zettlemoyer, and M.~D. Ernst, ``{NL2Bash: A Corpus and Semantic Parser for Natural Language Interface to the Linux Operating System},'' in \emph{Language Resource and Evaluation Conference (LREC)}, 2018.

\bibitem[Sharafaldin et~al.(2019)Sharafaldin, Lashkari, Hakak, and Ghorbani]{sharafaldin2019developing}
I.~Sharafaldin, A.~H. Lashkari, S.~Hakak, and A.~A. Ghorbani, ``{Developing Realistic Distributed Denial of Service (DDoS) Attack Dataset and Taxonomy},'' in \emph{IEEE International Carnahan Conference on Security Technology (ICCST)}, 2019, pp. 1--8.

\bibitem[Perdisci et~al.(2020)Perdisci, Papastergiou, Alrawi, and Antonakakis]{perdisci2020iotfinder}
R.~Perdisci, T.~Papastergiou, O.~Alrawi, and M.~Antonakakis, ``{IoTFinder: Efficient Large-Scale Identification of IoT Devices via Passive DNS Traffic Analysis},'' in \emph{IEEE European Symposium on Security and Privacy (EuroS\&P)}, 2020, pp. 474--489.

\bibitem[Zaker(2019)]{zaker_2019}
\BIBentryALTinterwordspacing
F.~Zaker, ``{Online Shopping Store - Web Server Logs},'' 2019. [Online]. Available: \url{https://www.kaggle.com/datasets/eliasdabbas/web-server-access-logs}
\BIBentrySTDinterwordspacing

\bibitem[lab({\natexlab{b}})]{labeledflows19}
``{LabeledFlows2019 Dataset},'' \url{https://www.kaggle.com/datasets/jsrojas/labeled-network-traffic-flows-114-applications}.

\bibitem[Hilmi et~al.(2020)Hilmi, Cahyanto, and Mustamiin]{hilmi2020}
\BIBentryALTinterwordspacing
M.~A.~A. Hilmi, K.~A. Cahyanto, and M.~Mustamiin, ``{Apache Web Server - Access Log Pre-processing for Web Intrusion Detection},'' 2020. [Online]. Available: \url{https://dx.doi.org/10.25126/jtiik.2022924107}
\BIBentrySTDinterwordspacing

\bibitem[MontazeriShatoori et~al.(2020)MontazeriShatoori, Davidson, Kaur, and Lashkari]{montazerishatoori2020detection}
M.~MontazeriShatoori, L.~Davidson, G.~Kaur, and A.~H. Lashkari, ``{Detection of DoH Tunnels using Time-series Classification of Encrypted Traffic},'' in \emph{IEEE Intl Conf on Dependable, Autonomic and Secure Computing, Intl Conf on Pervasive Intelligence and Computing, Intl Conf on Cloud and Big Data Computing, Intl Conf on Cyber Science and Technology Congress (DASC/PiCom/CBDCom/CyberSciTech)}, 2020, pp. 63--70.

\bibitem[Sedlar et~al.(2020)Sedlar, Kren, Štefanič Južnič, and Volk]{sedlar_2020_3687527}
\BIBentryALTinterwordspacing
U.~Sedlar, M.~Kren, L.~Štefanič Južnič, and M.~Volk, ``{CyberLab Honeynet Dataset},'' 2020. [Online]. Available: \url{https://doi.org/10.5281/zenodo.3687527}
\BIBentrySTDinterwordspacing

\bibitem[Zhu et~al.(2023)Zhu, He, He, Liu, and Lyu]{zhu2023loghub}
J.~Zhu, S.~He, P.~He, J.~Liu, and M.~R. Lyu, ``{Loghub: A Large Collection of System Log Datasets for AI-driven Log Analytics},'' in \emph{IEEE 34th International Symposium on Software Reliability Engineering (ISSRE)}, 2023, pp. 355--366.

\bibitem[Bremler-Barr et~al.(2022)Bremler-Barr, Meyuhas, and Shister]{bremler2022one}
A.~Bremler-Barr, B.~Meyuhas, and R.~Shister, ``{One MUD to Rule Them All: IoT Location Impact},'' in \emph{IEEE/IFIP Network Operations and Management Symposium (NOMS)}, 2022, pp. 1--5.

\bibitem[Hindy et~al.(2020)Hindy, Bayne, Bures, Atkinson, Tachtatzis, and Bellekens]{hindy2020machine}
H.~Hindy, E.~Bayne, M.~Bures, R.~Atkinson, C.~Tachtatzis, and X.~Bellekens, ``{Machine Learning based IoT Intrusion Detection System: An MQTT Case Study (MQTT-IoT-IDS2020 Dataset)},'' in \emph{International Networking Conference}.\hskip 1em plus 0.5em minus 0.4em\relax Springer, 2020, pp. 73--84.

\bibitem[Moustafa(2021)]{moustafa2021new}
N.~Moustafa, ``{A New Distributed Architecture for Evaluating AI-based Security Systems at the Edge: Network TON\_IoT Datasets},'' \emph{Sustainable Cities and Society}, vol.~72, p. 102994, 2021.

\bibitem[fwa()]{fwaf-dataset}
\BIBentryALTinterwordspacing
``{fwaf-dataset}.'' [Online]. Available: \url{https://www.kaggle.com/datasets/evg3n1j/fwaf-dataset}
\BIBentrySTDinterwordspacing

\bibitem[Consortium(2021)]{eu_spirit_consortium_2021_4767861}
\BIBentryALTinterwordspacing
E.~S. Consortium, ``{Dataset of EU SPIRIT Project},'' 2021. [Online]. Available: \url{https://doi.org/10.5281/zenodo.4767861}
\BIBentrySTDinterwordspacing

\bibitem[Dadkhah et~al.(2022)Dadkhah, Mahdikhani, Danso, Zohourian, Truong, and Ghorbani]{dadkhah2022towards}
S.~Dadkhah, H.~Mahdikhani, P.~K. Danso, A.~Zohourian, K.~A. Truong, and A.~A. Ghorbani, ``{Towards the Development of a Realistic Multidimensional IoT Profiling Dataset},'' in \emph{19th Annual International Conference on Privacy, Security \& Trust (PST)}, 2022, pp. 1--11.

\bibitem[Ferrag et~al.(2022)Ferrag, Friha, Hamouda, Maglaras, and Janicke]{ferrag2022edge}
M.~A. Ferrag, O.~Friha, D.~Hamouda, L.~Maglaras, and H.~Janicke, ``{Edge-IIoTset: A New Comprehensive Realistic Cyber Security Dataset of IoT and IIoT Applications for Centralized and Federated Learning},'' \emph{IEEE Access}, vol.~10, pp. 40\,281--40\,306, 2022.

\bibitem[Neto et~al.(2023)Neto, Dadkhah, Ferreira, Zohourian, Lu, and Ghorbani]{neto2023ciciot2023}
E.~C.~P. Neto, S.~Dadkhah, R.~Ferreira, A.~Zohourian, R.~Lu, and A.~A. Ghorbani, ``{CICIoT2023: A Real-Time Dataset and Benchmark for Large-Scale Attacks in IoT Environment},'' \emph{Sensors}, vol.~23, no.~13, p. 5941, 2023.

\bibitem[Karim et~al.(2023)Karim, Mubasshir, Rahman, and Bertino]{karim2023spec5g}
I.~Karim, K.~S. Mubasshir, M.~M. Rahman, and E.~Bertino, ``{SPEC5G: A Dataset for 5G Cellular Network Protocol Analysis},'' in \emph{13th International Joint Conference on Natural Language Processing and the 3rd Conference of the Asia-Pacific Chapter of the Association for Computational Linguistics (IJCNLP-AACL)}, 2023.

\bibitem[Maatouk et~al.(2023)Maatouk, Ayed, Piovesan, De~Domenico, Debbah, and Luo]{maatouk2023teleqna}
A.~Maatouk, F.~Ayed, N.~Piovesan, A.~De~Domenico, M.~Debbah, and Z.-Q. Luo, ``{TeleQnA: A Benchmark Dataset to Assess Large Language Models Telecommunications Knowledge},'' \emph{arXiv preprint arXiv:2310.15051}, 2023.

\bibitem[Durumeric et~al.(2015)Durumeric, Adrian, Mirian, Bailey, and Halderman]{durumeric2015search}
Z.~Durumeric, D.~Adrian, A.~Mirian, M.~Bailey, and J.~A. Halderman, ``{A Search Engine Backed by Internet-Wide Scanning},'' in \emph{Proceedings of the 22nd ACM SIGSAC Conference on Computer and Communications Security (CCS)}, 2015, pp. 542--553.

\bibitem[Zhu et~al.(2022)Zhu, Lu, Lyu, Pan, Jia, Cheng, Kang, Lv, Yang, Xue, et~al.]{zhu2022zoonet}
S.~Zhu, J.~Lu, B.~Lyu, T.~Pan, C.~Jia, X.~Cheng, D.~Kang, Y.~Lv, F.~Yang, X.~Xue \emph{et~al.}, ``{Zoonet: a Proactive Telemetry System for Large-Scale Cloud Networks},'' in \emph{Proceedings of the 18th International Conference on emerging Networking EXperiments and Technologies}, 2022, pp. 321--336.

\bibitem[Yu(2019)]{yu2019network}
M.~Yu, ``{Network Telemetry: Towards a Top-Down Approach},'' \emph{ACM SIGCOMM Computer Communication Review}, vol.~49, no.~1, pp. 11--17, 2019.

\bibitem[D’Alconzo et~al.(2019)D’Alconzo, Drago, Morichetta, Mellia, and Casas]{d2019survey}
A.~D’Alconzo, I.~Drago, A.~Morichetta, M.~Mellia, and P.~Casas, ``{A Survey on Big Data for Network Traffic Monitoring and Analysis},'' \emph{IEEE Transactions on Network and Service Management}, vol.~16, no.~3, pp. 800--813, 2019.

\bibitem[Qian et~al.(2024)Qian, Li, Wang, Zhou, and Shao]{qian2024netbench}
C.~Qian, X.~Li, Q.~Wang, G.~Zhou, and H.~Shao, ``{NetBench: A Large-Scale and Comprehensive Network Traffic Benchmark Dataset for Foundation Models},'' \emph{arXiv preprint arXiv:2403.10319}, 2024.

\bibitem[Jacobs et~al.(2022)Jacobs, Beltiukov, Willinger, Ferreira, Gupta, and Granville]{jacobs2022ai}
A.~S. Jacobs, R.~Beltiukov, W.~Willinger, R.~A. Ferreira, A.~Gupta, and L.~Z. Granville, ``{AI/ML for network security: The emperor has no clothes},'' in \emph{Proceedings of the 2022 ACM SIGSAC Conference on Computer and Communications Security}, 2022, pp. 1537--1551.

\bibitem[Engelen et~al.(2021)Engelen, Rimmer, and Joosen]{engelen2021troubleshooting}
G.~Engelen, V.~Rimmer, and W.~Joosen, ``{Troubleshooting an Intrusion Detection Dataset: the CICIDS2017 Case Study},'' in \emph{IEEE Security and Privacy Workshops (SPW)}, 2021, pp. 7--12.

\bibitem[Zhang et~al.(2021)Zhang, Xie, Bai, Yu, Li, and Gao]{zhang2021survey}
C.~Zhang, Y.~Xie, H.~Bai, B.~Yu, W.~Li, and Y.~Gao, ``{A Survey on Federated Learning},'' \emph{Knowledge-Based Systems}, vol. 216, p. 106775, 2021.

\bibitem[Li et~al.(2020)Li, Fan, Tse, and Lin]{li2020review}
L.~Li, Y.~Fan, M.~Tse, and K.-Y. Lin, ``{A Review of Applications in Federated Learning},'' \emph{Computers \& Industrial Engineering}, vol. 149, p. 106854, 2020.

\bibitem[Huang et~al.(2024{\natexlab{b}})Huang, Li, Du, Kang, Niyato, Kim, and Wu]{huang2024federated}
X.~Huang, P.~Li, H.~Du, J.~Kang, D.~Niyato, D.~I. Kim, and Y.~Wu, ``{Federated Learning-Empowered AI-Generated Content in Wireless Networks},'' \emph{IEEE Network}, 2024.

\bibitem[Schwartz et~al.(2020)Schwartz, Dodge, Smith, and Etzioni]{schwartz2020green}
R.~Schwartz, J.~Dodge, N.~A. Smith, and O.~Etzioni, ``{Green AI},'' \emph{Communications of the ACM}, vol.~63, no.~12, pp. 54--63, 2020.

\bibitem[Ray(2022)]{ray2022review}
P.~P. Ray, ``{A Review on TinyML: State-of-the-art and Prospects},'' \emph{Journal of King Saud University-Computer and Information Sciences}, vol.~34, no.~4, pp. 1595--1623, 2022.

\bibitem[Dutta and Bharali(2021)]{dutta2021tinyml}
L.~Dutta and S.~Bharali, ``{TinyML Meets IoT: A Comprehensive Survey},'' \emph{Internet of Things}, vol.~16, p. 100461, 2021.

\bibitem[Ciliberto et~al.(2018)Ciliberto, Herbster, Ialongo, Pontil, Rocchetto, Severini, and Wossnig]{ciliberto2018quantum}
C.~Ciliberto, M.~Herbster, A.~D. Ialongo, M.~Pontil, A.~Rocchetto, S.~Severini, and L.~Wossnig, ``{Quantum Machine Learning: A Classical Perspective},'' \emph{Proceedings of the Royal Society A: Mathematical, Physical and Engineering Sciences}, vol. 474, no. 2209, p. 20170551, 2018.

\bibitem[Dunjko and Briegel(2018)]{dunjko2018machine}
V.~Dunjko and H.~J. Briegel, ``{Machine Learning \& Artificial Intelligence in the Quantum Domain: A Review of Recent Progress},'' \emph{Reports on Progress in Physics}, vol.~81, no.~7, p. 074001, 2018.

\bibitem[Du et~al.(2023)Du, Zhang, Liu, Wang, Lin, Li, Niyato, Kang, Xiong, Cui, et~al.]{du2023beyond}
H.~Du, R.~Zhang, Y.~Liu, J.~Wang, Y.~Lin, Z.~Li, D.~Niyato, J.~Kang, Z.~Xiong, S.~Cui \emph{et~al.}, ``Beyond deep reinforcement learning: A tutorial on generative diffusion models in network optimization,'' \emph{arXiv preprint arXiv:2308.05384}, 2023.

\bibitem[Du et~al.(2024{\natexlab{b}})Du, Zhang, Liu, Wang, Lin, Li, Niyato, Kang, Xiong, Cui, et~al.]{du2024enhancing}
------, ``Enhancing deep reinforcement learning: A tutorial on generative diffusion models in network optimization,'' \emph{IEEE Communications Surveys \& Tutorials}, 2024.

\bibitem[Xu et~al.(2019)Xu, Uszkoreit, Du, Fan, Zhao, and Zhu]{xu2019explainable}
F.~Xu, H.~Uszkoreit, Y.~Du, W.~Fan, D.~Zhao, and J.~Zhu, ``{Explainable AI: A Brief Survey on History, Research Areas, Approaches and Challenges},'' in \emph{Natural language processing and Chinese computing: 8th cCF international conference}.\hskip 1em plus 0.5em minus 0.4em\relax Springer, 2019, pp. 563--574.

\bibitem[Dwivedi et~al.(2023)Dwivedi, Dave, Naik, Singhal, Omer, Patel, Qian, Wen, Shah, Morgan, et~al.]{dwivedi2023explainable}
R.~Dwivedi, D.~Dave, H.~Naik, S.~Singhal, R.~Omer, P.~Patel, B.~Qian, Z.~Wen, T.~Shah, G.~Morgan \emph{et~al.}, ``E{xplainable AI (XAI): Core Ideas, Techniques, and Solutions},'' \emph{ACM Computing Surveys}, vol.~55, no.~9, pp. 1--33, 2023.

\bibitem[Nascita et~al.(2024)Nascita, Aceto, Ciuonzo, Montieri, Persico, and Pescap{\'e}]{nascita2024survey}
A.~Nascita, G.~Aceto, D.~Ciuonzo, A.~Montieri, V.~Persico, and A.~Pescap{\'e}, ``{A Survey on Explainable Artificial Intelligence for Internet Traffic Classification and Prediction, and Intrusion Detection},'' \emph{IEEE Communications Surveys \& Tutorials}, 2024.

\bibitem[Kaddour et~al.(2022)Kaddour, Lynch, Liu, Kusner, and Silva]{kaddour2022causal}
J.~Kaddour, A.~Lynch, Q.~Liu, M.~J. Kusner, and R.~Silva, ``{Causal Machine Learning: A Survey and Open Problems},'' \emph{arXiv preprint arXiv:2206.15475}, 2022.

\bibitem[Sarker et~al.(2021)Sarker, Zhou, Eberhart, and Hitzler]{sarker2021neuro}
M.~K. Sarker, L.~Zhou, A.~Eberhart, and P.~Hitzler, ``{Neuro-Symbolic Artificial Intelligence},'' \emph{AI Communications}, vol.~34, no.~3, pp. 197--209, 2021.

\bibitem[Hitzler and Sarker(2022)]{hitzler2022neuro}
P.~Hitzler and M.~K. Sarker, \emph{{Neuro-Symbolic Artificial Intelligence: The State of the Art}}.\hskip 1em plus 0.5em minus 0.4em\relax IOS press, 2022.

\bibitem[Zheng et~al.(2018)Zheng, Xie, Dai, Chen, and Wang]{zheng2018blockchain}
Z.~Zheng, S.~Xie, H.-N. Dai, X.~Chen, and H.~Wang, ``{Blockchain Challenges and Opportunities: A Survey},'' \emph{International Journal of Web and Grid Services}, vol.~14, no.~4, pp. 352--375, 2018.

\bibitem[Pilkington(2016)]{pilkington2016blockchain}
M.~Pilkington, ``{Blockchain Technology: Principles and Applications},'' in \emph{Research handbook on digital transformations}.\hskip 1em plus 0.5em minus 0.4em\relax Edward Elgar Publishing, 2016, pp. 225--253.

\end{thebibliography}
\endgroup

\vskip -2\baselineskip plus -1fil
\begin{IEEEbiography}[{\includegraphics[width=1in,clip,keepaspectratio]{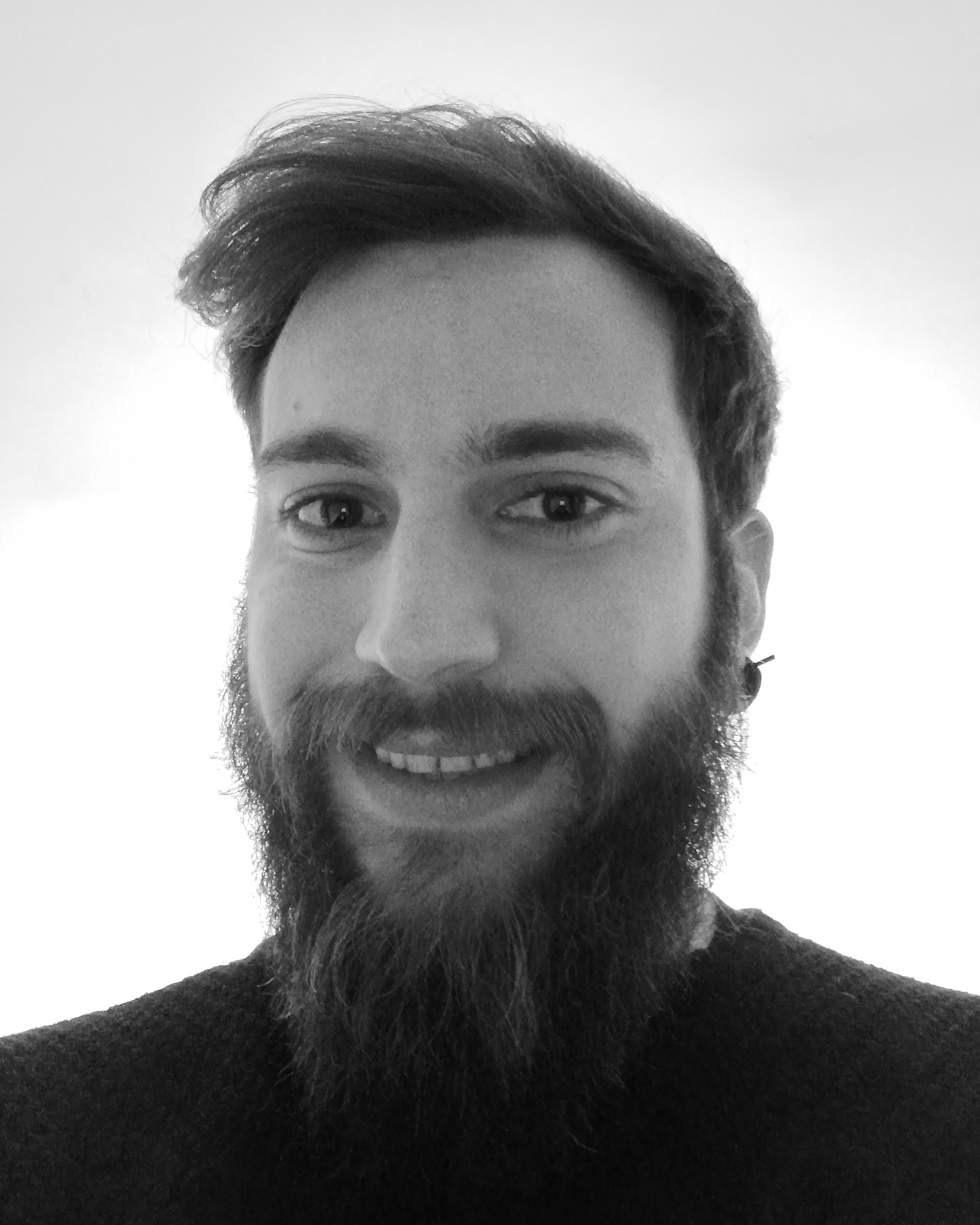}}] {Giampaolo Bovenzi} is an Assistant Professor at DIETI of the University of Napoli Federico II, since October 2023. He received his Ph.D.~degree at the same University in June 2022. His research interests focus on (anonymized and encrypted) traffic classification, network security (with a focus on IoT), and blockchain. He has co-authored more than 20 papers in international journals and conference proceedings.
\end{IEEEbiography}

\vskip -2\baselineskip plus -1fil
\begin{IEEEbiography}[{\includegraphics[width=1in,clip,keepaspectratio]{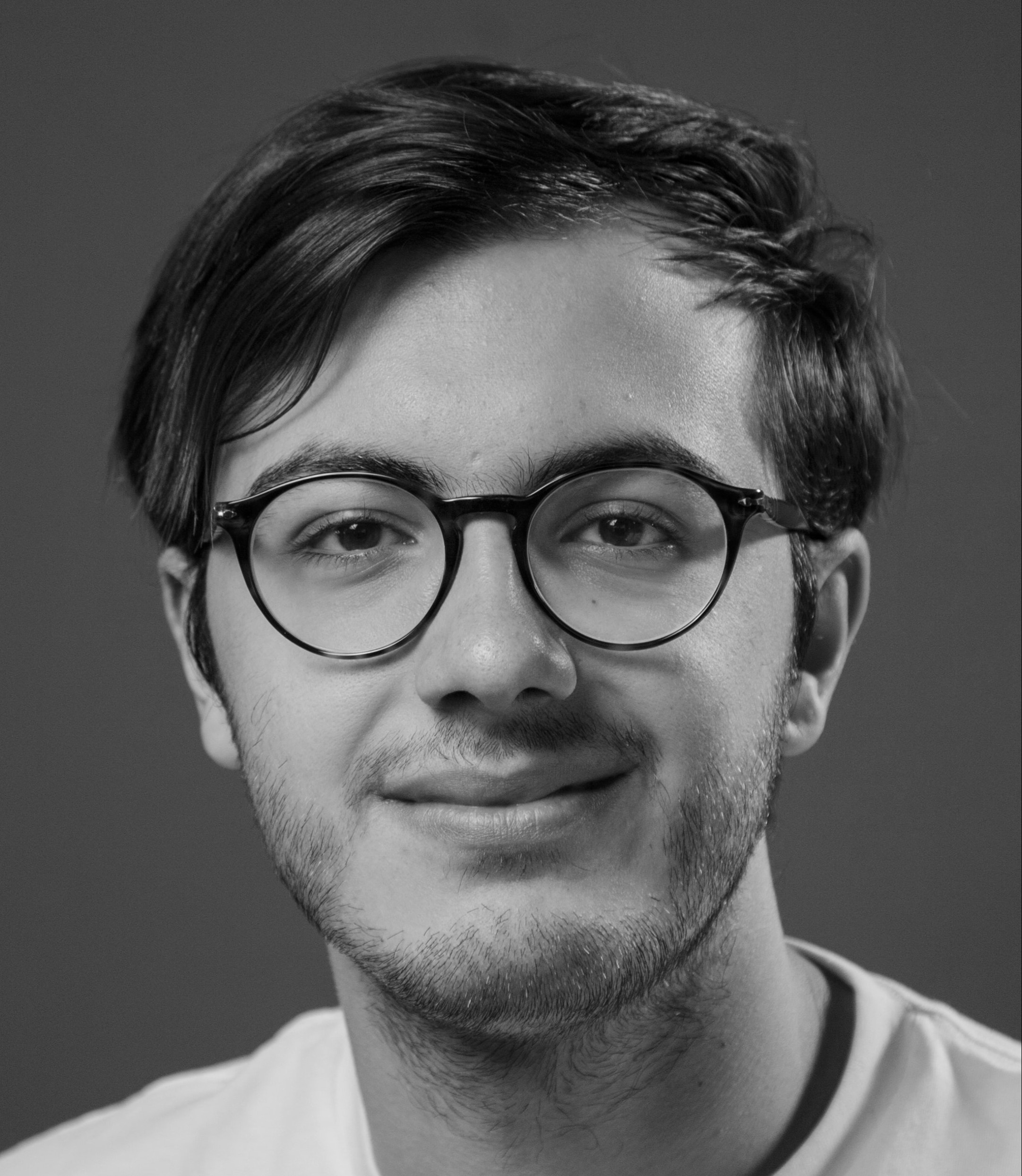}}]{Francesco Cerasuolo} is a Ph.D. student at DIETI of the University of Napoli Federico II. He received his M.S.~Laurea Degree in Computer Engineering in July 2022 from the same University. His research interests include traffic classification, machine and deep learning, and class incremental learning.
\end{IEEEbiography}

\vskip -2\baselineskip plus -1fil
\begin{IEEEbiography}[{\includegraphics[width=1in,clip,keepaspectratio]{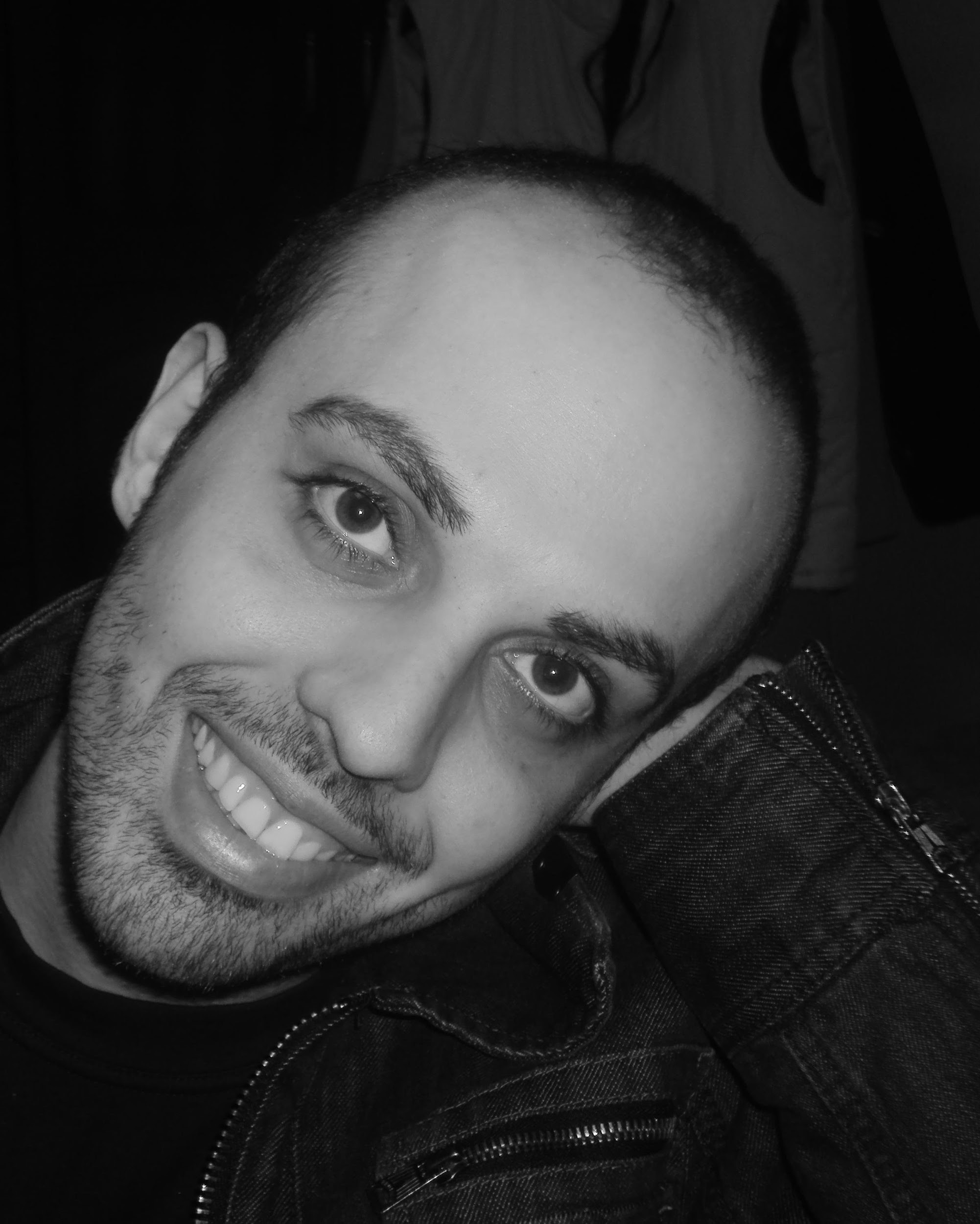}}] {Domenico Ciuonzo (S'11-M'14-SM'16)} is a Tenure-Track Professor at the University of Napoli Federico II. He holds a Ph.D.~from the University of Campania Luigi Vanvitelli.
He is the recipient of two Best Paper awards (IEEE ICCCS 2019 and Elsevier ComNet 2020), the 2019 {\sc IEEE AESS} Exceptional Service award, the 2020 {\sc IEEE Sensors Council} Early-Career Technical Achievement award and the 2021 {\sc IEEE AESS} Early-Career Award.
His research interests include data fusion, network analytics, IoT, and AI.
\end{IEEEbiography}

\vskip -2\baselineskip plus -1fil
\begin{IEEEbiography}
[{\includegraphics[width=1in,clip,keepaspectratio]{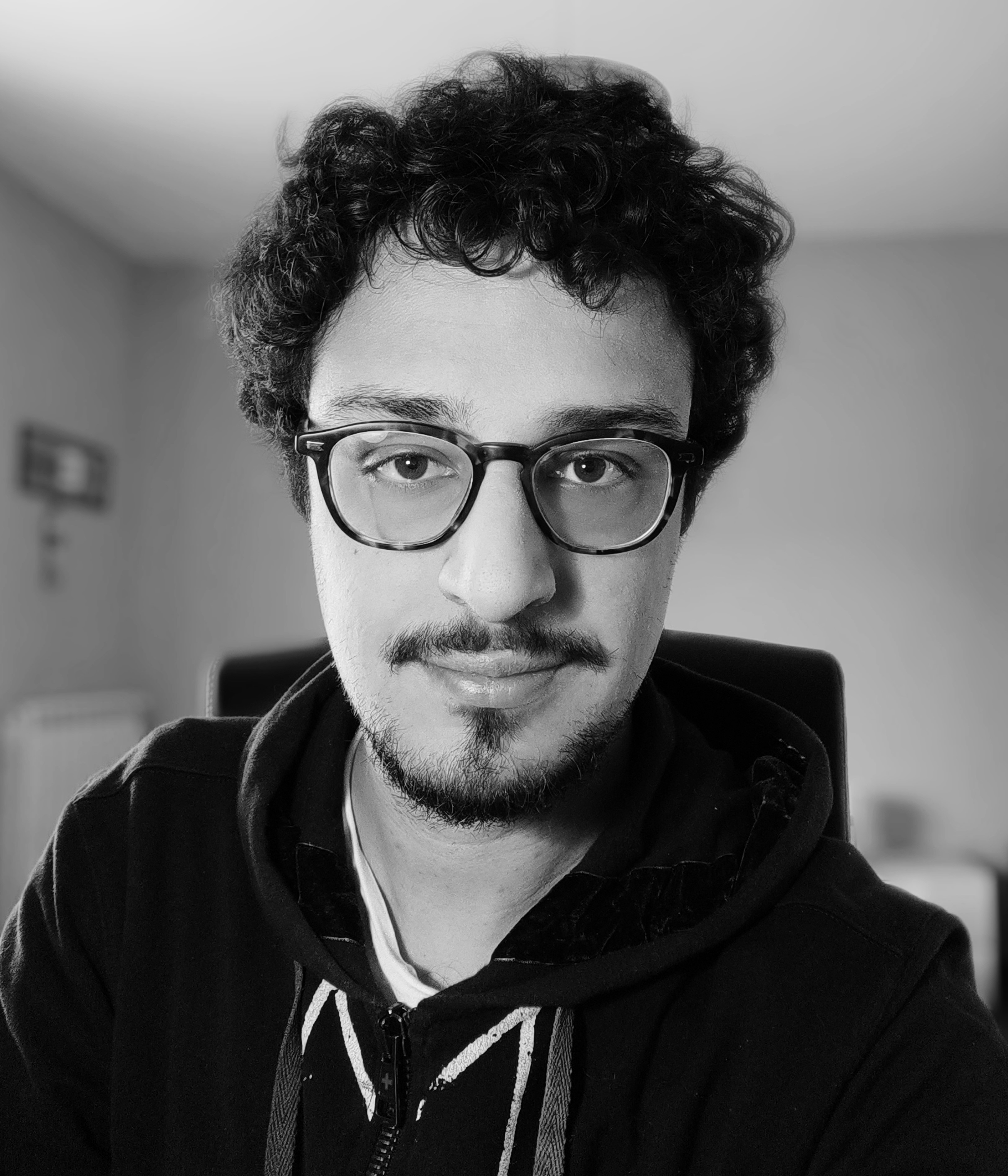}}]{Davide Di Monda} is a Ph.D.~student at IMT School for Advanced Studies Lucca and University of Napoli Federico II since December 2022. He received his M.S.~Laurea Degree (summa cum laude) in Computer Engineering in July 2022 from the University of Napoli Federico II. His research interests include cybersecurity, attack classification, and blockchain.
\end{IEEEbiography}

\vskip -2\baselineskip plus -1fil
\begin{IEEEbiography}
[{\includegraphics[width=1in,clip,keepaspectratio]{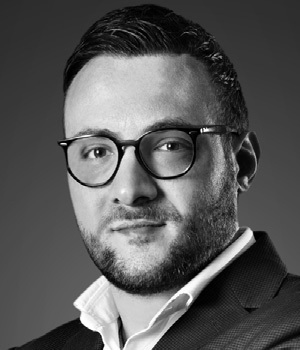}}]{Idio Guarino} is a postdoctoral researcher in the Department of Computer Science of the University of Verona since April 2024. He received his Ph.D.~in Information Technology and Electrical Engineering in February 2024 and his M.S.~degree in Computer Engineering in July 2020, both from the University of Napoli Federico II. His research focuses on analyzing network traffic via AI to develop innovative methodologies for network traffic classification and prediction.
\end{IEEEbiography}

\vskip -2\baselineskip plus -1fil
\begin{IEEEbiography}
[{\includegraphics[width=1in,clip,keepaspectratio]{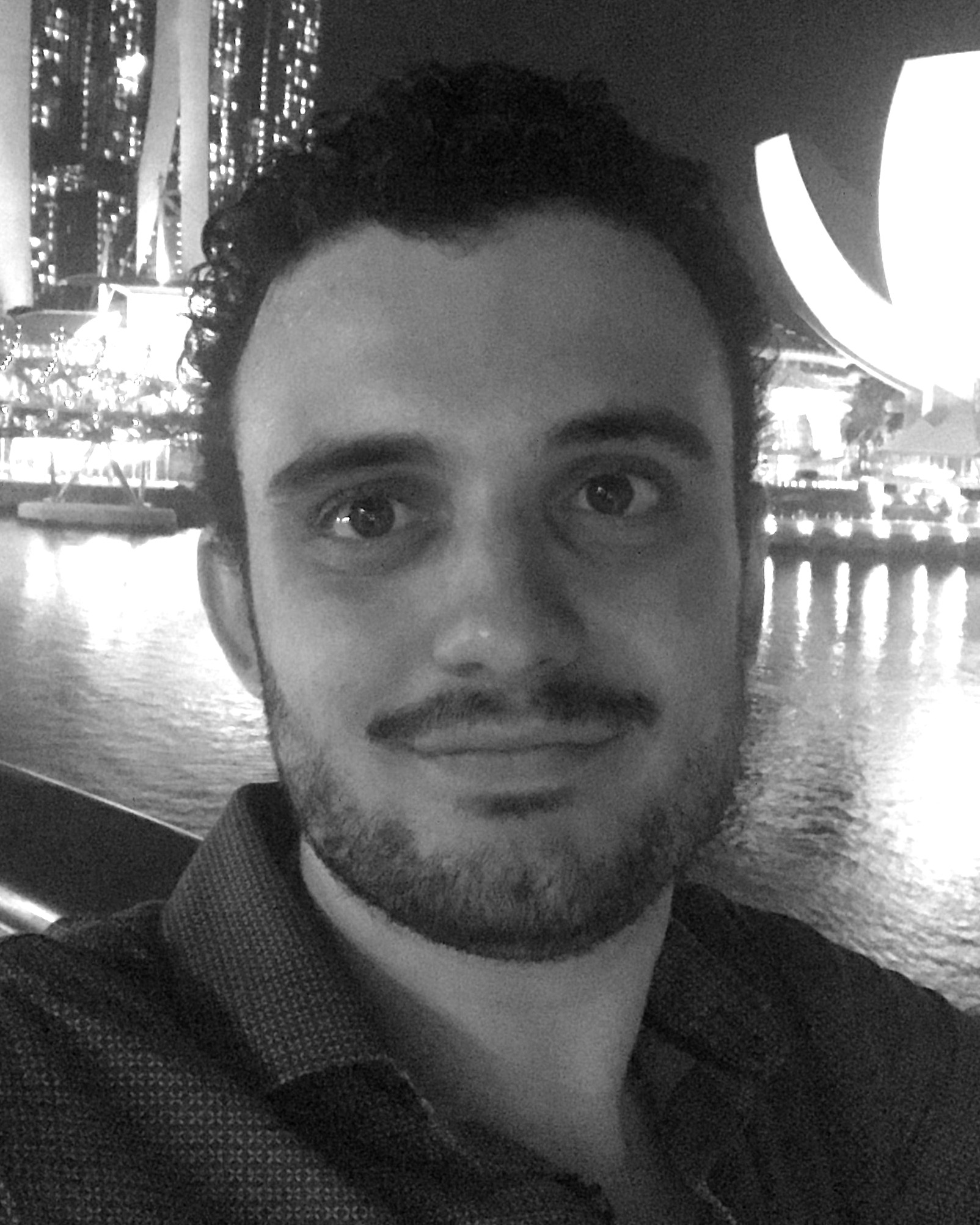}}]{Antonio Montieri} is an Assistant Professor at the University of Napoli Federico II, where he earned his Ph.D.~in Information Technology and Electrical Engineering in April 2020. His research focuses on network security, traffic classification, modeling and prediction, and explainable AI and generative AI for networking and traffic analysis. He has co-authored over 50 papers in leading international venues and received various awards, including the Computer Networks 2020 Best Paper Award.
\end{IEEEbiography}

\vskip -2\baselineskip plus -1fil
\begin{IEEEbiography}[{\includegraphics[width=1in,clip,keepaspectratio]{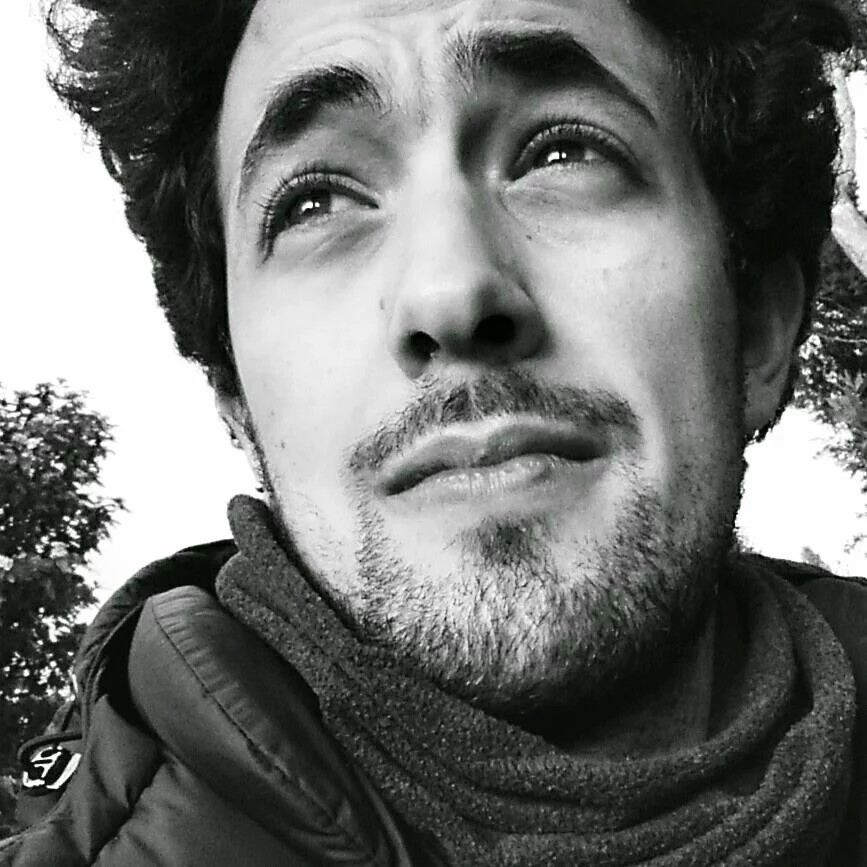}}] {Valerio Persico} is an Associate Professor at the University of Napoli Federico II, where he received the Ph.D.~in Computer and Automation Engineering in 2016. His work concerns network measurements, traffic analysis, cloud-network monitoring, and Internet path tracing. He has co-authored more than 70 papers within international journals and conference proceedings and is the recipient of several awards, including IEEE ISCC 2022, IEEE ICCCS 2019, and IEEE CSIM 2018 Best Paper awards.
\end{IEEEbiography}

\vskip -2\baselineskip plus -1fil
\begin{IEEEbiography}[{\includegraphics[width=1in,clip,keepaspectratio]{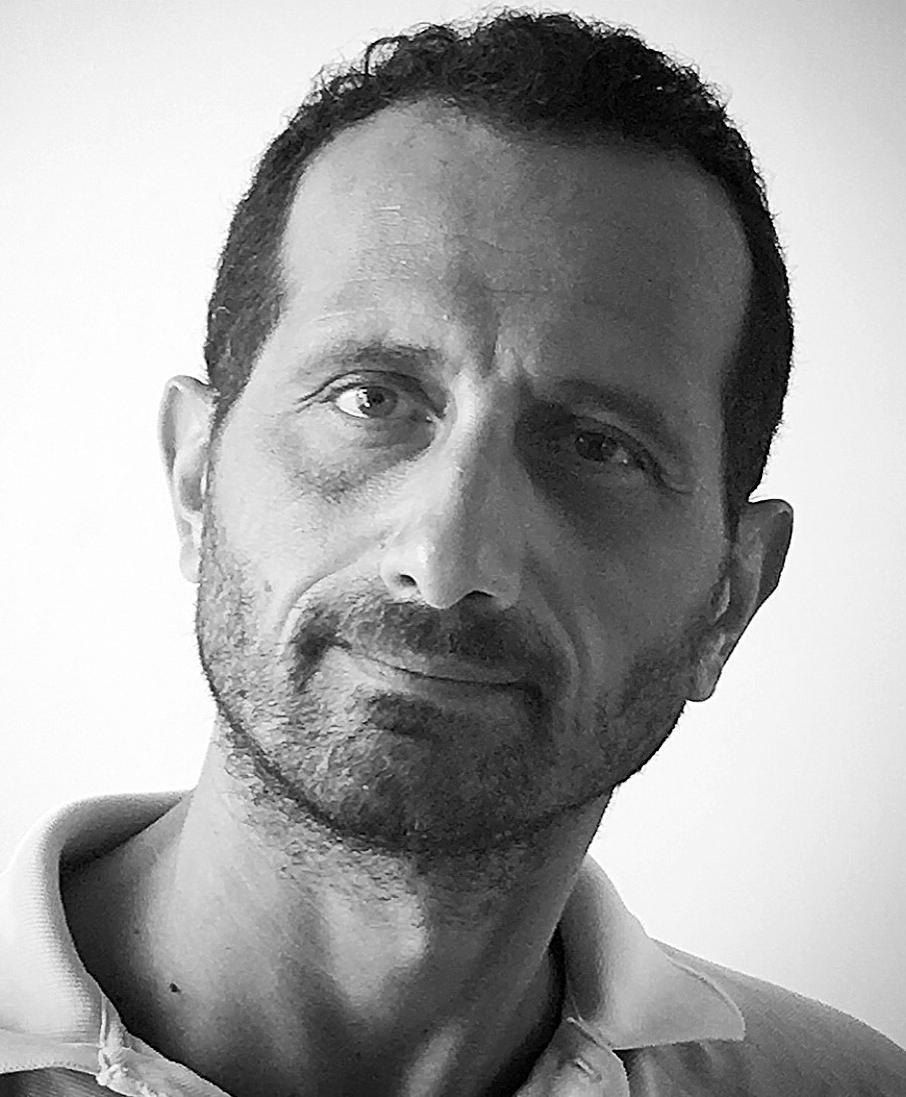}}] {Antonio Pescap\'e (SM’09)} is a Full Professor of Computer Engineering at the University of Napoli Federico II. His work focuses on Internet technologies, more precisely on measurement, monitoring, and analysis of the Internet, and on AI for networking. He has co-authored more than 200 conference and journal papers and he is the recipient of several research awards. Also, he has served as an independent reviewer/evaluator of research projects/project proposals co-funded by a number of governments and agencies.
\end{IEEEbiography}

\balance

\end{document}